\documentclass[12pt,MSc,wordcount,twoside,anon]{muthesis}
\usepackage{verbatim}
\usepackage{graphicx}
\usepackage{url} % typeset URL's reasonably
\usepackage{listings}
\usepackage{pslatex} % Use Postscript fonts
\usepackage{geometry}
\usepackage{multirow}
\usepackage{tablefootnote}
\usepackage{subcaption}
\usepackage{dirtytalk}
\usepackage{epigraph}
\usepackage{amsmath}%

\makeatletter
\def\@makechapterhead#1{%
  {\parindent \z@ \raggedright \normalfont
    \ifnum \c@secnumdepth >\m@ne
        \huge\bfseries \@chapapp\space \thechapter
        \par\nobreak
        \vskip 5\p@
    \fi
    \interlinepenalty\@M
    \Huge \bfseries #1\par\nobreak
    \vskip 10\p@
  }}
\def\@makeschapterhead#1{%
  {\parindent \z@ \raggedright
    \normalfont
    \interlinepenalty\@M
    \Huge \bfseries  #1\par\nobreak
    \vskip 20\p@
  }}
\makeatother

\begin{document}
% Uncomment the following lines to leave out list of figures, tables
% and copyright until final printing
%\figurespagefalse
%\tablespagefalse
%\copyrightfalse

\title{Automatic Detection of Deaths from Social Networking Sites}
\author{Nuhu Ibrahim}
\stuid{10723572}
\principaladviser{Riza Batista-Navarro, PhD}

\beforeabstract

\prefacesection{Abstract}
\abstracttitle
This dissertation analysed and discussed the differences between the linguistic characteristics and practices in pre-mortem social media contents and their post-mortem counterparts and reported machine learning classifiers that achieved high performance in automatically detecting deaths of social networking sites users from the posts associated with their profiles. A new dataset was developed by leveraging open-source platforms: Wikidata and Twitter. Machine learning models, both traditional: RF, KNN, LR and SVM, and deep learning: BiLSTM, CNN, and the state-of-the-art BERT were trained on features extracted using varieties of techniques: TF-IDF and famous pre-trained embeddings (Glove, Word2Vec, and Fasttext) to classify post-mortem contents from their pre-mortem counterparts. The results obtained showed that, for the models trained using the traditional machine learning algorithms, RF models outperform all others. For the models trained using the deep learning machine learning algorithms, the BiLSTM models outperformed the CNN models. In addition, TF-IDF consistently outperformed pre-trained word embeddings for all the trained traditional machine learning-based models, and Word2Vec consistently outperformed Glove and Fasttext for all the trained deep learning-based models. Overall, the state-of-the-art BERT outperformed all other models that were trained in this experiment. A comprehensive comparison between the similarities and differences of the linguistic practices in post-mortem tweets and their pre-mortem counterparts was conducted. It was found that although pre-mortem and post-mortem tweets express almost the same extent of positive sentiments, post-mortem tweets express higher negative sentiments while the pre-mortem tweets express higher neutral sentiments. Also, feelings that suggest negativity, i.e., sad, angry, surprise, and fear, are more dominant in post-mortem tweets than in pre-mortem tweets, and the feeling that suggests positivity, i.e., happy, is more dominant in pre-mortem tweets than in post-mortem tweets. It was also found that the number of words, personal pronouns, verbs, family words, religious words, death words, and swear words in post-mortem tweets are higher than those in pre-mortem tweets; whereas, the number of impersonal pronouns and informal words in pre-mortem tweets are higher than those in post-mortem tweets. Additionally, analytical thinking is more expressed in post-mortem conversations than in pre-mortem conversations. This experiment's significant contribution is the successful development of an incredibly high performing technique for automatically detecting deaths of users of social networking sites from the posts associated with their profiles. This technique is a potential solution that can be used as part of the tools required for the creation and adoption of an international standard for transferring digital estates to the next-of-kin of Internet users who die a sudden death; hence, reducing the risks of subscribers dying and leaving their digital estates that are supposedly important to their relatives or friends in the coffers of online-service providers.

\afterabstract

\prefacesection{Acknowledgements}
I want to express a special appreciation to my supervisor, Riza Batista-Navarro, PhD. Her incredible support, mentorship and guidance throughout this Masters program and even beyond is second to none.\\\\ 
I must also thank my amazing friends, Mashud Abdulsalam and Kabiru Sani, for helping me to manually annotate the collected dataset to enable me to measure the extent of confidence I can have in the automatically annotated dataset. I am sincerely grateful.
\afterpreface

% These include the actual text
\chapter{Introduction}
\label{cha:intro}

\epigraph{``Social media is changing the world, and we are all here witnessing it"}{Iam Somerhalder}

The \emph{Internet} is a medium and mechanism for global communication, interaction and collaboration between individuals and their computers, regardless of their geographical location \cite{leiner2009brief}. The Internet, which originally served as an interconnection of laboratories engaged in government researches, has since 1994 expanded to serve billions of users for numerous purposes all around the world. 

Internet World Stat \cite{internetgrowth2021} attributed the interminable evolution of the Internet to two key factors: social networking platforms/sites (SNP/SNS) and mobile technologies. These two innovations have changed the way people use the Internet. On SNSs, people have found a new way to communicate, interact and consume information; mobile technology, on the other hand, has made possible much cheaper, accessible and efficient access to the Internet; thereby, increasing the number of Internet users all around the world.

\begin{figure}[!ht]
    \begin{center}
        \includegraphics[width=15cm, height=10cm]{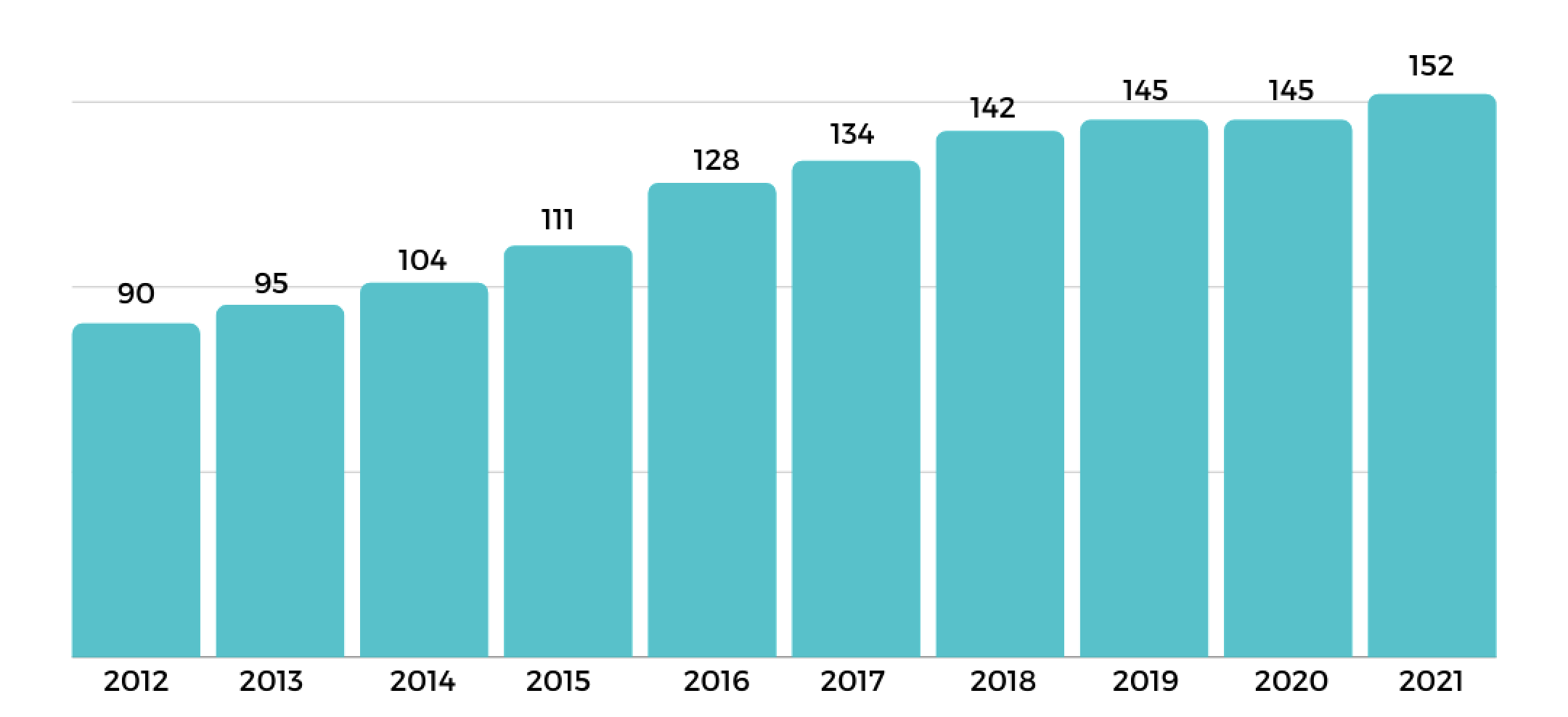}
    \end{center}
    \caption[Average minutes per day spent on SNSs (2012 - 2021)]{Average minutes per day spent on SNSs (2012 - 2021) (Taken from: FameMass \cite{famemass2021})}
    \label{chap1:internet_usage}  
\end{figure}

Boyd and Ellison \cite{boyd2007social} defined \emph{SNSs} as online platforms that allow individuals to “construct a public or semi-public profile” and connect with others within a “bounded system”. SNSs have continued to gain significant popularity from the late 20th through to the 21st century \cite{hillis2018digitalizing} and have grown to become an inextricable part of life as they keep people company between leaving their beds at dawn and returning to them at dusk. According to a statistic by TechJury \cite{techjury2021}, in 2020, 50.1\% of all the time spent on mobiles was on SNSs, and 83\% of internet users use at least one SNS. Figure \ref{chap1:internet_usage} shows the yearly increase in the average minutes per day spent on SNSs by internet users between 2012 and 2021.  

Due to the increasing relevance of SNS-based interactions to many people's social life, significant events including but not limited to graduations, marriage proposals, marriages, childbirths, etc., are now experienced in part through SNSs \cite{brubaker2011language}. This remains true even after the death of these SNSs' users, as friends, families and virtual connections often express their shock and grief on the deceased’s profile page following their death.

Buglass \cite{buglass2010grief} explained that though the terms \emph{grief}, \emph{mourning} and \emph{bereavement} are frequently used interchangeably, they have different meanings. He defined grief as an individual's response to loss that has ``emotional, physical, behavioural, cognitive, social and spiritual dimensions"; whereas, mourning as ``the outward and active expression of that grief"; and bereavement as ``the period after loss during which grief and mourning occur" \cite{buglass2010grief}.

Gathman \cite{gathman2014everybody} suggests that users may use SNSs for grieving or disclosing deaths to avoid the discomfort of having to individually announce the loss to every single person or tell the news over and over again. Dickinson \cite{dickinson2011shared} notes that allowing others to partake in grief, even if they are merely strangers, helps consoling those in grief. Besides, receiving consolations over SNSs also helps the bereaved feel that they are not alone \cite{katims2010grieving}.

Instagram, Facebook, and Twitter are among some of the most popular SNSs around the world \cite{chaffey2016global, mejova2015twitter}, with Twitter being one of the most utilised for academic research \cite{weller2014twitter}. This is because, ``Twitter data, in comparison with other SNSs such as Facebook, are more openly accessible and, a proportion of tweets can contain valuable metadata, including geospatial data, such as the precise latitude and longitude coordinates from which a tweet was posted" \cite{ahmed2017using}. 

\emph{Twitter} – an SNS that supports people to connect to others and follow their stream of posts, often referred to as ``Tweet", no more than 280 characters long \cite{twitterdevcounting}, as at the second quarter of 2021 (Q1'2021), has over 206 million \emph{monetisable daily active users (mDAU)}, with over 500 million tweets being sent every day \cite{twitterincq22021}. Twitter \cite{twitterincq22021} defines mDAU as ``people, organisations, or other accounts who logged in or were otherwise authenticated and accessed Twitter on any given day through twitter.com or Twitter applications that are able to show ads". 

\emph{Tweets} contain many structured and unstructured data that can be mined to gain intelligent insight into participants’ attitudes and behavioural responses in numerous situations \cite{chew2010pandemics}. Twitter allows researchers, developers or businesses to automatically retrieve Tweets and associated metadata for free through the Standard Application Programming Interface (API), or at a fee through the Premium or Enterprise APIs \cite{twitterdevapi}. This makes mining and analysing Twitter, and many other SNSs' data a promising research area as it presents significant opportunities to researchers to proffer solutions to some of the known most challenging problems, e.g., racism, crime, suicide, depression, etc., and it inhibits potentials to be a source of solutions to even unknown issues.

\section{Motivation}
Though the use of the Internet has been growing year on year since it became available to the public in the 1980s \cite{cerf1993internet}; however, the COVID-19 pandemic has caused a further conspicuous spike in the use of the Internet, and the data created, captured, copied, and consumed \cite{feldmann2020lockdown}. This data is continuously composing into larger series of digital assets for all Internet users \cite{walker2011cyberspace}. Digital assets may include digital images or videos; subscribed financial, cryptocurrency or cloud services; electronic bank and investment account statements; e-mail records and associated passwords; and SNSs' accounts \cite{perrone2012happens}. The combination of these digital assets forms a person's digital estate \cite{hopkins2013afterlife}. 

Sadly, despite this vast acceptance and use of the Internet, there is still no working standard for transferring digital estates to the next-of-kin of Internet users who die a sudden death. Currently, financial institutions, for instance, still depend on clients’ relatives to manually report their deaths before their estates get appropriately passed on. Also, most SNSs rely on people to report the death of a user \cite{jiang2018tending}. For example, Facebook supports its users to nominate another user who should be allowed access to change their profile after they die \cite{mccallig2014facebook}. But this takes months or sometimes forever, because some relatives are unaware of all the services that the deceased was subscribed to. 

Besides, some cloud platforms have also implemented measures to facilitate the transfer of data to others upon detection of the possibility of death. For example, Google implemented an Inactive Account Manager to help users to share parts of their account data or notify someone if they have been inactive for a certain period \cite{google2021}. This method is flawed because it would take some user-specified amount of time to finally conclude that a client is dead, and even though inactivity provides a helpful signal to detect if someone is dead, it is not reliable as inactivity could be caused by many different reasons other than death. 

The increased use of these Internet-based services (online financial and cloud services) because of the vast ``stay at home" COVID-19 national directives around the world despite regulators' failure to properly manage digital estates of the deceased has increased the risks of subscribers dying and leaving their digital estates that are supposedly important to their relatives or friends in the coffers of online-service providers. 

The growth in the use of SNSs encouraged a linguistic study of the wall posts of memorialised SNSs profiles where it was found that condolences wall posts show higher rates of negative emotion than their regular equivalents \cite{getty2011said, brubaker2012grief, brubaker2018describing}. These studies further suggest that social network profiles can be carefully studied and analysed to find generalisable conditions that could be used to detect whether the SNSs profiles’ owners are dead or still alive. Hence, shining a ray of hope that there could be a solution to the lack of a mechanism for automatically detecting service subscribers' deaths to facilitate the transfer of their digital estates to next-of-kins.

\section{Aim and Objectives}
This project aims to develop \emph{natural language processing (NLP)} based methods to analyze differences in the linguistic characteristics of pre- and post-mortem posts associated with deceased SNSs users' profiles. Then, different text classification models, both traditional- and deep learning-based machine learning algorithms, will be investigated and applied to achieve a real-time and automatic detection of deaths of SNSs users from the posts associated with their profiles. The investigation of these wide ranges of text classification models will encourage discovering their individual strengths and weaknesses; and, overall, the collective differences between the strength and effectiveness of traditional and deep learning machine learning algorithms.\\\\
To achieve the aim described above, the following objectives were set out: 

\begin{enumerate}
    \item Review related works on the use of NLP-based text mining for social media analytic and classification of annotated texts, and study privacy rights associated with social media data while taking note of the issue of informed consent;
    \item Mine deceased celebrities' SNSs data from open source repositories, mine pre- and post-mortem SNSs' posts associated with these data, and estimate agreement between the silver standard dataset (labelled using the dates posts were created and the date the celebrities died) and the manual annotations (labelled by human annotators by reading the content of the posts);
    \item Preprocess the dataset by applying a sequence of preprocessing steps to bring the texts into a normalised form and use NLP based methods to analyse the differences in the linguistic characteristics of posts associated with celebrities' before and after their demise;
    \item Extract features, perform and report exploratory analysis on the dataset, and train, evaluate and interpret the performance of different text classification models.
\end{enumerate}

\section{Research Questions}
\label{cha1:research_questions}
The following questions will be addressed in this dissertation:
\begin{enumerate}
    \item Are there differences between the linguistic characteristics and practices in pre- and post-mortem posts, and to what extent can humans and text classification models detect post-mortem posts from their pre-mortem counterparts?

    \item Between text classification models in either the traditional- or deep learning- based machine learning algorithms groups and between various feature engineering techniques (famous pre-trained word embeddings or Term Frequency-Inverse Document Frequency), which ones will perform better in classifying post-mortem posts from their pre-mortem counterparts?
    
    \item Will the famous state-of-art Bidirectional Encoder Representations from Transformers (BERT) perform better than the other text classification models (traditional- and deep learning machine learning-based) in classifying post-mortem posts from their pre-mortem counterparts?
\end{enumerate}

\section{Dissertation Structure}
Chapter 2 provides thorough background knowledge to the general area of text mining (TM), natural language processing (NLP) and text classification research. Additionally, it provides brief information about Twitter, Tweets, Twitter API, and Wikidata, a free open source knowledge base and its powerful query language, SPARQL. Chapter 3 presents a brief of SNSs as adopted for grieving, a review of previous works that have attempted to either identify the differences or perform a classification between pre- and post-mortem language, and the ethical considerations regarding using open source and social media data in this experiment. In Chapter 4, the methodologies used in the dissertation are explained. This includes data collection, data preparation, manual annotation and estimation of inter-annotator agreement (IAA), dataset splitting, feature extraction from the developed dataset, models selection from both traditional- and deep learning- based machine learning algorithms, and their application to the task of classifying post-mortem posts from their pre-mortem counterparts to enable the automatic detection of deaths of SNSs users. Further, Chapter 4 describes the development environment, including hardware, choice of the programming language and programming packages adopted. Chapter 5 shows a detailed discussion and interpretation of experimental results. Finally, Chapter 6 closes the dissertation with a conclusion and suggestions for future works.

\chapter[Background Knowledge]{Background Knowledge}
\label{chapter2}
This chapter intends to provide the reader with some background knowledge on the different platforms, concepts, frameworks, and algorithms used or discussed in various parts of this dissertation. The sources of data, i.e., platforms, from where the data used in the experiments were mined are briefly introduced in Sections \ref{sec2:wikidata} and \ref{sec2:twitter}. Section \ref{sec2:text_preprocessing} describes the relevance of text pre-processing and introduces the procedures for pre-processing standard texts and those for pre-processing texts from SNSs. Section \ref{sec2:npm_tm} provides background on NLP based text mining and describes the core tasks explored in the experiment: text classification, sentiment analysis, and emotion analysis. Further, Sections \ref{cha:machine_learning}, \ref{cha2:feature_engg} and \ref{sec2:bert} discuss the machine learning algorithms, techniques for feature engineering and state-of-the-art Bidirectional Encoder Representations from Transformers (BERT), respectively, that were employed to solve the task of text classification; in most cases, the strengths and limitations of some of these algorithms, tools and techniques are briefly discussed. Finally, Section \ref{sec2:evaluation_metrics} introduces numerous evaluation metrics that were used to measure the quality and effectiveness of the results obtained from the experiments.

\section{Wikidata and SPARQL}
\label{sec2:wikidata}
\begin{figure}[!ht]
    \begin{center}
        \includegraphics[width=15cm, height=10cm]{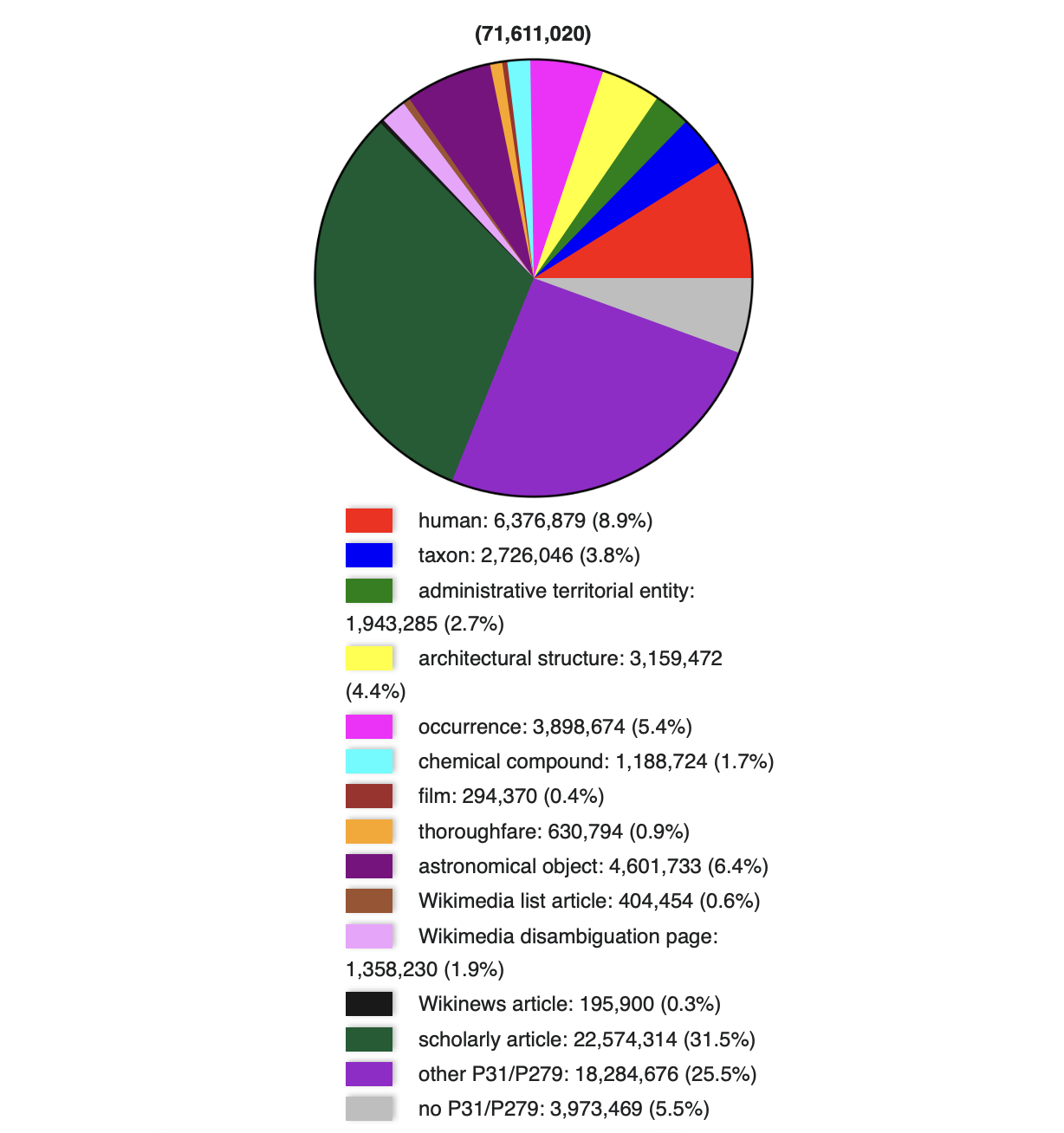}
    \end{center}
    \caption[Distribution of items on Wikidata as at February 2020]{Distribution of items on Wikidata as at February 2020 (Taken from: Wikidata)}
    \label{chap2:wikidata}  
\end{figure}

\emph{Wikimedia} launched \emph{Wikidata} in October 2012 to provide well-maintained and high-quality data for a significant purpose of being used within other Wikimedia projects \cite{wiki2021data}. Wikidata started with limited features that allowed editors only to create items and connect them to Wikipedia articles but has quickly evolved into a central data management platform of Wikipedia that has produced an open and extensive knowledge base with many exciting applications \cite{erxleben2014introducing}. As of August 2021, Wikidata contains more than 94 million data items managed by over 23 thousand active users worldwide. \footnote{See https://www.wikidata.org/wiki/Wikidata:Statistics} Figure \ref{chap2:wikidata} shows the distribution of data objects on Wikidata as of February 2020.

Wikidata is one of the essential websites today, and this is owed to the design decisions that characterised its approach; this includes open editing, community control, plurality (acceptance of conflicting data), multilingual data, easy access, secondary data (gathering facts published in primary sources), etc. \cite{vrandevcic2012wikidata}. The information collected by Wikidata can be browsed, queried, and even edited by applications in real-time through the Wikidata API. Another efficient way of accessing the information in Wikidata is through the Wikidata Query Service (WDQS). This powerful tool, WDQS, could be queried using SPARQL to gain insight into Wikidata's content.  

\emph{SPARQL}, short for “SPARQL Protocol and RDF Query Language”, is one of the most powerful and most widely used query languages for querying Resource Description Framework (RDF) databases, i.e., graph-shaped data model and Linked Open Data \cite{bielefeldt2018practical}. Unlike the typical Structured Query Language (SQL) for relational databases, SPARQL queries are not constrained to work only within a single database, i.e., SPARQL federated queries can be used to access multiple databases at the same time.

SPARQL manipulates data stores as a directed and labelled graph, internally expressed as triples consisting of subject, predicate, and object. Therefore, a SPARQL query consists a set of triple patterns in which either of the elements (the subject, object or predicate) can be represented with a wildcard. Below is a sample SPARQL query that attempts to extract from the English version of Wikidata the name, Twitter username \emph{(wdt:P2002)} and country of citizenship \emph{(wdt:P27)} of 10 instances \emph{(wdt:P31)} of humans \emph{(wd:Q5)} that are linked to the occupation \emph{(wdt:P106)},  musician \emph{(wd:Q639669)}.

\begin{lstlisting}[captionpos=b, caption={SPARQL query that extracts the name, Twitter username and country of origin of 10 musicians from the English version of Wikidata}, label=lst:sparql,frame=single, breaklines=true,]
    SELECT DISTINCT ?personLabel ?twitterUsername ?countryLabel 
    WHERE {
      ?person wdt:P31 wd:Q5 .
      ?person wdt:P106 wd:Q639669.
      ?person wdt:P2002 ?twitterUsername.
      ?person wdt:P27 ?country .
      SERVICE wikibase:label { bd:serviceParam wikibase:language ``[AUTO_LANGUAGE],en". }
    }
    LIMIT 10
\end{lstlisting}

\section{Twitter, Tweets, and Twitter API}
\label{sec2:twitter}
Purohit et al. \cite{purohit2013kind} described Twitter as a ``microblogging service (or platform) that provides a social network structure and a medium for information flow, where users post updates and subscribe to (referred to as ‘following’) other users to receive updates (microblogs)". Twitter, as of Q1'2021, reported having 206 million mDAU, with over 500 million tweets being sent every day \cite{twitterincq22021}. Twitter is popularly known to attract more researchers than other similar SNSs such as Facebook or Instagram because its data is more easily and openly accessible. Some key features in Twitter include:

\begin{enumerate}
    \item \textbf{Tweet: }a short post from a user of Twitter that is no more than 280 characters long. Tweets are usually made up of significant writing informalities, like abbreviations, phonetic substitutions, emoticons, emojis and ungrammatical structures; users use these to properly express themselves, notwithstanding the 280 characters restriction.
    \item \textbf{Username: }also known as \textbf{``handle"}, is a unique identifier for every user account on Twitter. It begins with the ``@" symbol and can be used to search for any user on the Twitter platform.
    \item \textbf{Hashtag: }is any contiguous characters, i.e., a word, with no space character in between that begins with the ``\#" symbol. Hashtag is a popular SNSs comprehensive convention used by users to widely discuss a topic; it is helpful for grouping conversations to a particular topic. 
    \item \textbf{Reply: }as the name signifies, it is a feature on Twitter that allows users to respond to an existing tweet with their own opinion or answer. Replies also can not be more than 280 characters long.
    \item \textbf{Mention: }allows an author of a tweet to call out other Twitter users in their tweet. Mentioning is done by using the ``@" symbol followed by a particular user's username. Users will usually get a notification when they are mentioned in other people's tweets.
    \item \textbf{Retweet: }when a user retweets a particular tweet, the tweet is forwarded to their followers. Retweeting on Twitter is similar to forwarding an email.
    \item \textbf{Followers: }the followers of a particular user are other users that have subscribed to receive the tweets, retweets, replies, or likes of that specific user in their feeds.
\end{enumerate}

Twitter allows researchers, developers or businesses to automatically retrieve Tweets and associated metadata for free through the Standard Application Programming Interface (API) or at a fee through the Premium or Enterprise APIs \cite{twitterdevapi}. These APIs also allow authorised researchers, developers or businesses to listen to and analyse public conversations and engage with people on Twitter.

\section{Text Pre-processing}
\label{sec2:text_preprocessing}
Pre-processing is the first step conducted after data collection and plays a vital role in text mining projects \cite{vijayarani2015preprocessing}. A pre-processing pipeline takes an input of raw text and returns cleaned tokens. Tokens, as used in this context, are single words, groups of words or delimiters (e.g., ``\&", ``@", ``\$" etc.) that serve as features in the analysis \cite{anandarajan2019text}. The stages along the pre-processing pipeline standardise the dataset into a normalised form that is predictable and analysable in the other text mining stages by removing unnecessary information, thereby reducing the number of dimensions in the dataset. Though this step reduces complexity in the dataset that benefits the text mining process by reducing the time and storage required for analysis, it also leads to a loss of information in the dataset. Therefore, there is a need to balance the retained information after pre-processing and the dimensionality reduction. Far more time is spent in a text mining task to collect, prepare and preprocess the text dataset than in the actual analysis \cite{dumais1998using} because a thorough and adequately carried out pre-processing makes the analysis process smoother and promises a more accurate result.

\subsection{Typical Text Pre-processing}
\label{sec2:typical_text_preprocessing}
The standard text pre-processing procedures includes Tokenisation, text data cleansing (case standardising and removal of numbers, punctuation and special characters), stop words removal, and stemming or lemmatisation \cite{anandarajan2019text}. They are described in detail below:

\begin{enumerate}
    \item \textbf{Tokenisation: }``is the process of breaking a stream of textual content up into words, terms, symbols, or some other meaningful elements called tokens" \cite{vijayarani2016text}. For example, the stream ``I love text mining and natural language processing (NLP)." may be tokenised into \{``I", ``love", ``text", ``mining", ``and", ``natural", ``language", ``processing", ``(", ``NLP", ``)", ``."\} Tokenisation is crucial because it helps to separate the content of a stream into individual entities so that they can be individually analysed to identify the meaningful ones. Also, many other stages in the pre-processing pipeline can only be carried out on already tokenised streams. 
    \item \textbf{Standardise Case: }is the process of converting all the tokens in the stream to their lowercase equivalent. This is done to ensure consistency in the dataset that will improve the accuracy of any analysis that would be further conducted. For instance, before case standardisation is completed, the words ``Language", ``LANGUAGE", and ``language" would be considered by most text mining tools as different tokens, and this will tamper with the accuracies and results obtained from these analyses. However, a downside to case standardisation is that words like ``US", i.e., the United States and ``us", i.e., a pronoun used by a speaker to refer to himself, herself or themselves would all be standardised into ``us" and this may also tamper the results in text analyses.
    \item \textbf{Removing numbers, punctuation and special characters: }following the conversion of tokens to consistent standardised cases, numerals (e.g., ``1", ``10", ``1.99", etc.), punctuation (e.g., ``,", ``!", ``.", etc.) and special characters (e.g., ``\$", ``£", ``@", etc.) are stripped from the text. 
    \item \textbf{Stop Words Removal: }A word is recognised as a stop word if it is commonly encountered in texts without dependency on a particular topic (e.g., conjunctions, prepositions, articles, etc.) \cite{uysal2014impact}. Since these words can be discovered in different texts irrespective of the topic, they are assumed to be irrelevant and are removed before the steps of the actual text analyses pipeline are conducted. 
    \item \textbf{Stemming: }is a procedure that reduces all words with the same stem (root) to the same form \cite{balakrishnan2014stemming}. For example, several forms of the root word ``come", e.g., ``comes", ``coming", ``came", or ``come" will be returned to the root word ``come".
    \item \textbf{Lemmatisation: }removes inflectional endings and returns the base or dictionary form of a word \cite{balakrishnan2014stemming}. Though stemming and lemmatisation are similar, lemmatisation is different. Unlike stemming, lemmatisation uses the part of speech of a word as used in a sentence as part of the decision-making criteria when attempting to reduce a word to its root form.
\end{enumerate}

\subsection{Tweets Pre-processing}
\label{sec2:tweets_preprocessing}
Tweets usually are short, and this means it will usually be required for users to include abbreviations, phonetic substitutions, emoticons, emojis and ungrammatical structures that torments text-processing tools \cite{sproat2001normalization}. Therefore, beyond typical text pre-processing steps such as tokenisation, case standardisation, removal of numbers, punctuation, special characters and stopwords, or stemming and lemmatisation, further steps for pre-processing Twitter data need to be taken. Some of these steps include:

\begin{enumerate}
    \item \textbf{Tweets anonymisation: }Twitter users are fond of mentioning others (adding their Twitter user- names) in their tweets. Researchers consider the removal of these usernames for two reasons; first is that it is desired to maintain the anonymity of the authors of tweets that are used in text analysis because of privacy concerns, and second is that it is not expected for text analysis tools to learn these usernames as part of features because they are not generalisable. After all, they are unique among all users.
    \item \textbf{Handling Hashtags: }a way to manage hashtags is to replace them with unique identifiers such as ``hash78909" in all their occurrences to be utilised as features \cite{gromann2017hashtag}. This is so that other special characters are processed without changing those in the hashtags. However, another widely used technique of handling hashtag is removing them. This is because they rarely significantly impact popular text analysis tasks such as sentiment analysis, emotion analysis, or text classification.
    \item \textbf{Handling URLs: }similar to hashtags, URLs can be replaced with unique identifiers such as ``URL12345" in all their occurrences so that they can be utilised as features \cite{gromann2017hashtag}. However, they are more frequently removed during pre-processing before text analysis because they have insignificant effects in numerous text analysis tasks such as sentiment analysis, emotion analysis, topic modelling or text classification.
    \item \textbf{Replacing abbreviations or slangs: }Twitter users often use abbreviations and slangs to express longer thoughts, notwithstanding the 280 characters restriction. However, these abbreviations and slangs torment text processing tools as their meanings are implicit or numerously represented within the text. Properly replacing these abbreviations with their complete forms or meanings may significantly affect the results obtained from text analysis \cite{dhuliawala2016slangnet}.
    \item \textbf{Removing accented characters: }accented characters are non-standard characters signs that change the sound of letters and words, e.g., ``Á", ``É", ``Í", ``Ó", ``Ú", ``Ý", etc. The performances of powerful text analysis tools are declined by accented characters \cite{franco2010accented}; hence, they are removed from the text or replaced with their closest equivalent in the conventional English characters.
    \item \textbf{Managing emojis: }Emojis are “picture characters” or pictographs that are widely adopted for simplifying the expression of emotions and for enriching communication on SNSs \cite{li2018multi}. Because Twitter users widely use them to express longer thoughts, notwithstanding the 280 characters restriction, their impact on text analysis and ways to manage them needs to be adequately studied. Prominent ways of managing emojis include: removing them from the text, replacing them with their meaning, or replacing them with unique identifiers. 
\end{enumerate}

\section{NLP Based Text Mining}
\label{sec2:npm_tm}
Andreas et al. \cite{hotho2005brief} noted that \emph{text mining} or \emph{knowledge discovery from text (KDT)} which is a significant area in Artificial Intelligence, was first mentioned in \cite{feldman1995knowledge}. It uses techniques from information retrieval (IR), information extraction (IE), and NLP while connecting them with data mining, machine learning, and statistics to perform machine-supported analysis on text. While an \emph{information retrieval system} is designed to analyse, process and store sources of information and retrieve those that match a particular user's requirements \cite{chowdhury2010introduction}, the goal of \emph{information extraction} methods is to extract specific information from text documents \cite{wilks1997information}. Besides, the general purpose of NLP is to achieve a better understanding of natural language by the use of computers \cite{kodratoff1999knowledge}.

\subsection{Text Classification}
\emph{Text classification} is a fundamental task with wide practical applicability (e.g., sentiment analysis, topic modelling, span detection, intent detection, etc.) in the Text Mining and NLP area due to the large number of text documents that we have to deal with daily. Text classification is a technique that assigns a set of known categories, often called ``class labels", to open-ended texts. We define the task of text classification as follows: given a set of documents $\{d_1, d_2, ..., d_n\}$ and a set of categories $\{c_1, c_2, ..., c_n\}$, text classification is the problem of assigning to each and every document one or more appropriate categories \cite{aggarwal2012survey}. Often, a specific training ($D_{train}$), documents with known class labels and test data ($D_{test}$), documents with hidden class labels are maintained. $D_{train}$ is then used to construct a classification model, which relates the features in the underlying records to one or more of the class labels. Further, the constructed classifier is used to predict the class labels of $D_{test}$. The predicted classes are then utilised to evaluate the efficiency of the constructed classifier by comparing them with the initially hidden category labels of $D_{test}$. 

Text classification can be performed both manually and automatically. The manual approach involves a human who interprets the content of texts and completes the assignment of appropriate categories. The manual method is known to yield good results due to humans underlying knowledge of context in language. Still, it is time-consuming and expensive when the documents that are to be classified are too large. The Automatic approach to text classification applies text mining, NLP, machine learning (ML), deep learning and any other appropriate artificial intelligence (AI) related techniques to automatically assign categories to text in a more efficient, cheaper and accurate way. 

Automatic text classification approaches are either rule-based, machine learning-based or hybrid. The rule-based text classification approaches construct a set of rules from the set of training documents and use these rules to assign class labels to the set of test documents \cite{vijayan2017comprehensive}. In contrast, the machine learning-based approaches learn to make classifications based on past observations instead of manually constructed rules \cite{kadhim2019survey}. This is achieved by using machine learning algorithms (see \ref{cha:machine_learning}) on an already labelled training dataset to learn the different associations between the pieces of texts in the training dataset and their associated categories. Before the activity of learning using machine learning algorithms takes place, feature engineering that is well explained in \ref{cha2:feature_engg} needs to have been conducted. However, the hybrid text classification approaches combine a machine learning-trained classifier with a rule-based system to improve the results further. A relative evaluation of different kinds of text classification methods may be found in \cite{hartmann2019comparing}.

Regarding categories, text classification approaches can be either binary text classification, multi-class text classification, or multi-label text classification. They are explained below:

\begin{enumerate}
    \item \textbf{Binary text classification: }is a kind of text classification where pieces of text in the dataset can only belong to two categories. By definition: If \emph{D} is a set of text documents $\{d_1,..., d_n\}$ and \emph{C} is a set of categories $\{c_1, ..., c_n\}$, a binary text classification assigns to all $d_i \in D $ a single category $c_i$, where $c_i \in C$ and $|C| = 2 $. Typical examples of problems in this category are: spam detection (i.e., spam or not) \cite{wu2017twitter}, sentiment analysis (i.e., negative or positive) \cite{mouthami2013sentiment}, classifying post-mortem contents on SNSs (i.e., post-mortem or pre-mortem) \cite{jiang2018tending, ma2017write}, etc. Binary classification tasks involve one class, the normal state, and another class, the abnormal state.
    \item \textbf{Multi-class text classification: }is a kind of text classification where pieces of text in the dataset can belong to any known specific \emph{n} number of categories, where \emph{n} is greater than 2 (i.e., $n > 2$). By definition: If \emph{D} is a set of text documents $\{d_1,..., d_n\}$ and \emph{C} is a set of categories $\{c_1, ..., c_n\}$, a multi-class text classification assigns to all $d_i \in D $ a category $c_i$, where $c_i \in C$ and $|C| > 2 $. Examples of problems in this category include: assigning textual customer complaints to multiple categories, detecting the type of hate in a hate speech, etc.
    \item \textbf{Multi-label text classification: }is a kind of text classification where there are two or more class labels, and one or more class labels may be assigned to any of the pieces of text in the dataset. By definition: If \emph{D} is a set of text documents $\{d_1,..., d_n\}$ and $C_i$ is a set of categories $\{c_1, ..., c_n\}$, a multi-label text classification assigns to all $d_i \in D $ some categories $C_j$, where $C_j \subseteq C_i $, $|C_i| >= 2 $ and $|C_j| >= 1 $. An example of problems in this category is detecting genres that a particular textual article belongs to when an article can belong to multiple genres.
\end{enumerate}

\subsection{Sentiment Analysis}
\emph{Sentiment analysis} or \emph{opinion mining} is the computational study of people's opinions, attitudes, responses and emotions toward entities; entities referred here may represent individuals, topics or events \cite{liu2012survey}. The two terminologies, sentiment analysis and opinion mining are often used interchangeably; however, some researchers have argued that they are slightly different \cite{tsytsarau2012survey}. Medhat et al. \cite{medhat2014sentiment} explained that opinion mining extracts and analyses people’s opinions about an entity while sentiment analysis identifies the sentiment expressed in a text then analyses it. Analysis of sentiments in texts is a binary text classification task because it attempts to decide the polarity, i.e., positive or negative, in a piece of text. 

Due to the dramatic growth of the Internet and SNSs, individuals, businesses and governments are increasingly depending on public opinions from these platforms for their decision-making processes. For example, businesses are now interested in knowing the opinions of customers about their products and services, governments are inclined to understand how citizens are reacting to policies and rules enforcement, and individuals want to know the opinion of a particular company's existing customers about the product or services they are considering to use.

\subsection{Emotion Analysis}
\label{cha2:emotion_analysis}
\begin{figure}[!ht]
    \begin{center}
        \includegraphics[width=15cm, height=10cm]{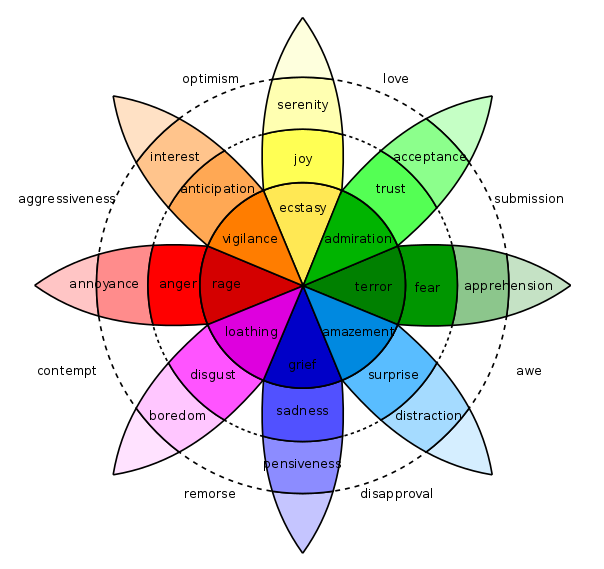}
    \end{center}
    \caption[Plutchik's emotions wheel]{Plutchik's emotions wheel (Taken from: Plutchik \cite{plutchik2001nature})}
    \label{chap2:Plutchik_emotion}  
\end{figure}

\emph{Emotion analysis} aims to detect and recognise the extent of different feelings (e.g., anger, disgust, fear, happiness, sadness, surprise, etc.) from texts. Though emotion analysis and sentiment analysis are similar, they are different in that sentiment analysis aims to detect whether the feelings represented in a text are either positive, negative, or neutral. Emotional categories are often divided into discrete emotion labels, e.g., Ekman's basic emotion model \cite{ekman1992argument}, which splits emotions into six classes, i.e., anger, disgust, fear, happiness, sadness and surprise, and Kellerman and Plutchik's bipolar emotion model \cite{kellerman1980emotion}, that divided emotions into eight classes, i.e., a superset of Ekman's model with two additional classes, trust and anticipation. Figure \ref{chap2:Plutchik_emotion} shows Plutchik's hybrid emotional wheel where the emotions are arranged into concentric circles, with the inner being the primary and the outer being more complex.

\section{Machine Learning}
\label{cha:machine_learning}
\emph{Machine Learning (ML)} is an area of \emph{artificial intelligence (AI)} and computer science concerned with the development of techniques that allow computers to ``learn” by the analysis of datasets \cite{hotho2005brief}. These datasets could be in the form of digitised human-labelled training data or other types of information obtained through machines' interactions with the real world. In all cases, the quality and size of the datasets are crucial to the predictions made by the machine learning model \cite{mohri2018foundations}. A typical example of a machine learning problem is how to use a finite set of randomly selected text documents with each labelled whether they contain hate speech, to accurately and efficiently predict whether other unseen text documents include hate speech. Obviously, the larger the size of the labelled dataset, the easier and more accurate the machine learning algorithm can learn. In addition, the quality and accuracy of a model also depend on the correctness of the labels that are assigned to the training dataset, the more the number of incorrect labels assigned to the text documents, the higher the possibility that the trained machine learning model will make wrong predictions about the presence or absence of hate speech in the unseen documents. Broadly, there are four major recognised categories of machine learning: supervised, unsupervised, semi-supervised and reinforcement learning. They are briefly described below:

\begin{enumerate}
    \item \textbf{Supervised learning: }also known as supervised machine learning, uses labelled datasets to train algorithms on classifying or predicting accurate outcomes. The problems solved using this category of machine learning usually have at least two distinct datasets, i.e., training and test dataset, where the training dataset is pre-labelled and used to fit the model to ensure that it learns the underlying associations between the different pieces of data in the dataset. Further, the trained model is then used to predict the labels of the test dataset; these predicted labels are then compared with the initially hidden labels to evaluate the model. Classification of spam in separated folders and detecting hate speech in text documents are examples of problems that can be solved using the supervised machine learning algorithms category.
    \item \textbf{Unsupervised learning: }also known as unsupervised machine learning, uses machine learning algorithms to analyse, organise, or cluster unlabelled datasets so that a human can make further sense of the outcome. The algorithms in this category are capable of discovering hidden patterns or data groupings without human intervention. This category of algorithms interests researchers because of their capability to make sense of unstructured data considering that the overwhelming majority of data in the world is unlabelled and unstructured. Understanding buying habits, grouping user logs or building a recommendation system are examples of problems that can be solved using the unsupervised machine learning algorithms category.
    \item \textbf{Semi-supervised learning: }is an approach to machine learning that combines labelled data and unlabelled data during training. It uses a smaller labelled dataset during training to guide learning on a larger unlabelled dataset. 
    \item \textbf{Reinforcement learning: }also known as reinforcement machine learning is a behavioral machine learning model that is similar to supervised machine learning but different because it is not trained on sample data set. Instead, it continuously learns using trials and error by comparing its decisions to the sequence of rules that are already clearly defined. Reinforcement learning has been applied in real-life industry to develop self-driving cars and NLP for text summarisation, question answering, and machine translation. 
\end{enumerate}

There have been substantial advancements in machine learning, but it boils down to two prevalent concepts: Traditional or Classical Machine Learning and Deep Learning. Although machine learning, deep learning, and traditional machine learning tend to be used interchangeably, the differences between them are worth noting. Machine learning, traditional machine learning and deep learning are all sub-fields of artificial intelligence; however, deep learning and traditional machine learning are sub-fields of machine learning. Deep learning and traditional machine learning algorithms only differ in how they learn. Deep learning automates much of the feature extraction processes, eliminates some of the manual human intervention required by traditional machine learning and enables the use of larger datasets. In contrast, traditional machine learning is more dependent on human intervention to learn, i.e., human experts may be needed to determine the set of features that the algorithm will use to analyse the underlying differences between data inputs.

\subsection{Traditional Machine Learning}
\label{cha2:traditional_learning}
Machine learning algorithms that use the fundamental algorithmic structures for solving problems are mostly considered \emph{traditional machine learning}. These machine learning algorithms are trained to learn the underlying differences between the data inputs in a dataset whose features are extracted by subject experts. More often, traditional machine learning models expect their input data to be in the form of structured data. Some machine learning algorithms that are considered to be examples of traditional machine learning algorithms are explained below.

\subsubsection{Random Forest (RF)}
\begin{figure}[!ht]
    \begin{center}
        \includegraphics[width=15cm, height=10cm]{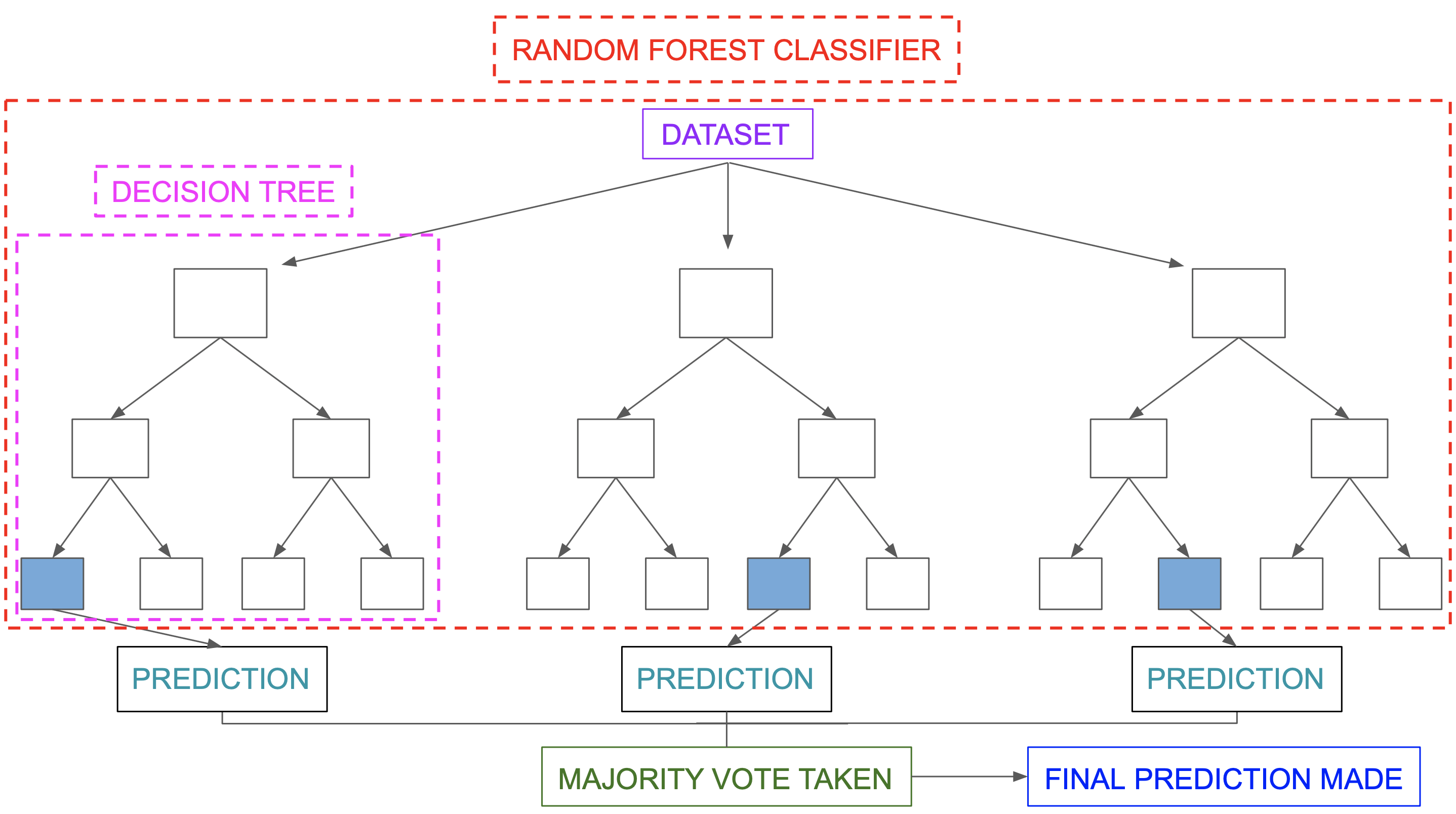}
    \end{center}
    \caption[Random Forest]{Random Forest (Taken from: Kashyap \cite{karan2021})\protect\footnotemark}
    \label{chap2:random_forest}  
\end{figure}

\emph{Random forests} or \emph{random decision forests} is a machine learning technique used to solve regression and classification problems. It is capable of solving text classification problems. It uses ensemble learning, i.e., a technique that combines multiple classifiers to solve complex problems. The main idea of RF is to use the random forest algorithm to generate random decision trees, then train them through bagging or bootstrap and finally aggregate and establish outcomes based on the predictions of the decision trees. Breiman \cite{breiman2001random} found convergence for RF as margin measures, \emph{(mg(X, Y))} as follows:

\begin{equation}
mg(X, Y) = av_k I(h_k(X) = Y) - \underset{j \neq Y}{maxav_k} I(h_k(X) = j)
\end{equation}
The margin function measures the extent to which the average number of votes at \emph{(X, Y)} for the correct class exceeds the average vote for any other class \cite{kulkarni2012pruning}. Figure \ref{chap2:random_forest} shows a visual description of the random forest algorithm.\\\\
\textbf{Limitation of Random Forests: }\\
Random forests run fast during training, even on large datasets compared to deep learning algorithms, but it is slow in making predictions from the trained model \cite{bansal2018social} because results are only gotten through voting from the predictions of all the decision trees that make up the forest. Thus, the number of trees in the forest must be reduced to achieve a fast prediction since more trees in the forest increases time complexity in the prediction step. Thus, it is an impractical choice in environments where predictions need to be made rapidly.

\footnotetext{See https://medium.com/analytics-vidhya/machine-learning-decision-trees-and-random-forest-classifiers-81422887a544}

\subsubsection{K-Nearest Neighbours (KNN)}
The \emph{K-Nearest Neighbours (KNN)} is a non-parametric machine learning technique that can be used for both classification and regression. The general concept of KNN is to classify data points based on those that are similar to them. For any data point \emph{x} whose category is to be predicted, the KNN algorithm finds the distance between \emph{x} and all the documents in the training set, identify the \emph{k} nearest neighbours of \emph{x}, and score the potential label of \emph{x} based on the labels of the \emph{k} neighbours. In regression, it may return the mean of the labels of the \emph{k} neighbours, and in classification, it may return the mode of the labels of the \emph{k} neighbours. Kowsari et al. \cite{kowsari2019text} described the decision rule of KNN as:

\begin{equation}
f(x) = arg\underset{j}{max}S(x,C_j)
= \sum_{d_i \in KNN} sim(x, d_i)y(d_i, C_j) 
\end{equation}

where \emph{S} is the score value with respect to \emph{$S(x, C_j)$}, i.e., the score value of candidate \emph{i} to the class of \emph{j}, and output of \emph{f(x)} is a label to the data point \emph{x}. To select the value of \emph{k} that is most appropriate for a dataset, the KNN algorithm runs several times with different values of \emph{k} until a value of \emph{k} that reduces the number of errors during training and also testing on unseen dataset if found. 

\begin{figure}
    \begin{center}
        \includegraphics[width=15cm, height=10cm]{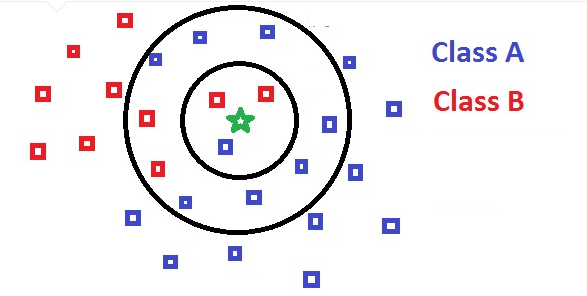}
    \end{center}
    \caption[K-Nearest Neighbours]{K-Nearest Neighbours (Reproduced from: Sanjay \cite{sanjay2021}) \protect\footnotemark}
    \label{chap2:knn}  
\end{figure}
\footnotetext{See https://towardsdatascience.com/knn-using-scikit-learn-c6bed765be75}

For example, as shown in Figure \ref{chap2:knn}, the data points in red and blue belong in classes \emph{A} and \emph{B}, respectively, while the data point in green is the data point of interest whose label is predicted. For \emph{k = 3}, the green data point will be predicted to be in class \emph{B} because the mode of the labels of the three nearest data points is class \emph{B}. However, if \emph{k = 5}, the green data point will be predicted to be in class \emph{A} because the mode of the labels of the five nearest data points is class \emph{A}.\\\\
\textbf{Limitation of KNN: }\\
KNN is easy to understand and implement and can easily handle multi-label problems. However, KNN is inefficient and slow when the dataset is large or contains high dimensional data points because the cost of calculating the distance between the new point and train points increases as the dataset becomes larger, and it becomes more challenging to find distances in high dimensions. In addition, KNN is sensitive to outliers in the dataset.

\subsubsection{Logistic Regression (LR)}
\emph{Logistic Regression} is a machine learning algorithm based on the concept of probability that can be used for classification problems. It predicts categorical dependent variables using a dataset of independent variables. The outcome of this prediction may be categorical or discrete, i.e., \emph{true} or \emph{false}, \emph{1} or \emph{0}, \emph{yes} or \emph{no}, \emph{A} or \emph{B}, etc. However, instead of producing a linear value by fitting a straight line or hyperplane, it uses the logistic function, i.e., sigmoid function (see equation \ref{cha2:logistic regression}), to squeeze the output of a linear equation (See equation \ref{cha2:logistic regression2}) to produce a probabilistic value that lies between 0 and 1.

\begin{equation}
\label{cha2:logistic regression2}
y^{(i)} = \beta_0 + \beta_1X_i^{(i)} + \beta_2X_2^{(i)} + \beta_3X_3^{(i)} + ... + \beta_pX_p^{(i)}
\end{equation}

\begin{equation}
\label{cha2:logistic regression}
logistic(y) = \dfrac{1}{ 1 + e^{(-y)}}
\end{equation}
\textbf{Limitation of Logistic Regression: }\\
Though logistic regression classifiers work well for predicting categorical outcomes, they require that each data point in the training dataset be independent \cite{huang2015unconstrained}. Also, sufficient training examples need to be available for all the categories for the logistic regression classifier to make correct predictions. 

\subsubsection{Support Vector Machine (SVM)}
\begin{figure}
    \begin{center}
        \includegraphics[width=15cm, height=10cm]{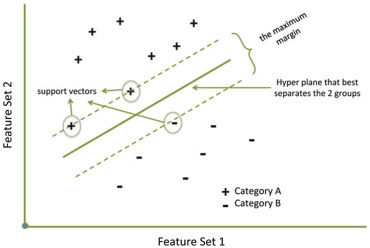}
    \end{center}
    \caption[Hyperplane and Support Vectors in SVM]{Hyperplane and Support Vectors in SVM (Reproduced from: Pisner \& Schnyer \cite{pisner2020support})}
    \label{chap2:support_vector}  
\end{figure}

\emph{Support Vector Machine} is a machine learning technique that can be used for both classification and regression problems. However, it is mostly used for classification problems; though initially designed for binary classification problems, it can now be extended for multi-class classification problems \cite{kowsari2019text}. The objective of the SVM is to find a hyperplane in an n-dimensional space, where n is the number of features that distinctly differentiates the data points base on their categories. \emph{Hyperplanes} are decision boundaries that help separate the datapoints based on their categories, i.e., data points on either side of the hyperplane belong in different classes \cite{noble2006support} (See Figure \ref{chap2:support_vector}). On the other hand, \emph{support vectors} are the closest data points to the hyperplane that influence the position and orientation of the hyperplane \cite{pisner2020support} (See Figure \ref{chap2:support_vector}). To find hyperplanes, the SVM algorithm plots the dataset features in an n-dimensional space where the value of each feature represents a coordinate; it then chooses the plane with the maximum margin, i.e., the maximum distance between data points of different categories. 

\begin{figure}
    \begin{center}
        \includegraphics[width=15cm, height=10cm]{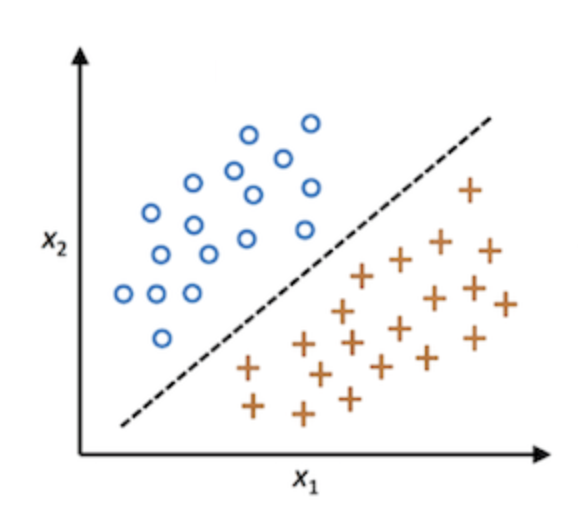}
    \end{center}
    \caption{SVM linearly separable vectors}
    \label{chap2:support_vector1}  
\end{figure}

\begin{figure}
    \begin{center}
        \includegraphics[width=15cm, height=10cm]{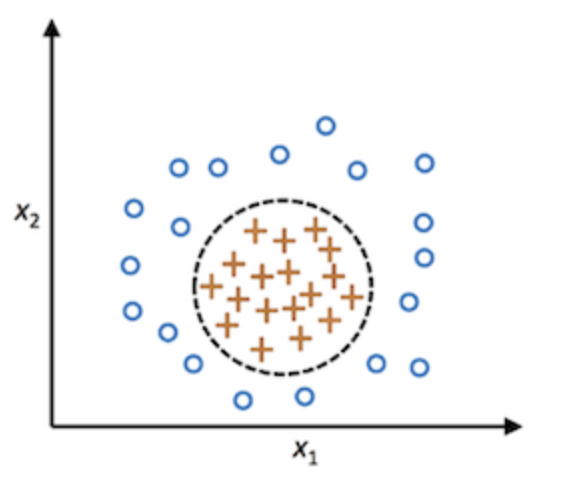}
    \end{center}
    \caption{SVM nonlinearly separable vectors}
    \label{chap2:support_vector2}  
\end{figure}

An SVM can be linear or nonlinear but is most commonly linear. Figure \ref{chap2:support_vector1} shows an example of a classification problem where the data points of different categories are located on different sides of the hyperplane and can be linearly separated. However, in a scenario where they can not be linearly separated (see Figure \ref{chap2:support_vector2}), the SVM applies the kernel method to transform the support vectors to a higher-dimensional space; therefore, converting the features from nonlinearly separable to linearly separable ones.

\subsection{Deep Learning}
\label{cha2:deep_learning}
Unlike traditional machine learning algorithms that require guidance or adjustments by human experts when they return inaccurate predictions, \emph{deep learning} algorithms can determine on their own if a prediction is accurate or not through their neural network. The design of an \emph{artificial neural network (ANN)} is inspired by the biological nervous system, such as the brain, leading to a learning process that’s far more capable than that of traditional machine learning models \cite{dongare2012introduction}. Deep learning can leverage labelled datasets to inform its algorithm, but it does not necessarily require a labelled dataset. It can ingest unstructured data in its raw form (e.g. text, images) and automatically determine the set of features that distinguish different categories of data from one another. Deep learning and neural networks are broadly applied and recognised for outstanding performance in computer vision, natural language processing, and speech recognition \cite{bhardwaj2018deep}. 

\begin{figure}
    \begin{center}
        \includegraphics[width=15cm, height=10cm]{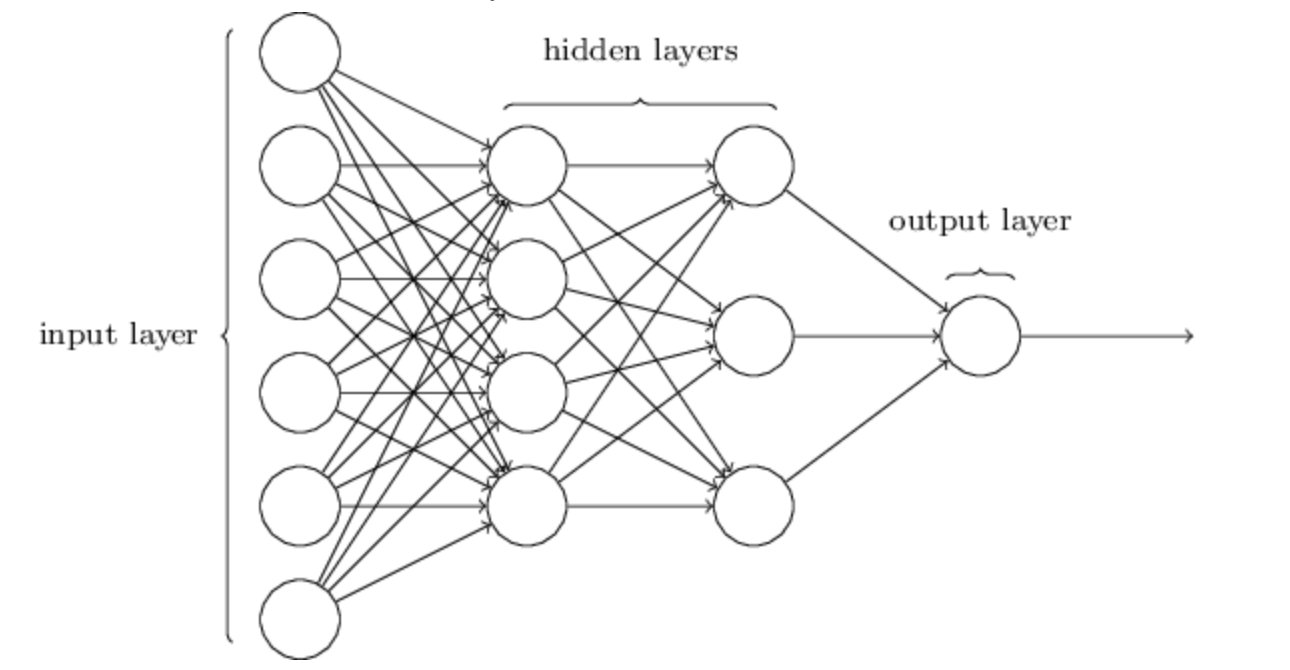}
    \end{center}
    \caption{A four-layer neural network with two hidden layer}
    \label{cha2:deep_learning2}  
\end{figure}

Deep neural networks learn through a connection of multiple layers, where every layer only receives connections from previous layers and only provides a connection to the next layer in the network \cite{kowsari2017hdltex}. Given a finite set of inputs, \emph{X} (where $X = \{x_1, ...., x_m\}$ and $|X| = m $), each input is multiplied by a weight (weights are $ \theta_1,...,\theta_m $), the sum of the weights multiplied by the inputs is computed (i.e., $ \sum_{i = 1}^m x_i \theta_i $), a bias, \emph{b}, is added and the result is passed through a non-linear activation function, \emph{g}, to produce an output, \emph{$ \overset{\wedge}{y} $} \cite{pamina2019survey} (See equation \ref{cha2:deep_learning_eqn}). Figure \ref{cha2:deep_learning2} depicts the structure of a four-layer neural network that has two hidden layers.

\begin{equation}
\label{cha2:deep_learning_eqn}
\overset{\wedge}{y} = g(b + \sum_{i = 1}^m x_i \theta_i)
\end{equation}
where \emph{$\overset{\wedge}{y}$} is the output, \emph{g} is the non-linear activation function, \emph{b} is the bias, and \emph{$\sum_{i = 1}^m x_i \theta_i$} is linear combination of inputs and weights.

\subsubsection{Bidirectional Long Short-Term Memory (BiLSTM)}
\begin{figure}[!ht]
    \begin{center}
        \includegraphics[width=15cm, height=10cm]{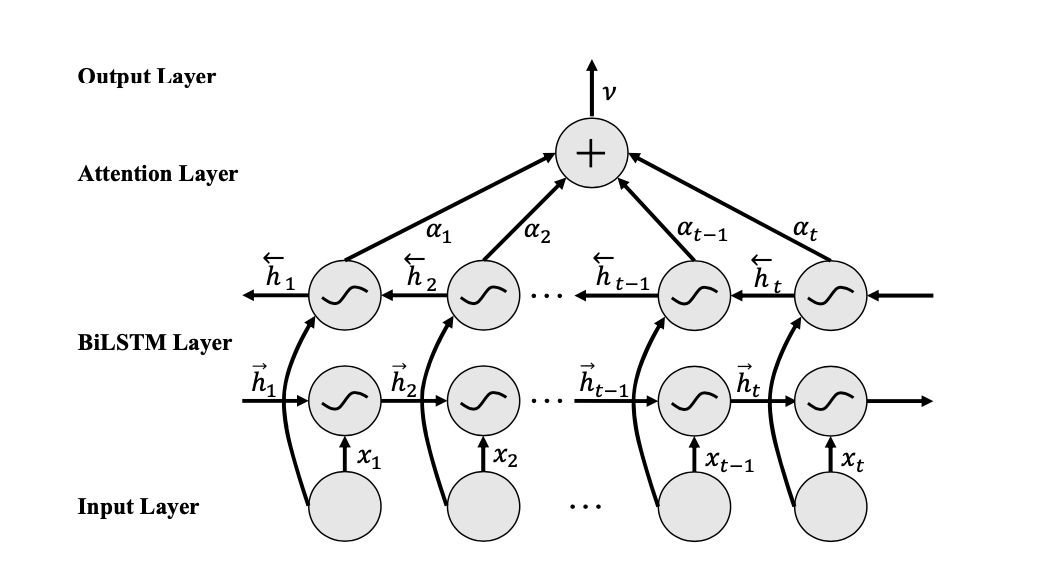}
    \end{center}
    \caption[Partial expansion of the BiLSTM network model with the attention mechanism along the time axis.]{Partial expansion of the BiLSTM network model with the attention mechanism along the time axis. (Taken from: Chen et al. \cite{chen2019research})}
    \label{cha2:bilstm}  
\end{figure}

Recurrent neural networks (RNNs) have been used by researchers for text mining and classification \cite{sutskever2011generating, mandic2001recurrent, wang2016combination} because it considers information of previous nodes in a comprehensive way which supports it to reach a profound semantic analysis of the underlying structure in a dataset. Despite its benefits, it is vulnerable to vanishing gradient problems when the network back propagates the errors of the gradient descent \cite{bengio1994learning}. Thus, \emph{LSTM} is a particular type of RNN that solves the vanishing gradient problem in RNN by using multiple gates to regulate the amount of information allowed into each node state. On the other hand, \emph{A Bidirectional LSTM}, or \emph{BiLSTM}, consists of two LSTMS: one taking inputs in a forward direction and the other in a backwards direction. This technique increases the available information in the BiLSTM network, improving the contextual knowledge that the network can use. Figure \ref{cha2:bilstm} shows a partial expansion of the BiLSTM network model with the attention mechanism along the time axis.

\subsubsection{Convolutional Neural Network (CNN)}
\emph{Convolutional neural network (CNN)} is a deep learning architecture initially built for image processing but is now commonly and effectively used for text document classification \cite{lecun1998gradient, zain2018convolutional, wang2018densely}. CNN have three main layers: convolutional layer, pooling layer, and fully-connected layer. The convolutional layer is the first layer of a convolutional network, followed by pooling layers or more convolutional layers, and the full-connected layer is the final layer. The convolutional layers are the building block of the network, as more computations are performed there; the pooling layers conduct dimensionality reduction to reduce the number of parameters in the input; and the fully-connected layer is responsible for performing the task of classification based on the features extracted from the previous layers by fully-connecting its layers directly to nodes in the previous layers.

\begin{figure}[!ht]
    \begin{center}
        \includegraphics[width=15cm, height=10cm]{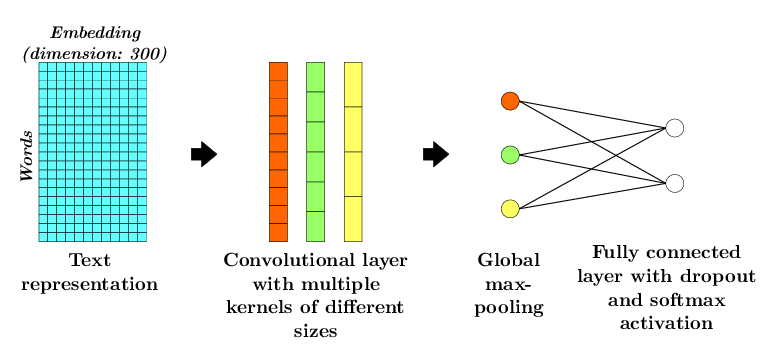}
    \end{center}
    \caption[Single-layer convolutional neural network model with different kernel sizes for text classification.]{Single-layer convolutional neural network model with different kernel sizes for text classification. (Taken from: Soll et al. \cite{soll2019evaluating})}
    \label{cha2:cnn}  
\end{figure}

A convolution is a basic application of a filter to input to obtain an activation. A feature map of activation that indicates the locations and strength of a detected feature in inputs like texts or images may be obtained through a repeated application of the same filter to input results. A usual problem for CNNs during text classification is the size of the feature space, e.g., the number of channels if it is an image or the dimension of a word-embedding in a text \cite{kowsari2017hdltex}. It is easier in image classification because images generally have few channels, e.g., only one channel in black and white images or at most three channels in RGB images; however, it may be in high dimensions for text classification applications because the size of the feature space may be large \cite{johnson2014effective}. Figure \ref{cha2:cnn} represents a single-layer convolutional neural network model with different kernel sizes that receive a word embedding of 300 dimensions in the input layer. 

\section{Feature Engineering}
\label{cha2:feature_engg}
Texts and documents are unstructured datasets; thus, \emph{feature engineering} is required to convert them into a structured feature space or vectorised format, i.e., numeric representations, so that machine learning algorithms can understand them. There are numerous techniques for feature extraction, e.g., Term Frequency-Inverse Document Frequency (TF-IDF) \cite{ramos2003using}, bag-of-words (BoW) \cite{harris1954distributional}, pre-trained word embeddings (e.g., Word2Vec, fasttext, GloVe), etc. Some of these techniques are briefly introduced below. 

\subsection{Term Frequency-Inverse Document Frequency (TF-IDF)}
\emph{Term Frequency-Inverse Document Frequency (TF-IDF)}, which is a combination of two values: \emph{term frequency (TF)} and \emph{inverse document frequency (IDF)}, is a statistical measure that estimates the relevance of a word to a document in a collection of documents. IDF was proposed by Jones \cite{jones1972statistical} as a novel method to be used together with the frequency of terms in documents, i.e., term frequency, to lessen the effect of common words, i.e., stop words in datasets. The weight of a word in a document in TF-IDF is mathematically represented as:

\begin{equation}
    \label{cha2:fasttext1}
    W(d,t) = TF(d, t) * log (\dfrac{N}{df(t)})
\end{equation}

where \emph{N} is the number of documents and \emph{df(t)} is the number of documents containing the term \emph{t} in the corpus. The first term, i.e., \emph{TF(d, t)}, improves the recall, while the second term, i.e, \emph{$log (\dfrac{N}{df(t)})$} improves the precision \cite{tokunaga1994text}. Even though TF-IDF tries to overcome the problem of common terms in the document using the IDF, a limitation is that it still cannot account for the similarity between the words in the document since each word is independently presented as an index \cite{kowsari2019text}. 

\subsection{Bag of Words (BoW)}
The \emph{bag-of-words (BoW)} model is a simplifying representation used in NLP and IR that has also been applied to several other domains like computer vision \cite{sivic2008efficient}. The model represents a text document by simply selecting parts of it based on some basic criteria, such as word frequency. This model ignores grammar, word order, or word context in a text document but keeps its multiplicity. Because the BoW model maintains the frequency of occurrence of each word, it is popularly used in document classification tasks where the occurrences of the words are used as features for training a classifier. In a BoW, texts are thought of like a bag of word vectors \cite{kowsari2017hdltex}. Below is an example of the BoW: \\\\
\textbf{Text Document: }\\
Emmanuel loves reading motivational quotes. Jane loves reading motivational quotes too. Jane also loves reading NLP books.\\\\
\textbf{Bag-of-Words (BoW) vocabulary: }\\
(``Emmanuel", ``loves", ``reading", ``motivational", ``quotes", ``Jane", ``too", ``also", ``NLP", ``books")\\\\
\textbf{Bag-of-Feature (BoF): }\\
(1, 3, 3, 2, 2, 2, 1, 1, 1, 1)\\

The drawback of the BoW is that, as new sentences that contain unknown words are added, the vocabulary would increase, and so is the size of the bag-of-feature. Also, the bag-of-feature retains no information about the grammars in the sentences nor context and order of words.

\subsection{Word Embedding}
\emph{Word embedding} is a technique used to represent words in a text in the form of real-valued vectors that embody the meaning of those words so that words that are closer in meaning in the text are expected to be closer in the vector space \cite{jurafskyspeech}. To obtain the word embeddings of a vocabulary, a set of language modelling or feature learning techniques may be used to map the words in the vocabulary to vectors of real numbers. Various word embedding methods that translate single words, i.e., ``unigrams" into understandable and relatable vectors that can be fed into deep learning or even traditional machine learning algorithms have been proposed; this dissertation will focus on three of the most common methods that have been successfully used for both deep learning and traditional machine learning techniques, i.e., Word2Vec, Glove, and Fasttext \cite{kowsari2017hdltex}.

One main limitation of using word vectors to represent text vocabulary is that words with many meanings in the vocabulary are represented using the same vector, i.e., homonyms are not adequately handled. For example, ``building" will have the same Word2Vec or Glove vector representation for both its occurrences in the sentences ``He is building his house" and ``He is building his muscle".

\subsubsection{Word2Vec}
\begin{figure}
    \begin{center}
        \includegraphics[width=15cm, height=10cm]{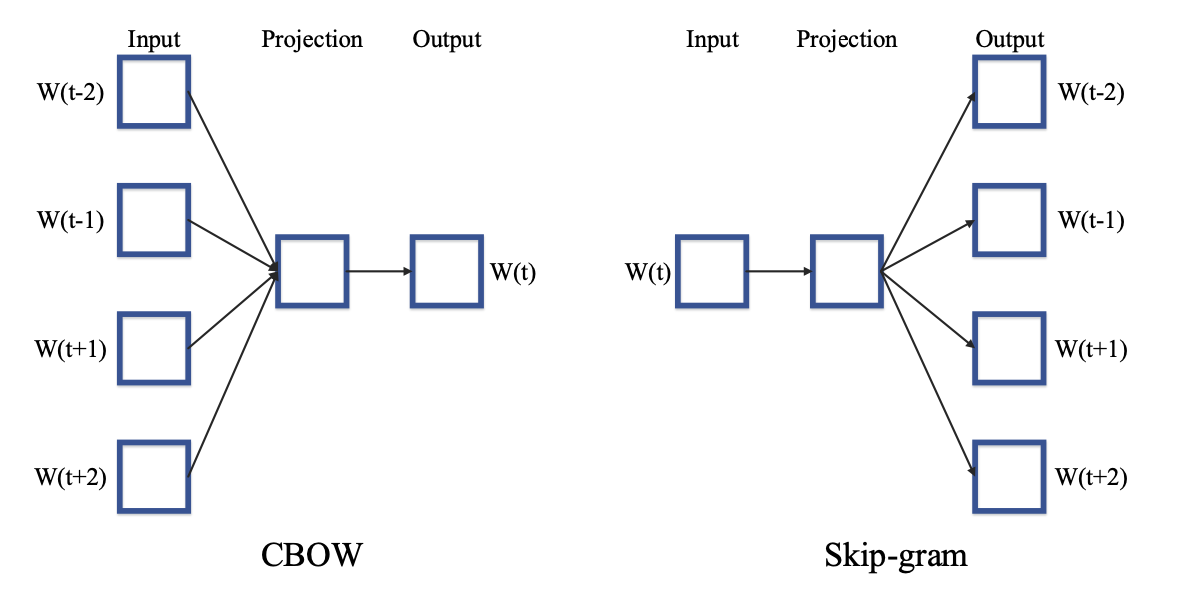}
    \end{center}
    \caption[The continuous bag-of-words (CBOW) architecture predicts the current word based on the context while the Skip-gram predicts surrounding words based on the given current word.]{The continuous bag-of-words (CBOW) architecture predicts the current word based on the context while the Skip-gram predicts surrounding words based on the given current word. (Taken from: Kowsari et al. \cite{kowsari2017hdltex})}
    \label{cha2:word2vec}  
\end{figure}

\emph{Word2Vec} by Mikolov et al. \cite{mikolov2013efficient} was developed to improve word embedding architecture. It uses statistical methods to efficiently create a high dimension vector for each word in a text corpus. It learns the word embedding in two ways: using the context of a word to predict a word, i.e., continuous bag of words (CBOW) or using a word to predict a target context, i.e., skip-gram. The first diagram in Figure \ref{cha2:word2vec} shows a simple CBOW model which attempts to find the word based on previous words, and the second diagram shows a Skip-gram which attempts to find words that might come in the vicinity of other words. Both methods are powerful tools for discovering relationships and similarities between words in a text corpus and representing them with real-valued vectors. For instance, they may assign the words ``man", ``woman", and ``girl" real-values that are close to themselves in the embedding vector space.

\subsubsection{Glove}
\begin{figure}
    \begin{center}
        \includegraphics[width=15cm, height=10cm]{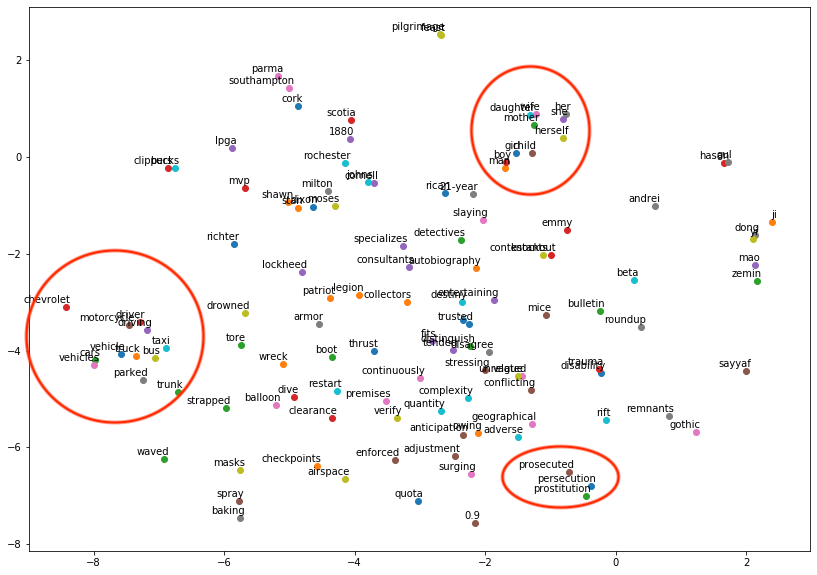}
    \end{center}
    \caption{GloVe: Global Vectors for Word Representation.}
    \label{cha2:glove}  
\end{figure}

\emph{Glove}, by Pennington et al. \cite{pennington2014glove}, is another powerful word embedding technique that extends the Word2Vec method. Similar to Word2Vec, Glove is an unsupervised learning algorithm that trains on aggregated global word-word co-occurrence or word-word context statistics. It represents each word in a text corpus with a high dimension vector based on its similarity and proximity with other words. Glove provides pre-trained word embeddings with 50, 100, 200, and 300 dimensions trained over big corpora. Figure \ref{cha2:glove} visualises the word distances over a sample data set using the t-SNE technique \cite{van2008visualizing}.

\subsubsection{Fasttext}
\emph{Fasttext}, by Facebook AI Lab, is another powerful and novel technique for word embedding that is an extension of the Word2Vec model. To take into consideration the morphology of words, Fasttext does not directly represent words in the text corpus with vectors; instead, it takes into consideration the internal structure of words by representing them as a bag of character n-grams that when added together produces feature vector representation of the word. Fasttext is a pre-trained word vector that supports 294 languages based on 300 dimensions. Suppose we have a dictionary of n-grams of size \emph{G} and given a word \emph{w}, which is associated as a vector representation \emph{z} to each n-gram \emph{g} and has output representation of \emph{v}. The scoring function is obtained as \cite{bojanowski2017enriching}: 

\begin{equation}
    \label{cha2:fasttext2}
    s(w,c) = \sum_{g \in g_w} z_g^Tv_c)
\end{equation}
where $g_w \in \{1,2,...,G\}$.

Fasttext captures the meaning of shorter words, allows the embedding to understand suffixes and prefixes, and works well with rare words as a word can be broken down into n-grams to get an embedding even if it wasn't seen during training.

\subsection{Bidirectional Encoder Representations from Transformers (BERT)}
\label{sec2:bert}
\begin{figure}
    \begin{center}
        \includegraphics[width=15cm, height=10cm]{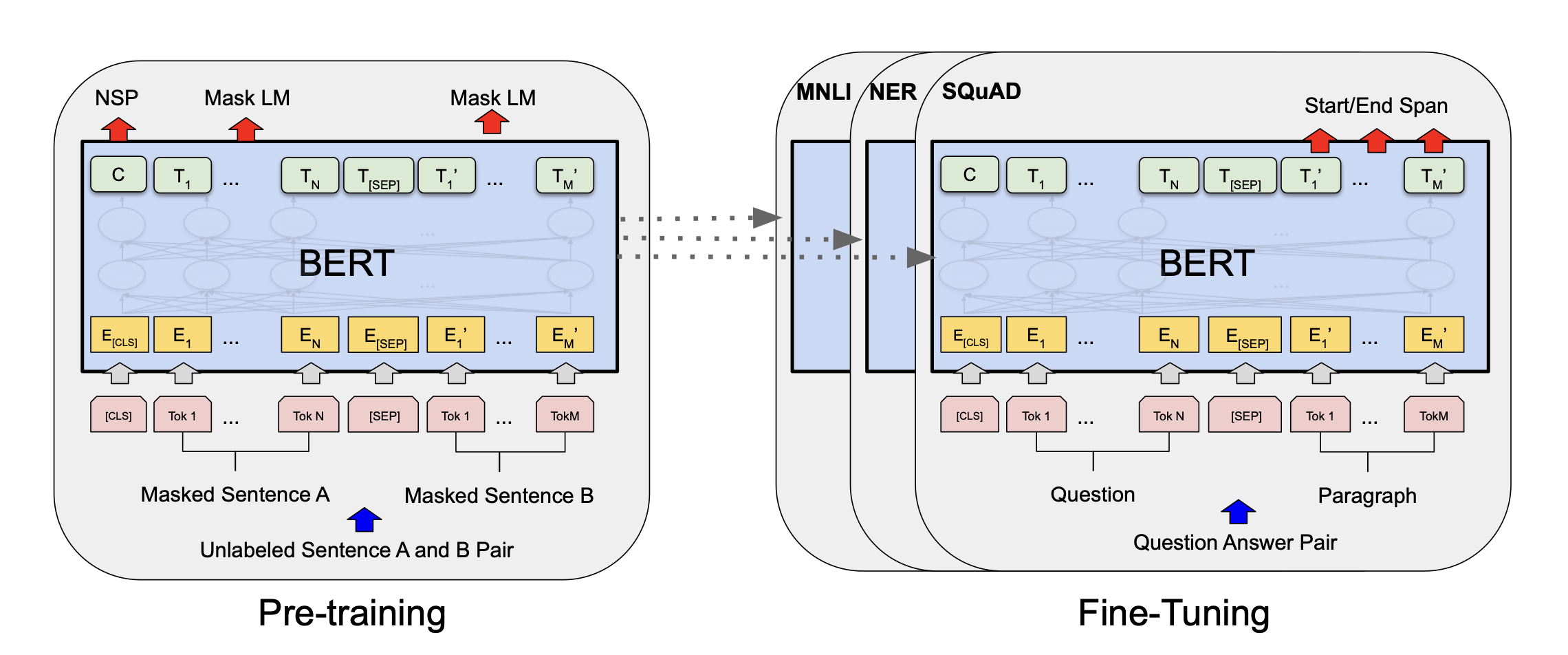}
    \end{center}
    \caption[Overall pre-training and fine-tuning procedures for BERT.]{Overall pre-training and fine-tuning procedures for BERT. Apart from the output layers, both architectures for pre-training and fine-tuning are same. Also, same pre-trained model parameters are used to initialize models for different down-stream tasks. (Taken from: Devlin et al. \cite{devlin2018bert})}
    \label{cha2:bert}  
\end{figure}

\emph{Bidirectional Encoder Representations from Transformers (BERT)}, considered to be the state-of-the-art technique on NLP \cite{gonzalez2020comparing}, is a transformer-based machine learning for NLP that has two steps in its framework: pre-training and fine-tuning \cite{devlin2018bert}. Unlike recent language representation models, e.g., Word2Vec \cite{goldberg2014word2vec}, Global Vectors for Word Representation (Glove) \cite{pennington2014glove}, fasttext \cite{young2019review}, etc., BERT is a bidirectional and unsupervised language representation that is pre-trained using only plain text corpus. So it can be fine-tuned with just a single additional output layer that can be applied to a wide range of tasks such as question answering, language inference, classification, etc., without substantially modifying the architecture based on the specificity of the task \cite{devlin2018bert}. The use of the concept of fine-tuning makes the final model fine-tuned from BERT on a specific task almost the same as BERT (see Figure \ref{cha2:bert}). 

In contrast to context-free models like Word2Vec or Glove, BERT considers the context that each word in the corpus occurred. For example, ``building" will have the same Word2Vec or Glove vector representation for both its occurrences in the sentences ``He is building his house" and ``He is building his muscle", but BERT will provide a contextual embedding for ``building" that will be unique according to the occurrences of the word in different sentences \cite{gonzalez2020comparing}. Though the concept of BERT is simple, it is empirically robust as it has obtained state-of-the-art results on eleven language processing tasks \cite{devlin2018bert}.

Two types of BERT models were primarily introduced, namely: $BERT_Base$ and $BERT_Large$. The $BERT_Base$ model has 12 layers of transformers block with a hidden size of 768 and a number of self-attention heads of 12 and has around 110M trainable parameters. On the other hand, $BERT_Large$ uses 24 layers of transformers block with a hidden size of 1024 and a number of self-attention heads of 16 and has around 340M trainable parameters \cite{devlin2018bert}. The size of $BERT_Base$ was chosen to ease comparison with the OpenAI GPT \cite{radford2018improving}.

\section{Evaluation Metrics}
\label{sec2:evaluation_metrics}
\emph{Evaluation metrics} help measure the quality and effectiveness of a machine learning model. There are numerous evaluation metrics available to assess a model; some of them are briefly explained below.

\subsection{Accuracy}
\emph{Accuracy} is the proportion of the total number of predictions that were correct. The limitation of accuracy is that it provides no information on False Negative \emph{(FN)} and False Positive \emph{(FP)}. Accuracy may be obtained by:

\begin{equation}
\label{cha2:accuracy}
Accuracy = \dfrac{(TP + TN)}{(TP + FP + FN + TN)}
\end{equation}
where \emph{TP} is True Positive, i.e., an outcome where the model correctly predicts the correct case, \emph{TN} is True Negative, i.e., an outcome where the model correctly predicts the negative class, \emph{FP} is False Positive, i.e., an outcome where the model incorrectly predicts the positive class, and \emph{FN} is False Negative, i.e., an outcome where the model incorrectly predicts the negative class. 

\subsection{Precision}
\emph{Precision}, also known as ``positive predictive value", is the proportion of positive cases that were correctly identified. The limitation of precision is that it does not evaluate \emph{TN} and \emph{FN}. Precision may be obtained by:

\begin{equation}
\label{cha2:precision}
Precision = \dfrac{(TP)}{(TP + FP)}
\end{equation}

\subsection{Recall}
\emph{Recall}, also known as sensitivity, is the proportion of actual positive cases which are correctly identified. The limitation of recall is that it does not evaluate \emph{TN} and \emph{FP}, and any classifier that predicts data points as positives is considered to have a high recall. Recall may be obtained by:

\begin{equation}
\label{cha2:recall}
Recall = \dfrac{(TP)}{(TP + FN)}
\end{equation}

\subsection{$F_\beta$-Score}
\emph{$F_\beta$-Score} is the most popular evaluation metrics for classifier evaluation \cite{lever2016points}; it combines recall and precision to obtain a full and balanced evaluation of models' performance by taking their harmonic mean. The $F_\beta$-Score may be obtained by:

\begin{equation}
\label{cha2:f_score_beta}
F_\beta-Score = \dfrac{(1 + \beta^2)(precision + recall)}{\beta^2 * precision + recall}
\end{equation}
But the most commonly used $\beta$ is 1, i.e., $F_1$, and may be obtained by:

\begin{equation}
\label{cha2:f_score_1}
F_1-Score = \dfrac{(2TP)}{(2TP + FP + FN)}
\end{equation}
The highest and lowest values of the F-measure are 1 and 0, respectively.

\subsection{Receiver Operating Characteristic (ROC)}
\emph{Receiver operating characteristic (ROC)} \cite{yonelinas2007receiver} curve is a graphical plot used to evaluate a classifier. The ROC curve is created by plotting the \emph{true positive rate (TPR)} against the \emph{false positive rate (FPR)} for different cut-off points of a parameter. The limitation of the ROC curve is that it fails to properly represent the performance of a classifier when there is class in-balance. The TPR and FPR may be obtained by:

\begin{equation}
\label{cha2:roc1}
TPR = \dfrac{(TP)}{(TP + FN)}
\end{equation}

\begin{equation}
\label{cha2:roc2}
FPR = \dfrac{(FP)}{(FP + TN)}
\end{equation}

\subsection{Area under the ROC Curve (AUC)}
\begin{figure}
    \begin{center}
        \includegraphics[width=15cm, height=10cm]{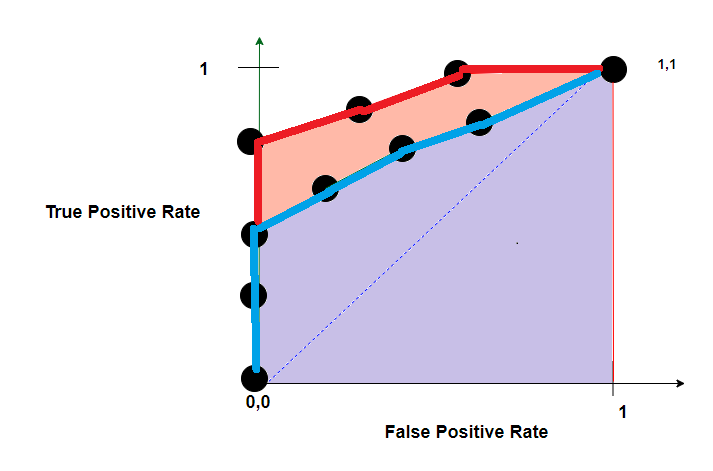}
    \end{center}
    \caption{ROC of two classifiers}
    \label{cha2:auc}  
\end{figure}

\emph{Area under the ROC Curve (AUC)} \cite{yonelinas2007receiver} measures the area underneath the ROC curve to evaluate the performance of a classifier. The highest and lowest values of AUC are \emph{1} and \emph{0}, respectively. The higher the AUC, the better the model predicts positive classes as positive and negative classes as negative. Figure \ref{cha2:auc} shows the ROC of two classifiers where one is coloured blue and the other, red. The AUC for the red ROC curve is greater than the AUC for the blue ROC curve. This means that the classifier with the red ROC curve is better.

\subsection{Inter-Annotator Agreement (IAA)}
\begin{figure}
    \begin{center}
        \includegraphics[width=15cm, height=7.5cm]{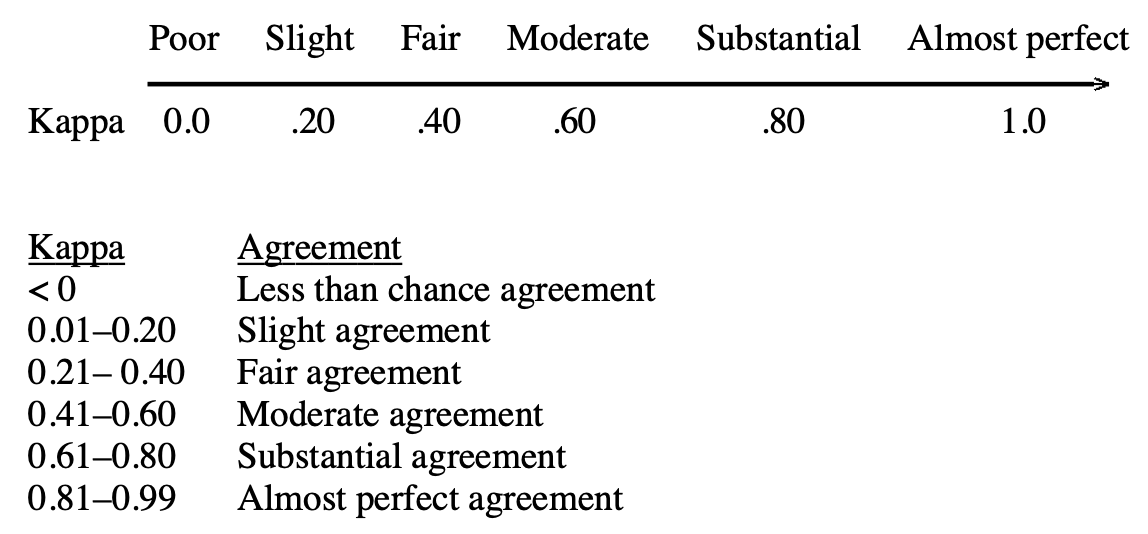}
    \end{center}
    \caption[Interpretation of kappa score.]{Interpretation of kappa score. (Taken from: Viera et al. \cite{viera2005understanding})}
    \label{cha2:kappa}  
\end{figure}

Manual annotation of texts is often employed in text mining and NLP on limited parts of datasets so that their results may be extrapolated by automated text classification on the overall dataset. Therefore, ensuring their reliability and sufficiency is important. Annotation agreement aims to estimate the reliability and adequacy of annotators' labelling by measuring the similarity of labelling decisions made by different annotators for different category labels. The Cohen's Kappa \cite{cohen1960coefficient}, often used in binary classification tasks because it assumes there are two annotators and it considers individual bias, may be estimated by:

\begin{equation}
\label{cha2:iaa1}
k = \dfrac{p_a - p_e}{1 - p_e}
\end{equation}
where;
\begin{equation}
\label{cha2:iaa2}
p_a = P(A1 = yes, A2 = yes) + P(A1 = no, A2 = no)
\end{equation}
and;
\begin{equation}
\label{cha2:iaa3}
p_e = P(A1 = yes) * P(A2 = yes) + P(A1 = no) * P(A2 = no)
\end{equation}
and $p_a$ is observed agreement, i.e., the proportion of times annotators agreed, $p_e$ is expected agreement, i.e., the proportion of times annotators are expected to agree by chance, \emph{A1} the first annotator, and \emph{A2} is the second annotator. If the annotators are in complete agreement, then \emph{k = 1}; if there is no agreement between the annotators, then \emph{k = 0}. Figure \ref{cha2:kappa} shows an interpretation of the kappa score \cite{viera2005understanding}.

\subsection{Word Cloud Analysis}
\begin{figure}
    \begin{center}
        \includegraphics[width=13cm, height=6.5cm]{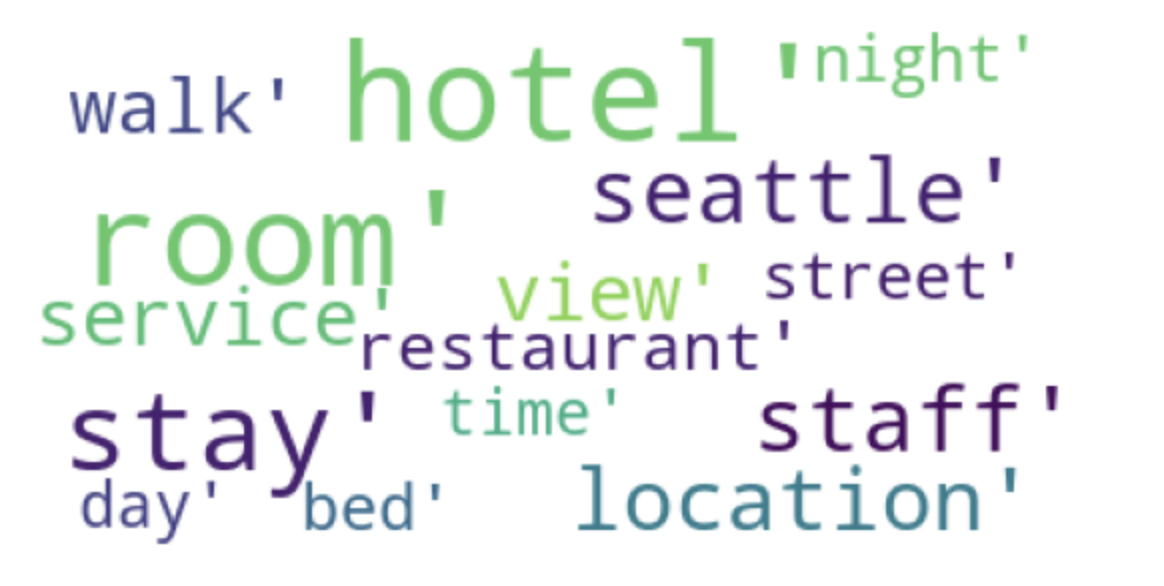}
    \end{center}
    \caption[Sample word cloud of 100 positive reviews from an hotel review dataset.]{Sample word cloud of 100 positive reviews from an hotel review dataset.}
    \label{cha2:word_cloud}  
\end{figure}

\emph{Word clouds} or \emph{tag clouds} are a visual representation of a text corpus using word frequency such that words that are more common in the text corpus appear more significant in the generated visual image. Considering the continuous increase in the use of the Internet and SNSs, which has led to the ever-increasing unstructured texts, there is a deer need for a fast and straightforward tool that will provide evaluators with exploratory textual analysis to help visually interpret or gain insight into the focus or most prominent items in a large and unstructured text corpus. Figure \ref{cha2:word_cloud} is a sample word cloud of 100 positive reviews from a hotel review dataset.

\chapter[Literature Review]{Literature Review}
\label{chapter3}
This chapter reviews current knowledge related to the research conducted in this dissertation and analyses the ethical considerations and concerns around collecting and using data in the conducted experiment. Section \ref{chap3:sns_for_grieving} reviews studies on the use of SNSs for grieving while putting into perspective: (a) the reasons for SNSs' relevance for grieving and memorialising the dead and (b) the reasons for the continuous adoption of online mourning rituals over the traditional mourning rituals. Further, in Section \ref{cha3:related_works}, a review of related works is conducted; and in Section \ref{cha3:research_relevance}, the relevance of this dissertation is debated. Finally, Section \ref{cha3:ethical_considerations} provides a thorough analysis of the ethical considerations and concerns around collecting and using data in this experiment.

\section{SNSs for Grieving and Memorialising the dead}
\label{chap3:sns_for_grieving}
Expressing grief and memorialising the dead has been done online since the early stages of the Internet \cite{brubaker2012grief}. Initially, individual websites were created and maintained for the dead, then numerous ``virtual cemeteries" where visitors can leave messages about a deceased became widely used \cite{roberts2000perpetual}. However, from the late $20^{th}$ through to the $21^{st}$ century, when SNSs gained significant popularity \cite{hillis2018digitalizing}, users started expressing grief and memorialising the deceased on their SNSs profile pages \cite{brubaker2012beyond} and creating groups on SNSs to memorialise the dead \cite{degroot2008facebook}. Brubaker \& Hayes \cite{brubaker2011we} noted that memorialising on SNSs profile pages is distinct from doing so on websites, groups, ``virtual cemeteries", etc., because SNSs profile pages are created by the deceased and not by their friends or families after death.
Moore et al. \cite{moore2019social} conducted a comprehensive study investigating what motivates mourners to utilise SNSs during bereavements. They found several reasons: ``(a) sharing information with family or friends and (sometimes) beginning a dialogue, (b) discussing the deceased’s death with others, (c) discussing death with a broader mourning community, and (d) commemorating and continuing connection to the deceased." Gathman \cite{gathman2014everybody} also suggests that users may use SNSs for grieving or disclosing deaths to avoid the discomfort of having to individually announce the loss to every single person or tell the news over and over again. Dickinson \cite{dickinson2011shared} notes that allowing others to partake in grief, even if they are merely strangers, helps consoling those in grief. Besides, receiving consolations over SNSs also helps the bereaved feel that they are not alone \cite{katims2010grieving}.

Though People still go through the usual physical grieving process, i.e., sometimes wearing black or dark colours to rituals, Carroll and Laundry \cite{carroll2010logging} noted that there had been a shift in the expression of grief. The bereaved are now discouraged from showing too many emotions, instead, motivated to limit their time in mourning. Only a little time is now used to grief, ``which is perhaps part of the reason why the use of SNSs as forums of grief expression has become increasingly more popular" \cite{hillis2018digitalizing}. This is similar to an observation by Romanoff \& Tenezio \cite{romanoff1998rituals} that traditional mourning rituals are overtime getting replaced, decreased or eliminated, and on the other hand, online mourning rituals are increasingly becoming more popular.

\section{Related Works}
\label{cha3:related_works}
SNSs have grown to become an inextricable part of life as they keep people company between leaving their beds at dawn and returning to them at dusk; the uncomplicated and usually free access to the ``big data" composed on these SNSs by the hundreds of millions of daily active users has motivated researchers to use large scale text mining and NLP approaches to understand, detect and gain intelligent insights into online participants’ attitudes and behavioural responses in numerous situations.

Large scale text mining and NLP approaches have been successfully used in many recent pieces of research to quantify and predict mental illness severity in online pro-eating disorder communities \cite{chancellor2016quantifying}, classify suicide letters \cite{pestian2010suicide}, classify and analyse reviews \cite{turney2002thumbs, pang2002thumbs}, understand customer perceptions \cite{ramaswamy2018customer, saxena2018analysing}, identify emotional distress and depression \cite{brubaker2012grief, de2013predicting, kotikalapudi2012associating, oxman1982language, moreno2011feeling, cheng2017assessing}, detect adverse reactions to drugs \cite{bollegala2018causality}, identify mothers at risk of postpartum depression \cite{de2013predicting}, detect abusive contents on SNSs \cite{chen2017harnessing}, detect fake news \cite{monti2019fake}, examine the impact of user personality traits on word of mouth \cite{adamopoulos2018impact}, classify deceased sites members \cite{ma2017write}, analyse the response of users of SNSs to terrorism \cite{mansour2018social}, detect traffic accidents from social media \cite{zhang2018deep}, and classify post-mortem contents on SNSs \cite{jiang2018tending, brubaker2018describing, ma2017write}.

Most related and similar to the experiment carried out in this research are the works of Ma et al. \cite{ma2017write} and Jian and Brubaker\cite{jiang2018tending, jiang2018describing}. Ma et al. \cite{ma2017write} used a combination of unigram features detected from LIWC, e.g., ``funeral", ``died", ``grief", etc., and n-gram features that were manually identified from the training dataset, e.g., ``went to heaven", ``with our Lord", ``in lieu of flowers", etc., that are not included in LIWC death category. They calculated the term frequency of these death words, i.e., both LIWC and non-LIWC detected words and divided the number of occurrences of each unigram and n-gram by the total number of words in the three previous updates about the subject whose profile is to be checked. These features are then prepared and fed into six different traditional machine learning algorithms: dummy classifier, logistic regression, support vector classifier, naive bayes, decision tree and random forest, to train models that can predict whether the owner of the profile is deceased or not. Similar to Ma et al. \cite{ma2017write}, Jian and Brubaker\cite{jiang2018tending, jiang2018describing} used features detected through LIWC but also used n-gram features (with n = 1, 2, or 3) detected from TF-IDF weights. They compared using features from either of these methods (i.e., TF-IDF or LIWC) and the combination of both. Features generated from these three feature engineering techniques (i.e., TF-IDF, LIWC, and TF-IDF \& LIWC) were then prepared and fed into four traditional machine learning algorithms: naive bayes, logistic regression, linear SVM, and boosted trees, to train models that can predict whether a text is pre-mortem or post-mortem. Other research publications used various text processing approaches, like statistical-based inter- and intra- analysis of features \cite{kotikalapudi2012associating}; rule-based estimation of semantic orientation in linguistic styles of writing \cite{oxman1982language, turney2002thumbs, jiang2018tending}; dictionary-based language analysis packages like the Language Inquiry Word Count (LIWC) \cite{brubaker2012grief, de2013predicting, moreno2011feeling, getty2011said}; traditional machine learning algorithms like naive bayes, maximum entropy classification, support vector machines, etc., \cite{pestian2010suicide, pang2002thumbs}, and deep learning-based approaches \cite{ramaswamy2018customer, zhang2018deep, monti2019fake}.

Most of these researches observed significant differences in text contents. They found that positive emotion provides a basis for detecting the expression of joy \cite{brubaker2012grief, turney2002thumbs, pang2002thumbs} and that negative emotions provide a basis for detecting sadness \cite{brubaker2012grief, pestian2010suicide, de2013predicting, getty2011said}. Specifically, Getty et al. \cite{getty2011said} and Jiang \& Brubaker \cite{jiang2018tending, brubaker2018describing} found that post-mortem posts show higher rates of negative emotion and has higher number of word counts, first person pronoun, second person pronouns and past tense than pre-mortem equivalents. Overall, these studies show great successes and potentials in using computational linguistics to analyse differences in annotated texts and inspire this research.

\section{Research Relevance}
\label{cha3:research_relevance}
To my knowledge, only a few works have attempted to classify online data based on their owners' living or death status. Ma et al. \cite{ma2017write} differentiated relinquished sites from deceased sites on CaringBridge \footnote{https://www.caringbridge.org}, an online platform similar to a blog that helps users support and follow patients' health journeys. However, this research was based on an online health community platform where the linguistic practices are different and not diverse as in practical SNSs. Another limitation of Ma et al. \cite{ma2017write} is that the results presented were derived from a limited dataset; only 388 and 202 profiles were used in training and testing, respectively.

Jian and Brubaker\cite{jiang2018tending, jiang2018describing} also used computational linguistic analysis on a dataset collected from the profiles of 13,200 deceased MySpace users in April 2010 and developed classifiers to detect mortality from profiles and comments. A limitation of this work is the efficient generalisability of the trained classifiers to other contexts, i.e., dataset from other SNSs and the changes in linguistic practices since the dataset was drawn from over eleven years ago. Though \cite{jiang2018tending, jiang2018describing} tested their classifiers to see how they will generalise to data from other platforms and time frames and achieved promising results, the new datasets they used to test their classifiers were from Facebook memorial groups and newspapers obituaries. The Facebook memorial groups' dataset contained messages posted in groups created by friends and family to memorialise the deceased. The dataset collected from newspaper obituaries was not from social media and consisted of more formal writing styles. Overall, similar to the dataset collected from MySpace, the limitation of these datasets (i.e., datasets collected from Facebook memorial groups and newspapers obituaries) is that they do not include pre-mortem contents posted by either the deceased or the survivors and are completely data in which friends or relatives are memorialising loved ones. 

The methodologies of Ma et al. \cite{ma2017write} and Jian and Brubaker\cite{jiang2018tending, jiang2018describing} were briefly discussed in Section \ref{cha3:related_works}, and in all cases \cite{ma2017write, jiang2018describing, jiang2018tending}, recent text mining and NLP discoveries that have achieved outstanding performances: deep learning algorithms like BiLSTM or CNN, pre-trained word embeddings for feature engineering like Word2Vec, Glove or Fasttext, and the state-of-the-art BERT were not explored.

This project takes a different approach by using SPARQL over WikiData to mine celebrities' data, i.e. status, whether deceased or alive; date of death if dead; and social media handles, which will be used to collect posts from their SNSs' profiles. This means that training classifiers will be done using posts collected from individual users’ profiles containing both pre-mortem and post-mortem contents that span through many years and contain recent contents, rather than just post-mortem contents from several years ago that may not generalise to recent linguistic practice. Additionally, recent text mining and NLP discoveries that have achieved outstanding performances, i.e., deep learning algorithms like BiLSTM or CNN, pre-trained word embeddings for feature engineering like Word2Vec, Glove or Fasttext, and the state-of-the-art BERT would be explored and compared with their traditional machine learning algorithms and feature engineering techniques counterparts.

\section{Ethical Considerations}
\label{cha3:ethical_considerations}
\subsection{Wikidata}
The data of Wikidata is available under a free license, exported using standard formats, i.e., JSON, RDF, and XML, and can be interlinked to other open data sets on the linked data web \cite{wiki2021data}. The free licence provided by Wikidata, i.e., \emph{CC0 1.0} \footnote{See https://creativecommons.org/publicdomain/zero/1.0/} allows third parties to ``copy, modify, distribute and perform the work, even for commercial purposes, all without asking permission" \cite{wiki2021data, cco1-ouniversal}.

In the value of peoples' dignity and safety, Wikidata ensures that the information that is stored about humans are treated with special consideration. Thus, ``instead of striving to provide all possible information about living people", Wikidata only provides information that it has high confidence in and that ``does not violate a person's reasonable expectations of privacy"; In addition, Wikidata support subjects of items data that are unhappy with any specific information in any items about themselves to request for the removal of such information \cite{wiki2021livingpersons}. 

\subsection{Twitter}
Twitter is different from other SNSs like Facebook because most Twitter contents are publicly accessible via the Twitter API and/or via data resellers. In contrast, most Facebook data is considered private and mostly only made available at an aggregate level. This shows that conducting research using Twitter data requires many ethical considerations; other than ethical implications, the public may react negatively when they feel that their data is being used in a way they are not comfortable with or that researchers are intruding on their privacy.

\emph{Informed consent}, which is the process of telling potential research participants about the key elements of a research study and what their participation will involve, may be challenge when using SNSs data because of the large number of users involved. Also, it may be impossible to reach all Twitter users to take informed consents as they may have stopped using their accounts or may not reply \cite{ahmed2017using}. However, a justification for continuing to research on Twitter data even without informed consent is that data reuse is permitted in Twitter's terms of service and the privacy policy as long as a researcher uses an authorised Twitter API. \\\\
The privacy policy says that:

\say{By publicly posting content, you are directing us to disclose that information as broadly as possible, including through our APIs, and directing those accessing the information through our APIs to do the same \cite{twitterprivacy}.}\\\\
And the Twitter terms of service says that:

\say{You agree that this license includes the right for Twitter to provide, promote, and improve the Services and to make Content submitted to or through the Services available to other companies, organisations or individuals for the syndication, broadcast, distribution, Retweet, promotion or publication of such Content on other media and services, subject to our terms and conditions for such Content use \cite{twittertos}.}

Even so, it is still important to thoroughly consider likely concerns around consent because previous research \cite{kramer2014experimental} was widely criticised for failure to obtain informed consent and was argued by \cite{kleinsman2015facebook} to be unethical. However, informed consent waivers are sometimes provided when some specific conditions are met: i.e., data is publicly available, non-sensitive, not involving vulnerable or dependent groups, and is anonymised \cite{manchestersocialmedia}. \footnote{\url{https://www.training.itservices.manchester.ac.uk/uom/ERM/ethics_decision_tool/story_html5.html}}

\chapter{Research Methodology}
\label{chapter4}
In this chapter, the methodologies employed to achieve the objectives of this research and to answer the research questions are discussed. Section \ref{cha3:data_collection} describes the procedures followed to extract the dataset used in this experiment from Wikidata using SPARQL and Twitter using the Twitter API. In Section \ref{cha4:data_preparation}, the strategies used for pre-processing the dataset, the unit of analysis used in this experiment and the technique for annotating and evaluating the annotations of the collected dataset are presented. Also, the methods used for extracting features from the text corpus are introduced in Section \ref{cha4:feature_engg}. Very importantly, the experimental design, i.e., steps employed for handling imbalance dataset, selection and evaluation of the machine learning algorithms, the selection of libraries and tuning of parameters for the text classification machine learning algorithms, sentiment analysis and emotion analysis tools, etc., are discussed in Section \ref{cha4:experimental_design}. Lastly, Section \ref{cha4:experimental_tools} briefly discusses the choice of project development environment and programming language, and highlights the essential libraries used in the experiment.

\section{Data Collection}
\label{cha3:data_collection}
As argued in Section \ref{cha3:research_relevance}, the datasets used in Ma et al. \cite{ma2017write} and Jian \& Brubaker\cite{jiang2018tending, jiang2018describing} have limitations. In Ma et al. \cite{ma2017write}, the dataset was mined from an online health community platform where linguistic practices are different and not as diverse as in practical SNSs. Also, in Jian \& Brubaker\cite{jiang2018tending, jiang2018describing}, there are two limitations: (a) changes in linguistic practices since the dataset was drawn from over eleven years ago and (b) dataset do not include pre-mortem contents posted by either the deceased or the survivors and are completely data in which friends or relatives are memorialising loved ones. Therefore, there is a need to mine new data that meets the requirements and standards of this experiment.

Twitter was selected as the SNS where the data used in this experiment, i.e., post- and pre-mortem tweets associated with deceased celebrities would be mined; this choice was influenced by the free, open and easy accessibility of Twitter's data. However, it won't be possible to get post- and pre-mortem tweets about deceased celebrities from Twitter without their Twitter usernames. Therefore, SPARQL would be used over Wikidata to mine deceased celebrities usernames, which would then be used to mine post- and pre-mortem tweets from Twitter.

\subsection{SPARQL Over WikiData for Extracting Deceased Celebrities Information}
To gain access and insight into the information on Wikidata, SPARQL was used to query the Wikidata Query Service (WDQS) that is made available by Wikidata through a webpage with URL ``https://query.wikidata.org" \footnote{https://query.wikidata.org} on their official website. To achieve this objective, i.e., extracting name, Twitter username, date of death, country of origin and date of birth of deceased celebrities, SPARQL was used to investigate the required properties needed in the full query to access the desired information. Below are the Wikidata properties and items that were used in the final query:

\begin{enumerate}
    \item \textbf{P31: }means ``instance of". It is used to reference the class of which a subject is a particular example and member, e.g., in  \emph{?person wdt:P31 wd:Q123}, it attempts to reference an item ?person that is an instance of item ``wd:Q123".
    \item \textbf{Q5: }means ``human". It is a common name of Homo sapiens, i.e., humans.
    \item \textbf{P570: }means ``date of death". It is used to reference date on which the subject died, e.g., in  \emph{wd:Q123 wdt:P570 ?dateOfBirth}, it attempts to reference the date of birth of item ``wd:Q123".
    \item \textbf{P2002: }means ``Twitter username". It is used to reference a subject's username on Twitter, e.g., in  \emph{wd:Q123 wdt:P2002 ?TwitterUsername}, it attempts to reference the Twitter username of the item ``wd:Q123".
    \item \textbf{P27: }means ``country of citizenship". It is used to reference a subject's country of citizenship, e.g., in  \emph{wd:Q123 wdt:P27 ?country}, it attempts to reference the country that recognises the item ``wd:Q123" as its citizen.
    \item \textbf{P569: }means ``date of birth". It is used to reference the subject's date of birth, e.g., in  \emph{wd:Q123 wdt:P569 ?dateOfBirth}, it attempts to reference the date that item ``wd:Q123" was born.
\end{enumerate}

Thus, the final SPARQL query that was used to mine the name, Twitter username, date of death, country of origin and date of birth of deceased celebrities from the English version of Wikidata was formed as:

\begin{lstlisting}[captionpos=b, caption={SPARQL query that extracts the name, Twitter username, date of death, country of origin and date of birth of celebrities from the English version of Wikidata}, label=lst:sparql2,frame=single, breaklines=true,]
    SELECT DISTINCT ?personLabel ?twitterUsername ?dateOfDeath 
        ?countryLabel ?dateOfBirth 
    WHERE {
      ?person wdt:P31 wd:Q5 .
      ?person wdt:P570 ?dateOfDeath.
      ?person wdt:P2002 ?twitterUsername.
      ?person wdt:P27 ?country.
      ?person wdt:P569 ?dateOfBirth
              
      SERVICE wikibase:label { bd:serviceParam wikibase:language
      ``[AUTO_LANGUAGE],en". }
    }
\end{lstlisting}

Executed on the $19^{th}$ of June 2021, this query returned 1,858 records that included even prominent celebrities that died in 2021, e.g., Kobe Bryant, DMX, Olympia Dukakis, etc. After running a fundamental exploratory analysis, it was observed that some celebrities appeared more than once in the result; therefore, duplicates were removed by using the celebrities Twitter usernames as a pivot, and this brings the total number of records to 1,639. Section \ref{cha5:exploratory_analysis} of Chapter \ref{chapter5} provides a comprehensive visualisation of the distribution of Twitter usernames mined from Wikidata. 

\subsection{Extracting Tweets About Deceased Celebrities Using Twitter API}
After obtaining the Twitter usernames and other information of deceased celebrities from Wikidata, enough information is available to extract post- and pre-mortem tweets of deceased celebrities from Twitter using the Twitter API. However, some initial steps were taken to ensure that the developed dataset is accurate and the experiment is ethical; they include: ensuring that celebrities whose information are used in the experiments are not minors, ensuring that celebrities' Twitter usernames are active using the user lookup endpoint on Twitter API, and ensuring that the dates of death of celebrities are beyond $21^{st}$ March 2006, i.e., the date when the first tweet was created (see Section \ref{cha5:exploratory_analysis} in Chapter \ref{chapter5} for comprehensive visualisation of the distribution of Twitter usernames by their status). In the context of this dissertation, the pre-mortem tweets of a deceased celebrity are tweets that mentioned the celebrity and were posted before the day the celebrity died; whereas, post-mortem tweets of a deceased celebrity are tweets that mentioned the celebrity and were posted after his or her death. To extract post and pre-mortem tweets about celebrities on Twitter, the v2 full-archive search endpoint that is currently only available via the Twitter Academic Research product track and that allows researchers to programmatically access public tweets from the complete archive dating back to the first tweet on $21^{st}$ March 2006 was used. For every individual celebrity, tweets that mentioned the celebrity, i.e., that contains the celebrity's Twitter username, were extracted.

\subsubsection{Extracting Pre-mortem Tweets}
Some decisions were made regarding extracting pre-mortem celebrities tweets:

\begin{enumerate}
    \item \textbf{Valid date span of pre-mortem tweets: }pre-mortem tweets of a celebrity were selected from the pool of tweets that mentioned the celebrity and were created between $21^{st}$ March 2006 and 3 days before the celebrity's death. To ensure that the developed text classification models are generalisable to changes in linguistic practices that span an extended period, precisely between 2006 when the first tweet was written till date, an algorithm was written to randomly select pre-mortem tweets from any dates between $21^{st}$ March 2006 and 3 days before the celebrity died (see algorithm \ref{lst:sparql2} implemented in python). Tweets are then selected from any randomly selected date range from the generated date ranges until the required number of tweets is obtained. This ensures that pre-mortem tweets for different celebrities span the entire period, i.e., 2006 when the first tweet was written till date, to ensure that the changing practices in linguistic characteristics on SNSs are captured in the trained classifiers.
    \item \textbf{Number of tweets: }a maximum of 150 tweets created within the valid date span of pre-mortem tweets was selected for every celebrity.
    \item \textbf{Tweets Language: }the scope of this project is to analyse and develop text classifiers for predicting deaths from texts in English; therefore, only tweets written in English were extracted from Twitter.
\end{enumerate}

\begin{lstlisting}[captionpos=b, caption={Algorithm in python used to extract pre-mortem tweets that span various dates to ensure that changes in linguistic practices are captured.}, label=lst:algorithm,frame=single, breaklines=true, basicstyle=\small,]
    # generates multiple dates between two interval dates at random
    def date_range(start, end, intv):
        dates = list()
    
        start = datetime.strptime(start, "%Y-%m-%d")
        end = datetime.strptime(end, "%Y-%m-%d")
    
        diff = (end  - start ) / intv
        for i in range(intv):
            dates.append((start + diff * i).strftime("%Y-%m-%d"))
        
        dates.append(end.strftime("%Y-%m-%d"))
    
        return dates
    
    # generates multiple begin and end dates from a list of dates
    def generate_random_to_from_dates(begin, end, no_of_dates = 5):
        comb_dates = itertools.permutations(date_range(begin, end, no_of_dates), 2)
    
        random_dates = []
        for dates in comb_dates:
            random_dates.append(sorted(dates))
    
        return random_dates
\end{lstlisting}

Following the complete extraction of pre-mortem tweets, the total number of pre-mortem tweets extracted was 79,431 from 1,340 valid celebrity Twitter usernames with at least one pre-mortem tweet (see Section \ref{cha5:exploratory_analysis} in Chapter \ref{chapter5} for comprehensive visualisations of the extracted pre-mortem tweets).

\subsubsection{Extracting Post-mortem Tweets}
Some decisions were made regarding extracting post-mortem celebrities tweets:

\begin{enumerate}
    \item \textbf{Valid date span of post-mortem tweets: }post-mortem tweets of a celebrity were selected from the pool of tweets that mentioned the celebrity and were created between the day of their death and seven days after.
    \item \textbf{Number of tweets: }a maximum of 150 tweets created within the valid date span of post-mortem tweets was selected for every celebrity.
    \item \textbf{Tweets Language: }the scope of this project is to analyse and develop text classifiers for predicting deaths from texts in English; therefore, only tweets written in English were extracted from Twitter.
\end{enumerate}

Following the complete extraction of post-mortem tweets, the total number of post-mortem tweets extracted was 46,180 from 1,124 valid celebrity Twitter usernames with at least one post-mortem tweet (see Section \ref{cha5:exploratory_analysis} of Chapter \ref{chapter5} for comprehensive visualisations of the extracted post-mortem tweets).

\section{Data Preparation}
\label{cha4:data_preparation}
\subsection{Data Pre-processing}
\label{cha3:data_processing}
\begin{figure}
    \begin{center}
        \includegraphics[width=15cm,height=8cm]{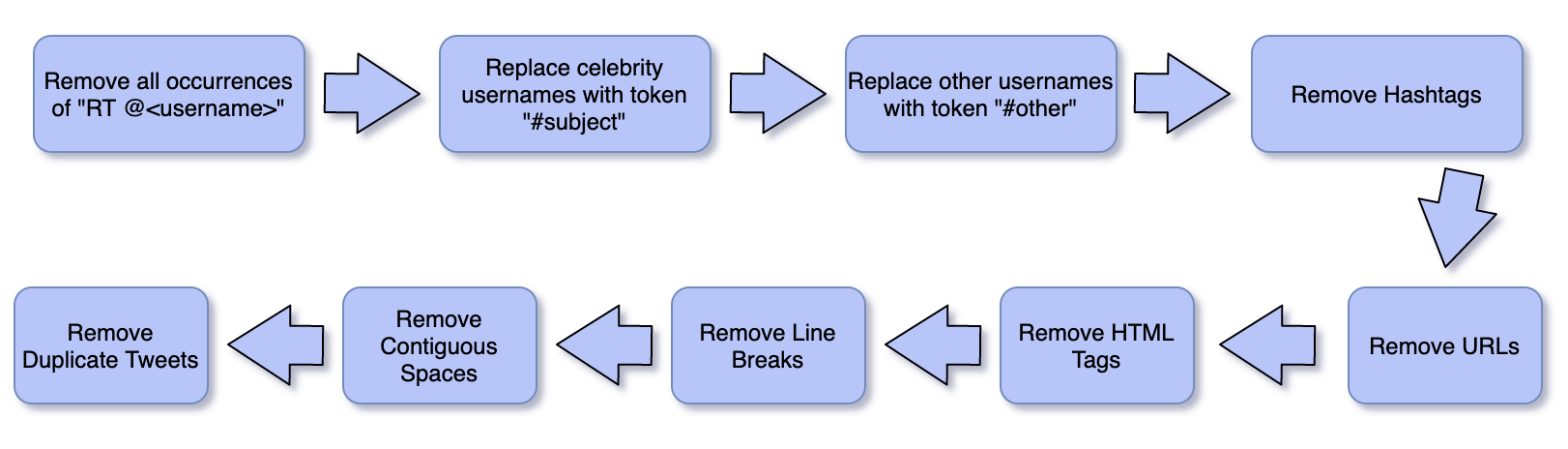}
    \end{center}
    \caption{Text pre-processing procedures applied to extracted tweets to remove unnecessary information before manual annotation}
    \label{chap4:text_preprocessing1}  
\end{figure}

\begin{figure}
    \begin{center}
        \includegraphics[width=15cm,height=10cm]{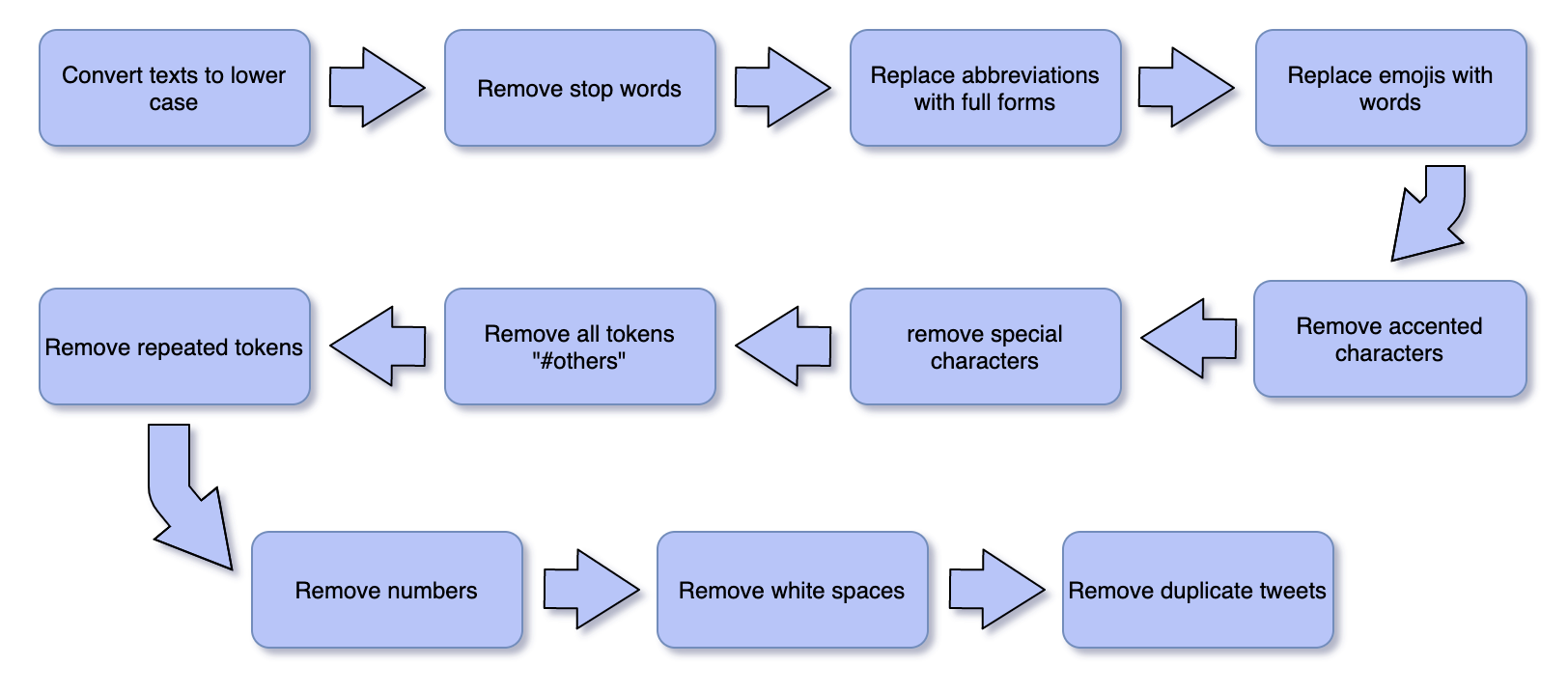}
    \end{center}
    \caption{Text pre-processing procedures applied to extracted tweets to normalise the dataset before text classification}
    \label{chap4:text_preprocessing2}  
\end{figure}

Tweets are generally short, and this means it will usually be required for Twitter users to include abbreviations, phonetic substitutions, emoticons, emojis and ungrammatical structures that torments text-processing tools \cite{sproat2001normalization}. To normalise the developed dataset that is composed of tweets, the text pre-processing pipeline was applied to remove unnecessary information that are of no use to the next text mining tasks. As discussed in Section \ref{cha3:data_processing}, both typical text pre-processing and tweet pre-processing procedures were applied; this is because beyond typical text pre-processing steps, other steps are required to handle the peculiar noise contained in tweets. Pre-processing was done in two stages: the first, to remove unnecessary information and make it conducive for human annotators to annotate the dataset; second, to bring the dataset to a normalised form for further text analysis: text classification, sentiment analysis and emotion analysis tasks. Figures \ref{chap4:text_preprocessing1} and \ref{chap4:text_preprocessing2} show the different stages of steps from typical text pre-processing and tweets pre-processing procedures, respectively, that were applied to remove unnecessary information from the dataset. 

As may be observed in Figure \ref{chap4:text_preprocessing1}, where pre-processing is performed to aid human annotation, the first step, which is removing retweets signature, removes text of the form ``RT $@<username>$:" that are added by Twitter at the beginning of tweets to signify that a particular tweet is a reply to another tweet. Afterwards, for all tweets where the subject celebrity is mentioned, the celebrity's username mentions were replaced by the token ``\#subject". This step is performed to (a) anonymise the dataset in line with ethical requirements, (b) prevent human annotators from deciding whether a celebrity is deceased or not from their knowledge, rather, from the linguistic characteristics in the text (c) prevent text-classifiers from learning information in the text that is unique to particular celebrities and is not generalisable. After celebrities usernames have been replaced, all other usernames were then replaced with the token ``\#others" to maintain a complete representation of the tweets that will not impair the understanding of human annotators; these will later be removed from the tweets before the text classification pipeline is applied. Other steps include removing hashtags, URLs, HTML tags, contiguous spaces, unnecessary line breaks and trailing spaces. 

In Figure \ref{chap4:text_preprocessing2}, it will be observed that other pre-processing steps were applied after human annotation was complete; this is to aid the task of text classification. For instance, because the scope of this experiment is on English based texts, all the texts were converted to lower cases to ensure consistency in the dataset. Other pre-processing steps applied include removing stop words, replacing emojis with terms, replacing abbreviations with their full form equivalent, removing accented characters, removing all special characters and correcting spelling errors.

\subsection{Unit of Analysis}
The unit of analysis in this project is a collection or aggregation of extracted tweets from Twitter that are either post- or pre-mortem but containing the Twitter username of a subject celebrity between a selected time of analysis. In this experiment, the desired maximum size of a collection or aggregation was chosen to be ten, i.e., every unit of analysis consists at most ten exclusive post or pre-mortem tweets about a subject celebrity. Since there are two distinct and unique sets of tweets extracted from Twitter for every celebrity, i.e., $Tweets_{deceased}$ and $Tweets_{alive}$, the tweets in each set were ordered by their creation time, i.e., the date they were posted on Twitter, and then automatically divided into collections, i.e., ``chunks" where each collection contains ten tweets. However, if the number of tweets in either of $Tweets_{deceased}$ and $Tweets_{alive}$ is not exactly divisible by ten, there may be at most a single collection obtained from this process composed of less than ten tweets. Thus, each of these collections is regarded as the unit of analysis in this experiment. The units of analysis were further handled in different ways depending on the task in consideration. For instance, during manual annotation, they were concatenated and delimited using a special token ``\_*\_" to help the human annotator recognise the beginning and end of a tweet; but during text classification, they were concatenated without any delimiter. Examples are shown below:\\\\
\textbf{Sample unit of analysis: }\\
{[\\
    ``this is the first sample tweet mentioning the celebrity @subject\_celebrity",\\
    ``this is the second sample tweet mentioning the celebrity @subject\_celebrity",\\
    ``this is the third sample tweet mentioning the celebrity @subject\_celebrity",\\
    ``this is the fourth sample tweet mentioning the celebrity @subject\_celebrity",\\
    ``this is the fifth sample tweet mentioning the celebrity @subject\_celebrity",\\
    ...\\
]}\\\\
\textbf{Sample unit of analysis during manual annotation: }\\
    ``this is the first sample tweet mentioning the celebrity \#subject \_*\_ this is the second sample tweet mentioning the celebrity \#subject \_*\_ this is the third sample tweet mentioning the celebrity \#subject \_*\_ this is the fourth sample tweet mentioning the celebrity \#subject \_*\_ this is the fifth sample tweet mentioning the celebrity \#subject..."\\\\
\textbf{Sample unit of analysis during text classification: }\\
    ``this is the first sample tweet mentioning the celebrity \#subject this is the second sample tweet mentioning the celebrity \#subject this is the third sample tweet mentioning the celebrity \#subject this is the fourth sample tweet mentioning the celebrity \#subject this is the fifth sample tweet mentioning the celebrity \#subject..."

\subsection{Dataset Labelling (Developing a Silver Standard Dataset)}
\label{cha4:silver_standard}
Collecting and manually annotating text mining and NLP datasets to develop gold standard datasets is laborious, time-consuming and expensive \cite{filannino2015gold}. This has encouraged researchers to question whether these gold standard datasets can be satisfactorily replaced with automatically annotated data, i.e., silver standard datasets. In this research, thousands of tweets were collected and later compacted into thousands of units of analysis; presumably, it will be time-consuming and laborious for human annotators to manually read through the tweets in each unit of analysis, comprehend them and decide if they convincingly communicate whether or not the celebrity it was written about is deceased. Thus, this research adopts the approach of automatically annotating the unit of analysis to develop a silver standard dataset.

To achieve this, all the units obtained using tweets extracted from Twitter that were posted after the death of celebrities were assigned a label, ``Deceased" to denote that they are post-mortem, and all the units obtained using tweets extracted from Twitter that were posted before the death of celebrities were assigned a label, ``Alive" to denote that they are pre-mortem. This resulted in 8,494 and 5,178 units of analysis that are pre-mortem and post-mortem, respectively (see Section \ref{cha5:exploratory_analysis} of Chapter \ref{chapter5} for comprehensive visualisations of the silver standard dataset).

\subsection{Inter-Annotator Agreement (IAA) and Evaluation of Silver Standard}
\label{cha4:iaa}
In Section \ref{cha4:silver_standard}, we discussed the motivation for developing a silver standard dataset: it is laborious, time-consuming and expensive to develop a gold standard. However, a great concern when using an automatically annotated dataset, i.e., silver standard dataset, is the correctness and accuracy of the annotations. Therefore, a sample of about 10\% of the silver standard was randomly selected out of 13,672 units to be manually annotated by humans. Also, to maintain a balance between the two categories, i.e., pre-mortem and post-mortem tweets, out of the units that were randomly selected, 50\% were from those developed from tweets extracted before the death of celebrities, and the other 50\% were from those developed from tweets extracted after the death of celebrities. Further, the human annotations would then be used to evaluate the accuracy and correctness of the silver standard dataset.

\subsubsection{Annotators Recruitment}
Two human annotators were recruited to manually annotate the randomly selected samples. Because the task of reading and understanding a text to denote whether it convincingly suggests that the person it was written about is deceased or alive is basic and straightforward, these annotators were only required to have a good understanding of the English Language; thus, prior knowledge of text mining or NLP is not needed.

\subsubsection{Manual Annotation Guideline}
A comprehensive and easy to understand annotation guideline was written to help human annotators gain a high-level understanding of the experiment, the task they are to complete, and how important their input is. This is to motivate them and ensure that they understand the relevance of their task to the success of the entire research (see Appendix \ref{appendix:annotation guide} for the annotation guideline document). Most importantly, it was highlighted to the annotators to annotate a unit as ``Deceased" if at least one tweet in the collection suggests that the subject is deceased and ``alive" if no tweet in the collection indicates that the subject is deceased.

\subsubsection{Measurement of IAA}
Following the annotation of the selected sample of the silver standard dataset by two human annotators, Cohen’s Kappa \cite{cohen1960coefficient} was used to estimate the agreement between these two annotators and between the two annotators and the silver standard dataset. The results of these evaluations are presented and discussed in Section \ref{cha5:iaa} of Chapter \ref{chapter5}.

\section{Feature Engineering}
\label{cha4:feature_engg}
As discussed in Section \ref{cha2:feature_engg}, machine learning algorithms expect numeric features in two dimensions, where the rows are the data instances and the columns are the features to be learnt. Thus, to perform machine learning on text, these texts need to be transformed into vector representations; hence, feature engineering is applied to extract a numerical multidimensional feature space that describes documents in the corpus on which machine learning algorithms can be applied. In this experiment, multiple feature engineering techniques are applied: TF-IDF and pre-trained word embeddings; their implementations are explained below.

\subsection{Term Frequency-Inverse Document Frequency (TF-IDF)}
TF-IDF was applied to the datasets using scikit-learn \footnote{https://scikit-learn.org} to obtain a suitable data representation for machine learning models. Scikit-learn provides a transformer called the \emph{TfidfVectorizer} for vectorising documents with TF–IDF scores. The TfidfVectorizer expects an input that is a sequence of filenames, file-like objects, or strings that contain a collection of raw documents and produces as output, a sparse matrix representation in the form of ((doc, term), tfidf) where each key is a document and term pair, and the value is the TF–IDF score. Section \ref{cha5:feature_importance} in Chapter \ref{chapter5} provides a comprehensive visualisation of the top vectors generated from TF-IDF.

\subsection{Pre-trained Word Embeddings}
TF-IDF's limitation is that it cannot account for the similarity between the words in a document since each word is independently presented as an index \cite{kowsari2019text}. This problem is solved by general-purpose word embeddings learned with GloVe, Word2vec, Fasttext, etc. These general-purpose word embeddings map every word in a corpus to a numerical vector so that the distance between the vectors reflects the semantic difference between the words. For instance, the words ``man",  ``woman",  and ``girl"  would be assigned real-values that are closer to themselves in the embedding vector space than the words ``cat", ``car", ``ball". In this experiment, three popularly used pre-trained word embeddings, i.e., Glove, Word2Vec, and Fasttext, were used to generate features to be learned by machine learning algorithms.

\subsubsection{Glove}
Glove provides pre-trained word embeddings with 25, 50, 100, 200, and 300 dimensions trained on big corpora: Wikipedia, Twitter, or Common Crawl. In this research, the Glove word embedding with 100 dimensions that was trained on 6 billion tokens and 400 thousand vocabularies was used to extract features from the dataset for the machine learning algorithms to learn from.

\subsubsection{Word2Vec}
Word2Vec provides a pre-trained word embedding with 300 dimensions, containing vectors for 3 million words and phrases, and trained on the Google News dataset (about 100 billion words) \cite{mikolov2013efficient}. The Word2Vec word embedding with 300 dimensions was used in this experiment to extract features from the dataset for the machine learning algorithms to learn from.

\subsubsection{Fasttext}
Fasttext, a pre-trained word embedding trained by Facebook AI Lab provides pre-trained word vectors with 300 dimensions for 294 languages trained on Wikipedia. The Fasttext word embedding with 300 dimensions was used in this experiment to extract features from the dataset for the machine learning algorithms to learn from.

\subsection{Contextualised Embeddings}
A limitation of context-free models like Word2Vec or Glove is that words with many meanings in the vocabulary are represented using the same vector, i.e., homonyms are not adequately handled. BERT solves this problem by considering the context that each word in the corpus occurred. For example, ``building" will have the same Word2Vec or Glove vector representation for both its occurrences in the sentences ``He is building his house" and ``He is building his muscle", but BERT will provide a contextual embedding for `building" that will be unique according to the occurrences of the word in different sentences \cite{gonzalez2020comparing}. Bert-base-uncased that was trained on lower-cased English text was fine-tuned on the dataset developed in this experiment to perform text classification.

\section{Experimental Design}
\label{cha4:experimental_design}
After data collection, data preparation and feature engineering, the next step is to train machine learning models capable of predicting whether some collection of tweets associated with a celebrity's Twitter username are pre-mortem or post-mortem. Overall, four traditional machine learning algorithms and two deep learning algorithms were trained using different techniques of feature engineering: TF-IDF and pre-trained word embeddings. In addition, the state-of-the-art BERT was fine-tuned on the dataset to perform text classification. Also, sentiment analysis and emotion analysis were performed to recognise the extent of different feelings in the dataset. This section describes the structure and steps taken to conduct these experiments.

\subsection{Handling Imbalance in the Dataset}
\label{cha4:handling_imbalance_dataset}
The problem of an imbalanced dataset occurs in classification, where the number of observations belonging to one category is significantly lower than the number of other classes. In a binary classification problem, we describe a dataset as imbalanced when the data observations in one category are heavily under-represented compared to the other category with the majority data observation. As mentioned in Section \ref{cha4:silver_standard}, after applying automatic annotation methods, 8,494 and 5,178 units belonging to the pre-mortem and post-mortem categories were obtained, respectively. It would be observed that there is a significant disparity in the number of data observations in each category and the post-mortem data observations are heavily under-represented compared to the pre-mortem category. In this experiment and similar to other classification problems, e.g., spam detection, hate speech detection, crime detection, e.t.c, the main challenge in imbalance dataset problem is that the underrepresented categories are often more useful than the over-represented category \cite{ramyachitra2014imbalanced}, but standard classifiers tend to learn more from the majority class. Thus, even if the performance is good during training, there is a high chance it will be poor during testing as the trained model would be unable to predict the minority class that is mostly the most important. The performance of classification algorithms is significantly affected by imbalanced datasets where the extracted features from a dataset are complex, e.g., features generated from text documents \cite{japkowicz2002class}.

Two common approaches have been proposed to overcome the imbalanced dataset problem: oversampling minority class, i.e., increasing the number of training samples with the minority category \cite{ling1998data} and under-sampling majority class, i.e., decreasing the number of training samples with the majority category \cite{kubat1997addressing}. In this research, the approach of random under-sampling \cite{prusa2015using} that has the effect of randomly reducing the number of examples in the majority category until an equal number of examples are reached for each category was applied. This approach is more suitable because even though there is a class imbalance, the samples in the minority category, i.e., post-mortem category, are sufficient to fit a model properly. After balancing the dataset, the number of units in both the post-mortem and pre-mortem categories was balanced to 5,178 (see Section \ref{cha5:exploratory_analysis} of Chapter \ref{chapter5} for comprehensive visualisation of the distribution of dataset size by category before and after balancing).

\subsection{Machine Learning Models Selection}
Model selection is the process of choosing one among many candidate models or selecting the proper level of flexibility for a predictive modelling problem \cite{james2017introduction}. The availability of easy-to-use machine learning libraries like scikit-learn \footnote{https://scikit-learn.org}, Keras \footnote{https://keras.io}, Pytorch \footnote{https://pytorch.org}, Huggingface \footnote{https://huggingface.co}, etc., have made it relatively straightforward to fit many different machine learning models on a given dataset. Therefore, the challenge of applying machine learning algorithms becomes how to choose among the range of available models that can be used for the problem. There are many competing concerns when performing model selection beyond model performance, such as complexity, availability of data, availability of tools, good Internet connection, etc. However, there is abundant availability of resources in this experiment: the dataset is rich, there are easy-to-use machine learning libraries for implementation that takes away complexity, there is availability of good Internet connection as the experiment is conducted primarily in England, and there are available online tools, e.g., Google Colaboratory \footnote{http://colab.research.google.com}, that takes away worries about the intensive computational requirements for training or implementing complex deep learning models and feature engineering techniques. 

In a situation where the concerns mentioned above: data, complexity, tools, etc., are not limitations for the exploration of many different machine learning methodologies together with varieties of feature engineering techniques, the best approach is to train a reasonable number of machine learning models (both traditional and deep learning approaches) together with varieties of feature engineering techniques that are remarkable and have been reported to perform well on text classification problems. To this end, many different traditional machine learning algorithms: RF, KNN, LR, SVM, and deep learning machine learning algorithms: BiLSTM and CNN, in combination with different techniques of feature engineering: TF-IDF, Word2Vec, Glove and Fasttest, will be evaluated on the dataset in this experiment. Finally, the state-of-the-art NLP technique, BERT, based on transformer machine learning, will be fine-tuned on the dataset using the pre-trained bert-base-uncased made easily available by Huggingface. Table \ref{table:maodel_selection} summarises the selected models and the techniques of feature engineering applied to the dataset to perform the classification of post-mortem collections of tweets from their pre-mortem counterparts. 

\begin{table}[]
\begin{tabular}{|p{4cm}|c|p{7cm}|}
\hline
\multicolumn{1}{|c|}{\textbf{Concept}} & \textbf{Model}  & \multicolumn{1}{c|}{\textbf{Feature}}          \\ \hline
\multirow{5}{*}{\textbf{Traditional}} 
        & \multicolumn{2}{|c|}{Baseline.} \\ \cline{2-3} 
        & NB  & TF-IDF, Glove, Word2Vec and Fasttext. \\ \cline{2-3} 
        & KNN    & TF-IDF, Glove, Word2Vec and Fasttext. \\ \cline{2-3} 
        & LR     & TF-IDF, Glove, Word2Vec and Fasttext. \\ \cline{2-3} 
        & SVM    & TF-IDF, Glove, Word2Vec and Fasttext. \\ \hline
\multirow{3}{*}{\textbf{Deep Learning}}                & CNN & Glove, Word2Vec and Fasttext          \\ \cline{2-3} 
        & BiLSTM & Glove, Word2Vec and Fasttext          \\ \cline{2-3} 
        & BERT   & bert-base-uncased                              \\ \hline
\end{tabular}
\protect\caption{\label{table:maodel_selection}Summary combination of machine learning algorithms with feature engineering techniques}
\end{table}

The use of many different approaches will offer an opportunity to implement, explore and compare many machine learning strategies and feature engineering techniques, and evaluate their efficiencies and applicability to the task at hand.

\subsection{Models Evaluation}
One of the major requirements in machine learning is training computational models that can optimally predict outputs of input samples, i.e., training data, and generalise well after training to predict correct outputs of previously unseen data \cite{mitchell1997mcgraw}; these two crucial requirements (good prediction on training dataset and good generalisation on unseen patterns) are conflicting and are also known as the Bias and Variance dilemma \cite{kononenko2007machine}. Poor generalisation to previously unseen data may be characterised by over-training, as the model just memorises the training examples and would not give correct outputs for patterns not in the training dataset \cite{reitermanova2010data}. In this experiment, the resampling method (precisely hold-out cross-validation) that seeks to estimate a model's performance on out-of-sample, i.e., previously unseen testing data, is employed as a technique to decide the quality and effectiveness of the machine learning algorithms. The hold-out cross-validation, decisions for splitting dataset and evaluation metrics are comprehensively described below.

\subsubsection{Hold-out Cross-Validation}
A popular technique that belongs in the resampling method class and is used to balance between minimal Bias and minimal Variance of the trained model is \emph{cross-validation}. The basic idea of cross-validation is to divide the dataset into two subsets so that one can be used for training and the other for evaluating the performance of the trained model.

On the other hand, \emph{Hold-out cross-validation} is a widely used type of cross-validation that separates the dataset into three mutually disjoint subsets: training, validation and testing. The model is trained on the training subset, while the validation subset is periodically used to evaluate the model's performance during training to avoid over-training. Training is stopped when the model's performance on the validation subset is good enough or not improving. After training, the testing subset is used to gain a confident estimate of the models' performance. The hold-out cross-validation technique is used in this experiment to validate the performance of classifiers with different feature engineering techniques to avoid the problem of over-fitting or over-training during the training phase.

\subsubsection{Dataset Splitting}
\label{cha4:splitting_the_dataset}
Correctly splitting a dataset into three disjoint subsets: training, validation, and testing is critical to the quality of the final model when using the hold-out cross-validation technique. If the dataset is poorly split, the patterns in the subset datasets will not sufficiently cover the underlying data patterns, leading to a poor quality final model. Since both categories, i.e., post- and pre-mortem, in the dataset are comprised of units containing tweets about all celebrities, the dataset was not split based on the criteria of units but based on the criteria of the celebrities the units are about.

In other words, the Twitter usernames of all the celebrities whose tweets were successfully extracted from Twitter were collected and used to decide the units in the dataset used for training, validation or testing. Thus, if a particular celebrity is selected to be in the testing subset, all their units both from the post- and pre-mortem categories must be used only in testing, and this is the same for those in the other two data subsets: training and validation. This is to ensure that there is no leak of information to the machine learning algorithms during training. Hence, the testing dataset is a collection of totally unknown and unseen post- and pre-mortem units belonging to celebrities who were not involved in the training or validation process. Division of the usernames was done in the following proportions: 68\% training, 17\% validation and 15\% testing (see Section \ref{cha5:exploratory_analysis} in Chapter \ref{chapter5} for comprehensive visualisation of the distribution of celebrities and dataset for training, validation, and testing).

\subsubsection{Evaluation Metrics}
\label{cha4:evaluation_metrics}
The performances of all the machine learning algorithms trained on multiple different feature engineering techniques using the hold-out cross-validation technique were evaluated using numerous available evaluation metrics: accuracy, precision, recall, F1-Score, ROC and AUC.

\subsection{Training Machine Learning Models}
To train machine learning models, decisions need to be made about the libraries to be used, how the set of optimal hyperparameters will be manipulated, the technique used to parse the extracted features to the machine learning model, etc. The decisions employed in this experiment during the training of machine learning models are briefly described below.

\subsubsection{Traditional Machine Learning Models}
For all the implemented traditional machine learning algorithms: Dummy Classifier, KNN, LR, SVM and RF, the implementations of the scikit-learn library was adopted to extract the features based on the TF-IDF feature extraction technique and to fit, test and evaluate the models. For the pre-trained word embeddings, for all the units in the dataset, a simple mean of the word vectors representing each unit's text was taken and used as features for the traditional machine learning algorithms.

The scikit-learn \emph{DummyClassifier} that makes predictions using simple rules is useful as a simple baseline for making comparisons with major trained classifiers. Thus, the DummyClassifier that uses a most frequent selection strategy was first applied to obtain a baseline to enable a thorough comparison with other classifiers. The random forest classifier was trained with 200 estimators; this low number of estimators is chosen to prevent the model from taking too much time when making predictions because results are only gotten through voting from the predictions of all the decision trees that make up the forest. In the KNN classifier, three and five were selected as the values for \emph{k}, i.e., the number of nearest neighbours that should be considered in predicting the category of a particular data sample. On the other hand, the logistic regression classifier was applied using its default parameters because it performs well for binary classification without tuning when the number of features is less than the number of data points, i.e., when the width of the two-dimensional feature representation is lower than the height. For SVM, the \emph{radial basic function (RBF)} kernel was used and the values of \emph{``gamma"} and \emph{``C"} were tuned. The \emph{``gamma"} parameter that defines how far the influence of a single training example reaches is selected to be automatic, and the \emph{``C"} parameter that serves as a regularisation parameter in the SVM was given a range between \emph{1} and \emph{10}. A lower \emph{C} value ranges generally lead to more support vectors, which may increase prediction time; however, it may lead to increased fitting time.

For all of the classifiers in the traditional machine learning category, the training and validation data subsets were combined, and scikit-learn's Grid Search that allows the tuning of classifier parameters and cross-validation was used. Five folds cross-validation was applied during training to ensure that the sub-model that best fits the training dataset with minimum loss or error is selected as the final model in each training combination. These final models were then evaluated using the testing data subset to determine how the models perform on previously unseen data. The results are reported and discussed in Sections \ref{cha5:classification_results} and \ref{cha5:discussion} of Chapter \ref{chapter5}, respectively.

\subsubsection{Deep Learning Machine Learning Models}
The implementation of CNN model in Keras, that of LSTM in the Pytorch, and that of BERT in Pytorch and Hugginface were adopted with some further tuning for the development, training and testing of the models. To evaluate the performance of the models, the scikit-learn library's ``classification\_report", ``accuracy\_score", ``roc\_auc\_score", ``roc\_curve", and ``auc" were used to get various estimates of the trained models' performance on the testing data subset. In all training scenarios for both BiLSTM and CNN, a vocabulary of the corpus was generated where each unique word is identified by a token, a pre-trained vector based on the pre-trained word embedding and the vocabulary was developed, and each training example is converted into a sequence of tokens based on the developed vocabulary; during training, these would then be fed into the algorithms with the training examples, in turn, to be used to learn the underlying relationship between vectors representing different training examples. 

The CNN model was defined with two \emph{1D} convolutional layers each followed by a \emph{1D} max-pooling layer and, lastly, a flattening layer that produces the predictions of the CNN model. Two \emph{1D} convolutional layers give the CNN model a better chance of learning the underlying representations in the features generated from the dataset. The LSTM was tuned into BiLSTM in Pytorch by setting \emph{``bidirectional"} to True and combining the results of LSTMs in two opposite directions. In the CNN model, \emph{``binary\_crossentropy"} was used as the loss function, \emph{``relu"} activation function was used in the \emph{1D} convolutional layers, while \emph{``sigmoid"} activation function was applied after flattening to get an estimate in the form of a probability that determines the categories of the data examples. In the BiLSTM model, the \emph{``negative log-likelihood loss``} was used as the loss function, and the \emph{``log softwax``} that is a generalisation of the ``sigmoid`` was used to get an estimate in the form of a probability that determines whether the training examples belong to one category or the other. In the BERT model, \emph{``BinaryCrossEntropy"} was used as the loss function since the problem is a binary classification problem, and the \emph{``sigmoid"} function was used to get an estimate in the form of probability that determines the categories of the data examples.

In all training scenarios, all the models, i.e., CNN, BiLSTM, and BERT, were trained for 30 epochs with different learning rates adjusted depending on how fast the algorithm is learning. However, early stopping and a checkpoint were implemented to ensure that training stops when five consecutive epochs do not improve the model's performance and that the model with the best validation loss during all the epochs is saved and returned to be used to evaluate the training data subset. The results are reported and discussed in Sections \ref{cha5:classification_results} and \ref{cha5:discussion} of Chapter \ref{chapter5}, respectively.

\subsection{Linguistic Characteristics and Practices}
The collected post-mortem and pre-mortem tweets were analysed separately using various techniques and tools to detect and discuss the differences between the embedded linguistic characteristics and practices in them. The techniques and tools employed to analyse these differences are described below.

\subsubsection{Sentiment Analysis}
\label{cha4:sentiment_analysis}
As earlier discussed in Section \ref{cha3:related_works} of Chapter \ref{chapter3}, previous related works observed that positive emotions are a basis for detecting the expression of joy \cite{brubaker2012grief, turney2002thumbs, pang2002thumbs} and that negative emotions provide a basis for detecting sadness \cite{brubaker2012grief, pestian2010suicide, de2013predicting, getty2011said}. Additionally, in Getty et al. \cite{getty2011said} and Jiang \& Brubaker \cite{jiang2018tending, brubaker2018describing}, it was found that post-mortem posts show higher rates of negative emotion than pre-mortem equivalents. In line with the observations of these works, the sentiments expressed in the collected pre-mortem and post-mortem tweets were separately estimated using Valence Aware Dictionary and Sentiment Reasoner (VADER) \cite{hutto2014vader} \footnote[10]{https://pypi.org/project/vaderSentiment}, a parsimonious rule-based model for sentiment analysis of social media text. The results are visualised in Section \ref{cha5:sentiment_analysis} of Chapter \ref{chapter5} to enable a comprehensive comparison between the sentiments expressed in pre-mortem tweets and those expressed in post-mortem tweets. 

\subsubsection{Emotion Analysis}
\label{cha4:emotion_analysis}
Beyond detecting sentiments, i.e., positive, negative or neutral from texts as done in sentiment analysis, emotion analysis takes a little step further to detect and recognise the extent of different feelings (e.g., anger, disgust, fear, happiness, sadness, surprise, etc.) from texts. Although other related research did not carry out emotion analysis on the collected pre-mortem and post-mortem contents, this experiment went further to recognise the extent of different feelings in the collected pre-mortem and post-mortem tweets; this is so that we would be able to know the exact kinds of feelings, whether anger, fear, happiness, sadness, surprise, etc., that dominates pre-mortem and post-mortem contents. Text2emotion \footnote[12]{https://pypi.org/project/text2emotion/} was used to extract five different feelings, happy, sad, angry, surprise and fear, from the pre-mortem and post-mortem contents, and the results are visualised and compared in Section \ref{cha5:emotion_analysis} of Chapter \ref{chapter5}.

\subsubsection{Language Inquiry and Word Count (LIWC)}
\label{cha4:liwc}
\emph{Linguistic Inquiry and Word Count (LIWC)} \cite{pennebaker2001linguistic} is a text analysis application program that estimates the linguistic, psychological or topical categories that proportion of words or tokens in a particular text belongs. Similar to Getty et al. \cite{getty2011said} and Jiang \& Brubaker \cite{jiang2018tending, brubaker2018describing}, this experiment used LIWC to separately analyse the collected tweets in each category, i.e., pre-mortem or post-mortem; the results will then be aggregated in order to detect the differences in linguistic practices used in the collected categories of tweets, e.g., use of first person or second person pronouns, adverb, preposition, conjunction, etc. The results are reported and discussed in Section \ref{cha5:liwc} of Chapter \ref{chapter5}. 

\subsection{Additional Tests, Ablation Study and Error Analysis}
Further experiments that involved altering some of the parameters and settings of the machine learning models were performed, followed by a thorough evaluation. Some alterations include: using different maximum sequence lengths for CNN and BERT, using other pre-trained transformers for BERT, e.g., \emph{``bert-large-uncased"}, altering the learning rates and the number of epochs for the deep learning models, etc. In addition, an ablation study that involves removing some ``feature(s)" of the deep learning algorithms to see how that affects their performance was performed and followed with thorough evaluations. Lastly, an error analysis was conducted to investigate the misclassified data observations in the testing subset. The observations from these additional tests and ablation study, and error analysis are presented in Section \ref{cha5:additional_tests} and \ref{cha5:error_analysis} of Chapter \ref{chapter5}, respectively.

\section{Experimental Tools}
\label{cha4:experimental_tools}
Google Colaboratory (Google Colab) \footnote{https://colab.research.google.com} that offers support to researchers to abstract load of heavy works from personal machines, was used as the experiment environment to mitigate the risk of unavailability of reliable and optimal computing power. The Google Colab, primarily based on Python, allows researchers and developers to write, execute and share codes without requiring any initial setup. In addition, numerous python packages are supported and are pre-installed on the cloud servers. Also, for the not pre-installed libraries, the Google Colab allows users to install these libraries as long as they are compatible with the version of python provided. Though the Google Colab platform offers free access to computing resources, including Graphics Processing Units (GPUs) and Tensor Processing Units (TPUs) that are necessary for training deep learning machine learning algorithms, this experiment made use of the Google Colab Pro that is only available upon premium subscription; this is to further ensure the reliable availability of high standard GPUs and TPUs. Several third-party python libraries were used in this experiment; they are briefly described in Table \ref{cha4:table_for_tools}.

\begin{table}[!ht]
    \begin{tabular}{|p{0.5cm}|p{2.5cm}|p{10cm}|}
        \hline
        \textbf{S/N} & \textbf{Package} & \textbf{Description} \\
        \hline
        1 & Request \tablefootnote{https://pypi.org/project/requests} & Allows users to send HTTP/1.1 requests extremely easily. \\
        \hline
        2 & Pymongo \tablefootnote{https://pypi.org/project/pymongo} & Contains tools for interacting with MongoDB database from Python.  \\
        \hline
        3 & Panda \tablefootnote{https://pandas.pydata.org} & Offers data structures and operations for manipulating numerical tables and time series. \\
        \hline
        4 & Numpy \tablefootnote{https://numpy.org} & Provides support for large, multi-dimensional arrays and matrices, along with a large collection of high-level mathematical functions to operate on these arrays. \\
        \hline
        5 & Matplotlib \tablefootnote{https://matplotlib.org} & Provides an object-oriented API for embedding plots into applications using general-purpose GUI toolkits. \\
        \hline
        6 & Seaborn \tablefootnote{https://seaborn.pydata.org} & Is based on matplotlib. It provides a high-level interface for drawing attractive and informative statistical graphics.\\
        \hline
        7 & Transformers \tablefootnote{https://huggingface.co} & Facilitates experiment with large-scale pre-trained models and helps to deploy them in downstream NLP tasks such as text classification and question answering. \\
        \hline
        8 & Fasttext \tablefootnote{https://fasttext.cc} & Library for learning of word embeddings and text classification created by Facebook's AI Research lab.\\
        \hline
        9 & Scikit-learn \tablefootnote{https://scikit-learn.org} & Is a open source machine learning library for python that is mostly used for traditional machine learning algorithms. \\
        \hline
        10 & Pytorch \tablefootnote{https://pytorch.org} & Developed by Facebook's AI research lab, is an open source machine learning library based on the Torch library, used for applications such as computer vision and natural language processing. \\
        \hline
        11 & Keras \tablefootnote{https://keras.io} & An open-source software library that provides a Python interface for artificial neural networks. \\
        \hline
        12 & Vader \tablefootnote{https://pypi.org/project/vaderSentiment} & A lexicon and rule-based sentiment analysis tool that is specifically attuned to sentiments expressed in social media, and works well on texts from other domains. \\
        \hline
        13 & Text2Emotion \tablefootnote{https://pypi.org/project/text2emotion} & A python package for extracting happy, angry, sad, surprise and fear emotions from text contents. \\
        \hline
    \end{tabular}
    \protect\caption{\label{cha4:table_for_tools}Some third-party Python packages used in the experiments.}
\end{table}
\chapter{Experimental Results and Discussion}
\label{chapter5}
This chapter presents and discusses the results obtained in the experiments conducted in this research. In Section \ref{cha5:exploratory_analysis}, the results obtained from conducting critical exploratory data analysis using visualisation methods are presented and discussed. Section \ref{cha5:iaa} presents the results obtained from estimating inter-annotator agreement using Cohen's Kappa \cite{cohen1960coefficient} between the two recruited human annotators and between the human annotators and the automatic annotation. In Section \ref{cha5:experimental_results}, the results obtained from training machine learning models to classify post-mortem contents from their pre-mortem counterparts are presented, and the classification error obtained by the trained models are analysed; the results obtained from using sentiment and emotion analysis to detect the similarities and differences between the linguistic characteristics and practices in post-mortem and pre-mortem contents were also presented. Finally, in Section \ref{cha5:discussion}, answers to the research questions are proposed based on the results observed from the conducted experiments.

\section{Exploratory Data Analysis}
\label{cha5:exploratory_analysis}
This experiment employed critical exploratory data analysis by using data visualisation methods to perform initial investigations on the collected data to discover patterns, spot anomalies, summarise main characteristics, check assumptions and test hypotheses. Overall, this is to help determine the best way to manipulate the collected dataset to get accurate answers to the research questions and establish confidence regarding whether the techniques considered for text analysis are appropriate, ethical and feasible. A significant concern about this experiment is determining whether the collected dataset is ethical, i.e., does not contain information about minors (people aged 17 years and below), captures the changes in linguistic practices over a long period, i.e., the dataset is about a wide range of people from different background and age group and includes pre-mortem contents that were posted by either the deceased or the survivors and are not completely data in which friends or relatives are memorialising loved ones. 

\begin{figure}
    \begin{center}
        \includegraphics[width=18cm,height=10cm]{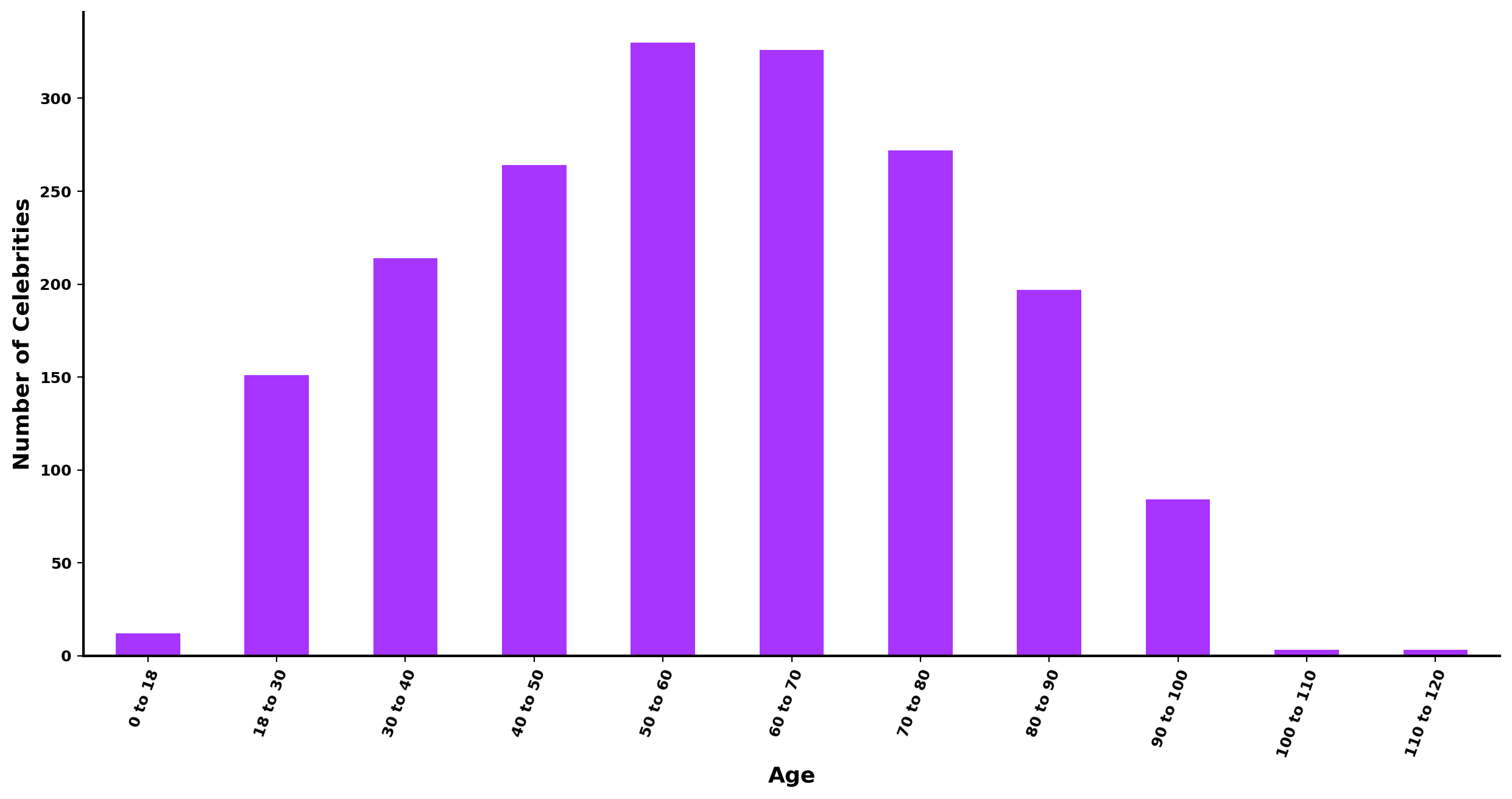}
    \end{center}
    \caption{Distribution of Celebrities by Age}
    \label{chap5:dist_by_age}  
\end{figure}

To ensure that the data extracted from Twitter does not contain information about minors, the Twitter usernames and other information about celebrities collected from Wikidata were distributed by age and visualised (see Figure \ref{chap5:dist_by_age}). It was observed that out of the 1,639 unique Twitter usernames that were extracted from Wikidata, about 12 are for minors. These Twitter usernames and related information were then removed from the collection. 

\begin{figure}
    \begin{center}
        \includegraphics[width=18cm,height=10cm]{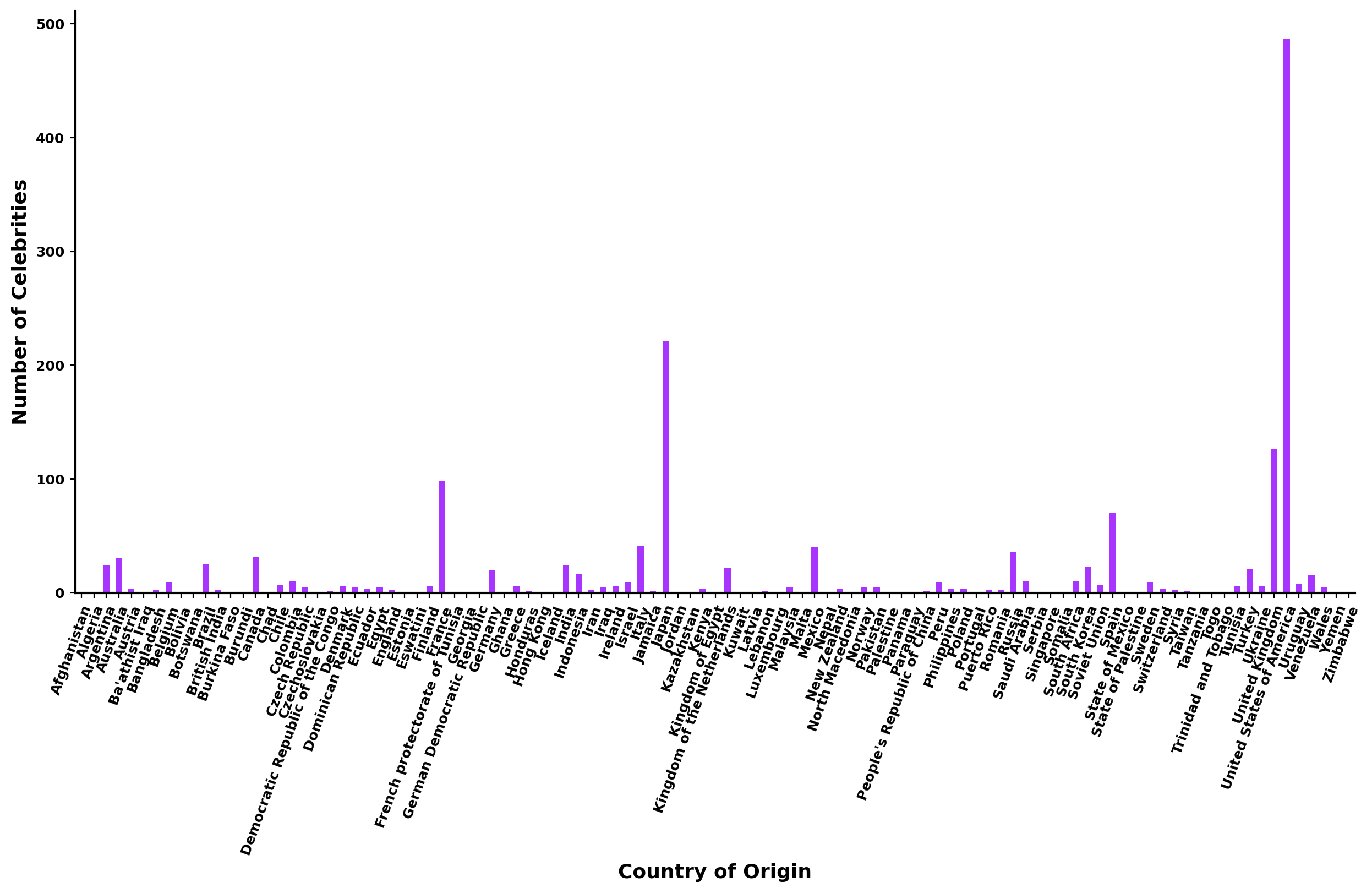}
    \end{center}
    \caption{Distribution of Twitter Usernames by Country of Origin}
    \label{chap5:country_of_origin}  
\end{figure}

\begin{figure}
    \begin{center}
        \includegraphics[width=18cm,height=10cm]{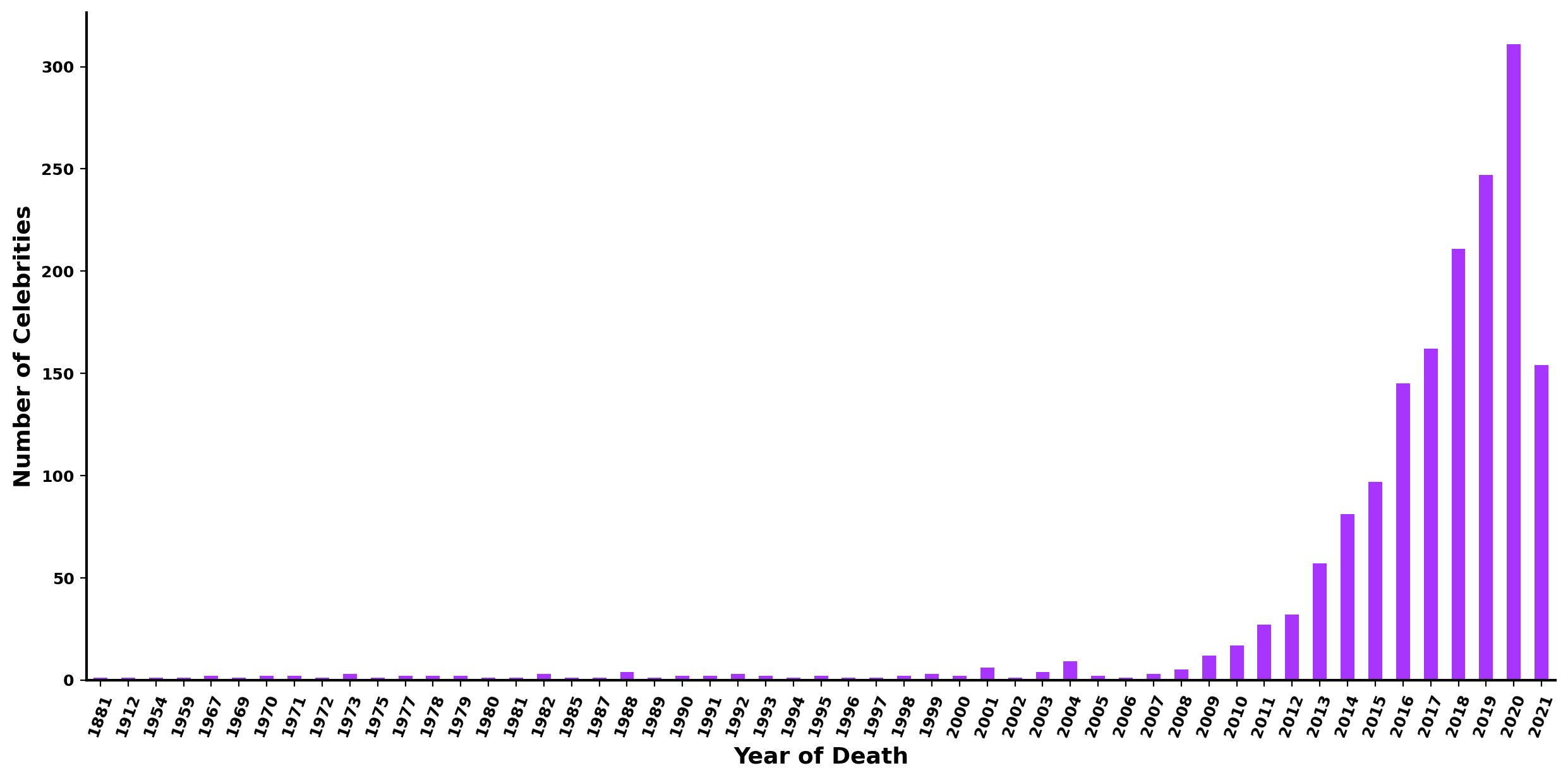}
    \end{center}
    \caption{Distribution of Twitter Usernames by Year of Death}
    \label{chap5:year_of_death}  
\end{figure}

To check whether the dataset covers a wide range of people from different backgrounds and age groups to create confidence regarding whether the trained models will be able to generalise and remain accurate in other contexts, the Twitter usernames collected from Wikidata were distributed by country of origin and year of death (see Figure \ref{chap5:country_of_origin} and \ref{chap5:year_of_death}). It was observed that even though the data is dominated by nationals of a few countries, e.g., United States, Japan, etc., it includes data about celebrities from so many other countries, and no one particular country dominates over half of the sample space; also, the year of death of the celebrities spread accurately across more than 14 years from 2006 till date. 

\begin{figure}
    \begin{center}
        \includegraphics[width=18cm,height=10cm]{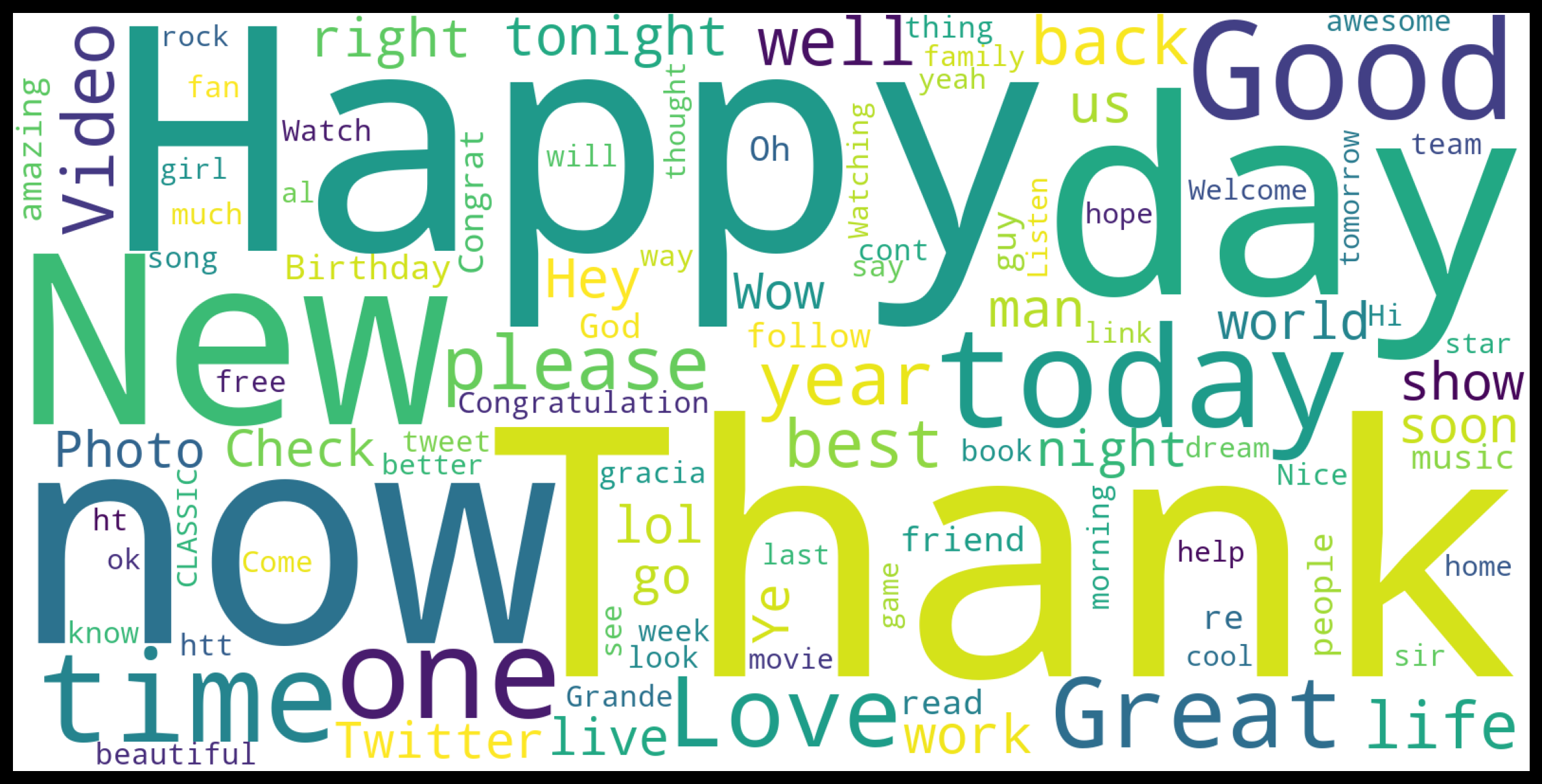}
    \end{center}
    \caption{Word Cloud Analysis of Pre-mortem Tweets}
    \label{chap5:word_cloud_pre_mortem}  
\end{figure}

\begin{figure}
    \begin{center}
        \includegraphics[width=18cm,height=10cm]{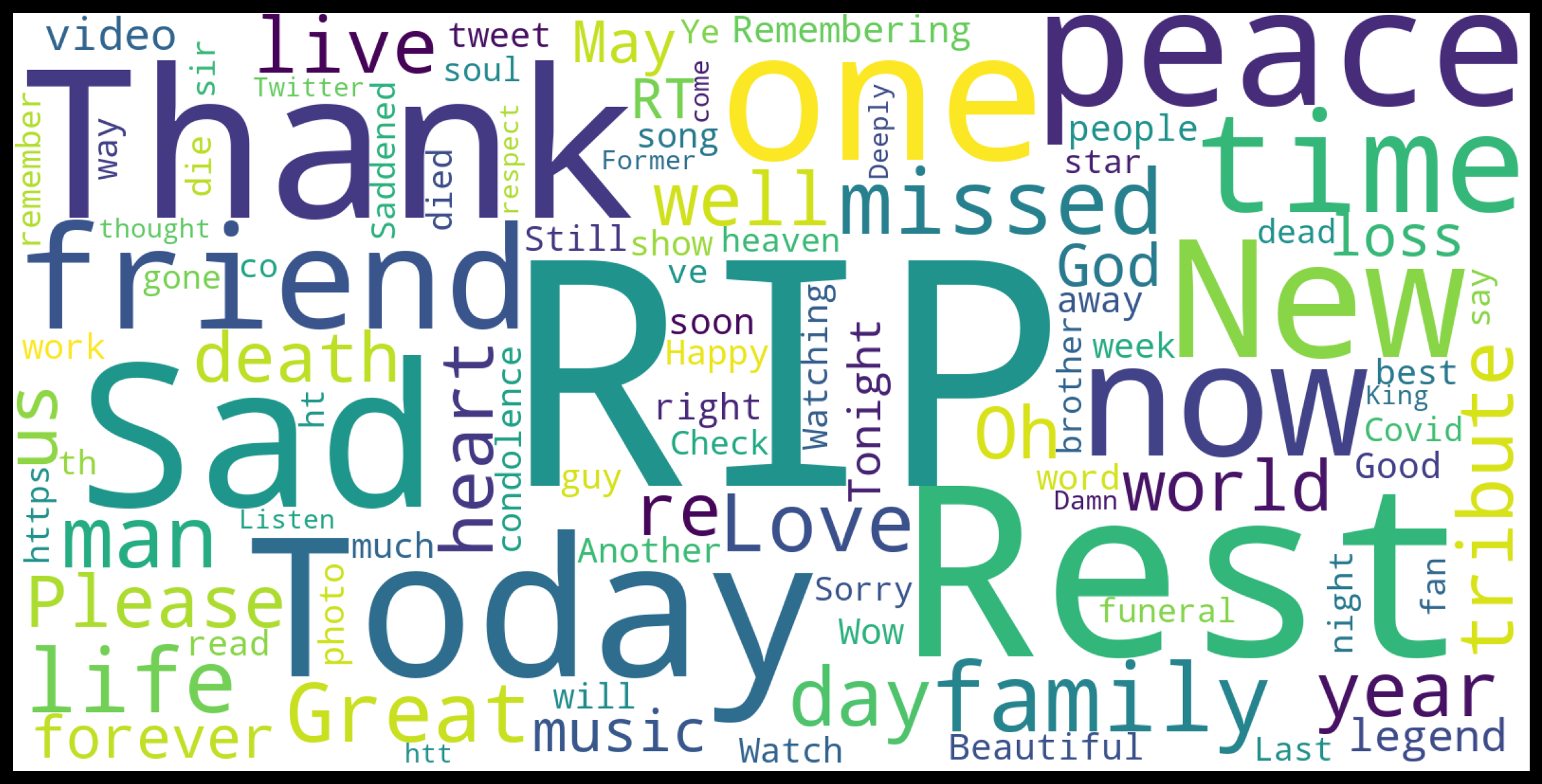}
    \end{center}
    \caption{Word Cloud Analysis of Post-mortem Tweets}
    \label{chap5:word_cloud_post_mortem}  
\end{figure}

Further, checking whether the collected contents from Twitter include pre-mortem contents that were posted about the deceased before their death couldn't have been easily achieved as there are thousands of data samples in the dataset, and manually reading through them will be time-consuming and laborious. To quickly gain a general overview that instils enough confidence, a word cloud analysis of the collected dataset was conducted, and two distinct word cloud representations were generated for the dataset subsets that were labelled pre-mortem, and the other subset labelled post-mortem. Figures \ref{chap5:word_cloud_pre_mortem} and \ref{chap5:word_cloud_post_mortem} show the word clouds generated for the pre-mortem and post-mortem subsets, respectively; as may be observed, the pre-mortem word cloud is dominated by words that are used in humans daily activities, e.g., ``happy", ``celebration", ``good", ``congratulation", ``love", ``work", ``great", ``life", etc., while its post-mortem counterpart is dominated by words that are used in grieving or memorialising the dead, e.g., ``RIP", ``sad", ``death", ``missed", ``gone", ``die", etc. 

\begin{figure}
    \begin{center}
        \includegraphics[width=18cm,height=10cm]{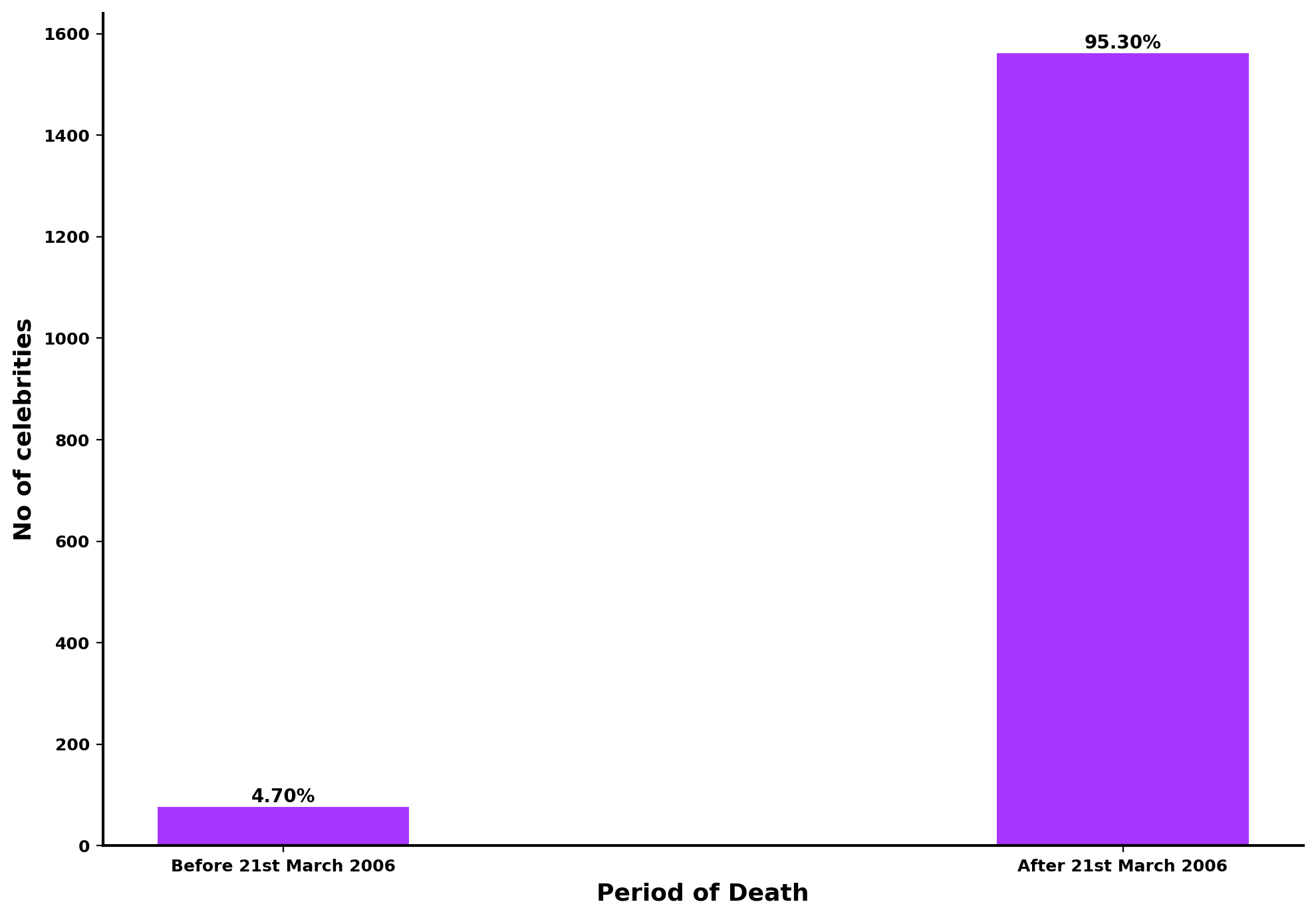}
    \end{center}
    \caption{Distribution of Celebrities by Period of Death (Before or after 21st March 2006 when the first tweet was created)}
    \label{chap5:dist_period_of_death}  
\end{figure}

\begin{figure}
    \begin{center}
        \includegraphics[width=18cm,height=10cm]{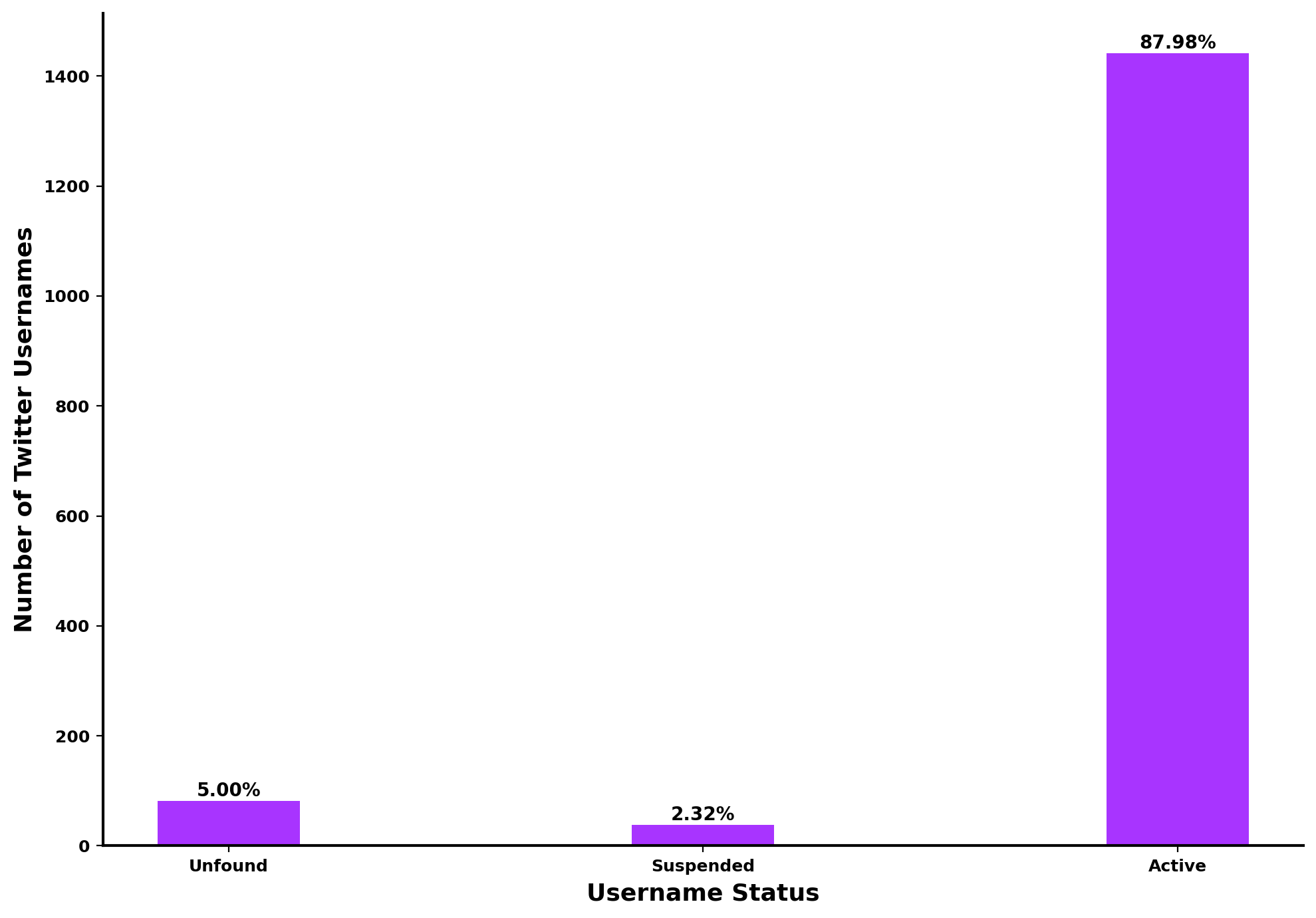}
    \end{center}
    \caption{Distribution of Celebrities Twitter Usernames by Status (Suspended, Active, Unfound)}
    \label{chap5:username_status}  
\end{figure}

Additionally, other fundamental analyses were conducted to gain confidence regarding the proportion of collected data that is accurate, correct, and useful in the experiment. For instance, because Twitter was only established in 2006 and received the first Tweet on $21^{st} March$ 2006, only Twitter usernames of users that died after this date can be utilised in this experiment as those are the only samples whose pre-mortem posts can be extracted. Figure \ref{chap5:dist_period_of_death} shows a distribution of the collected Twitter usernames by the period of their death; it would be observed that slightly over 96\% of the Twitter usernames are appropriate for further research as they belong to celebrities that died after $21^{st} March$ 2006, whose pre-mortem tweets can be extracted from Twitter. It was also relevant to check the proportion of collected Twitter usernames that were applicable for further analysis. The Twitter User lookup API was used to check the status of all the Twitter usernames and a distribution of Twitter usernames statuses, whether suspended, active or unfound, is shown in Figure \ref{chap5:username_status}; it would be observed that around 5\% and slightly over 2\% were unfound and suspended, respectively and slightly over 87\% are active. This creates confidence as 87\% is significant and acceptable for further analysis.

\begin{figure}
    \begin{center}
        \includegraphics[width=18cm,height=10cm]{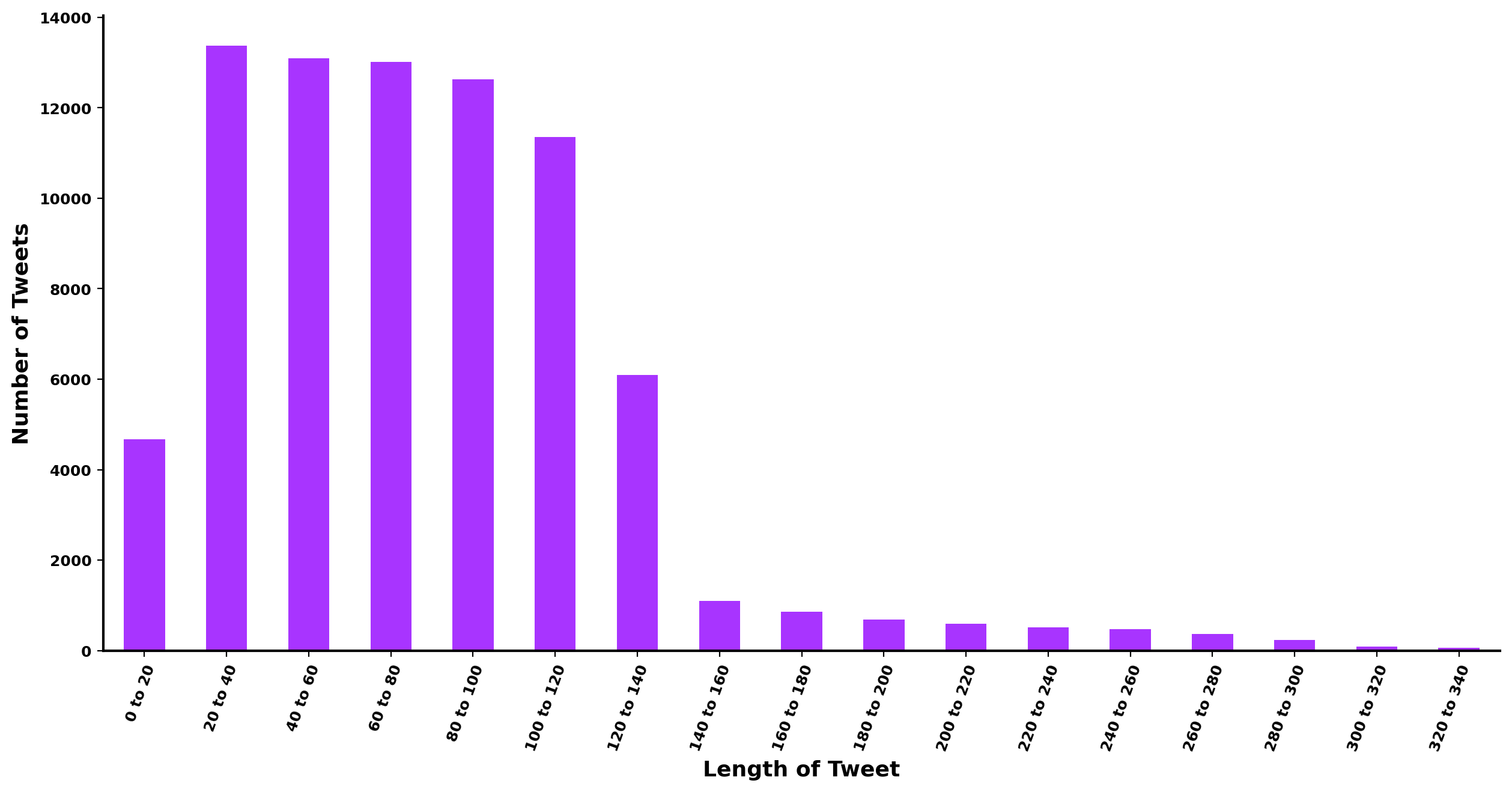}
    \end{center}
    \caption{Distribution of Pre-mortem Tweets by Length}
    \label{chap5:pre_mortem_lenght}  
\end{figure}

\begin{figure}
    \begin{center}
        \includegraphics[width=18cm,height=10cm]{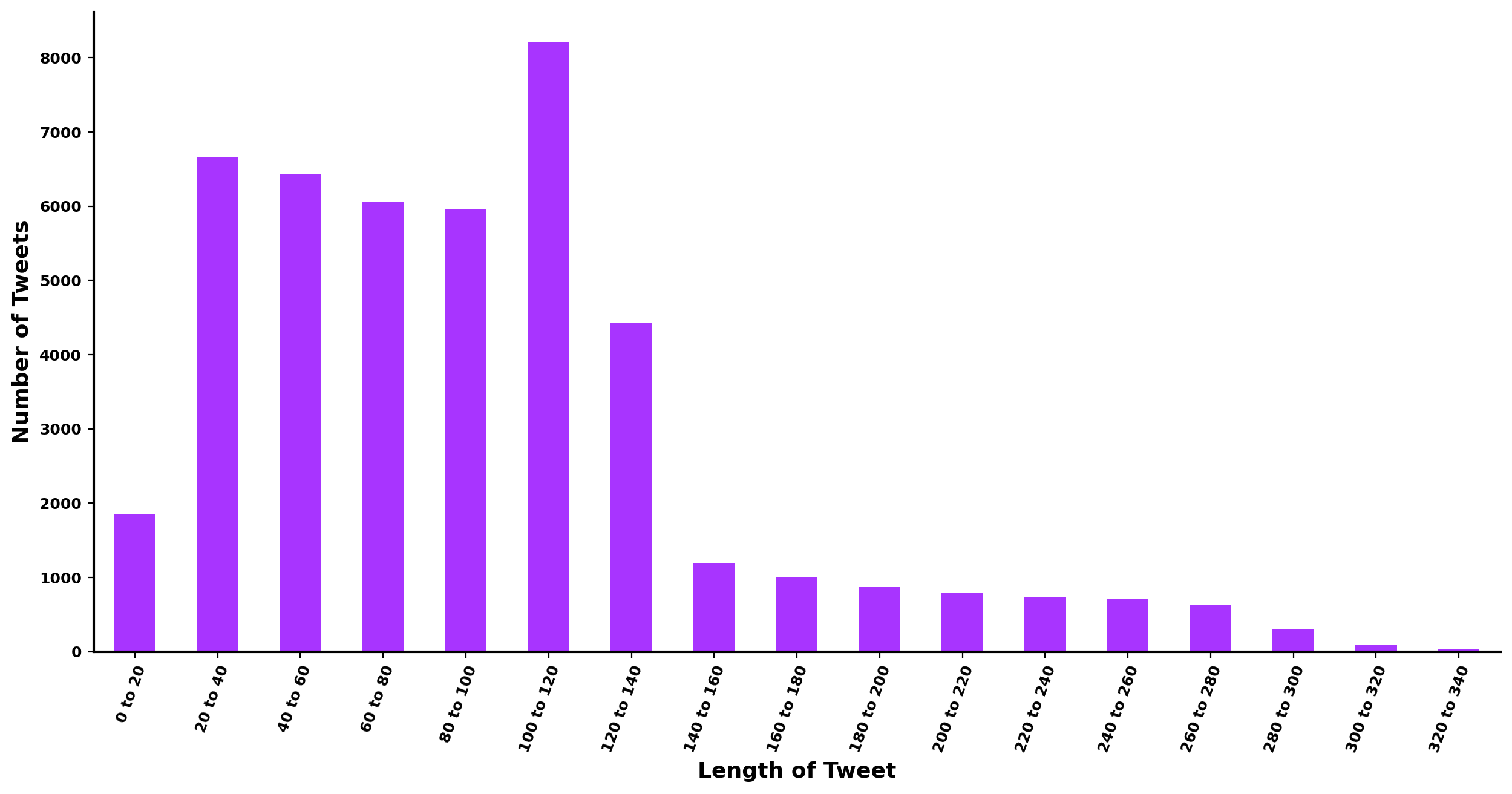}
    \end{center}
    \caption{Distribution of Post-mortem Tweets by Length}
    \label{chap5:post_mortem_lenght}  
\end{figure}

\begin{figure}
    \begin{center}
        \includegraphics[width=18cm,height=10cm]{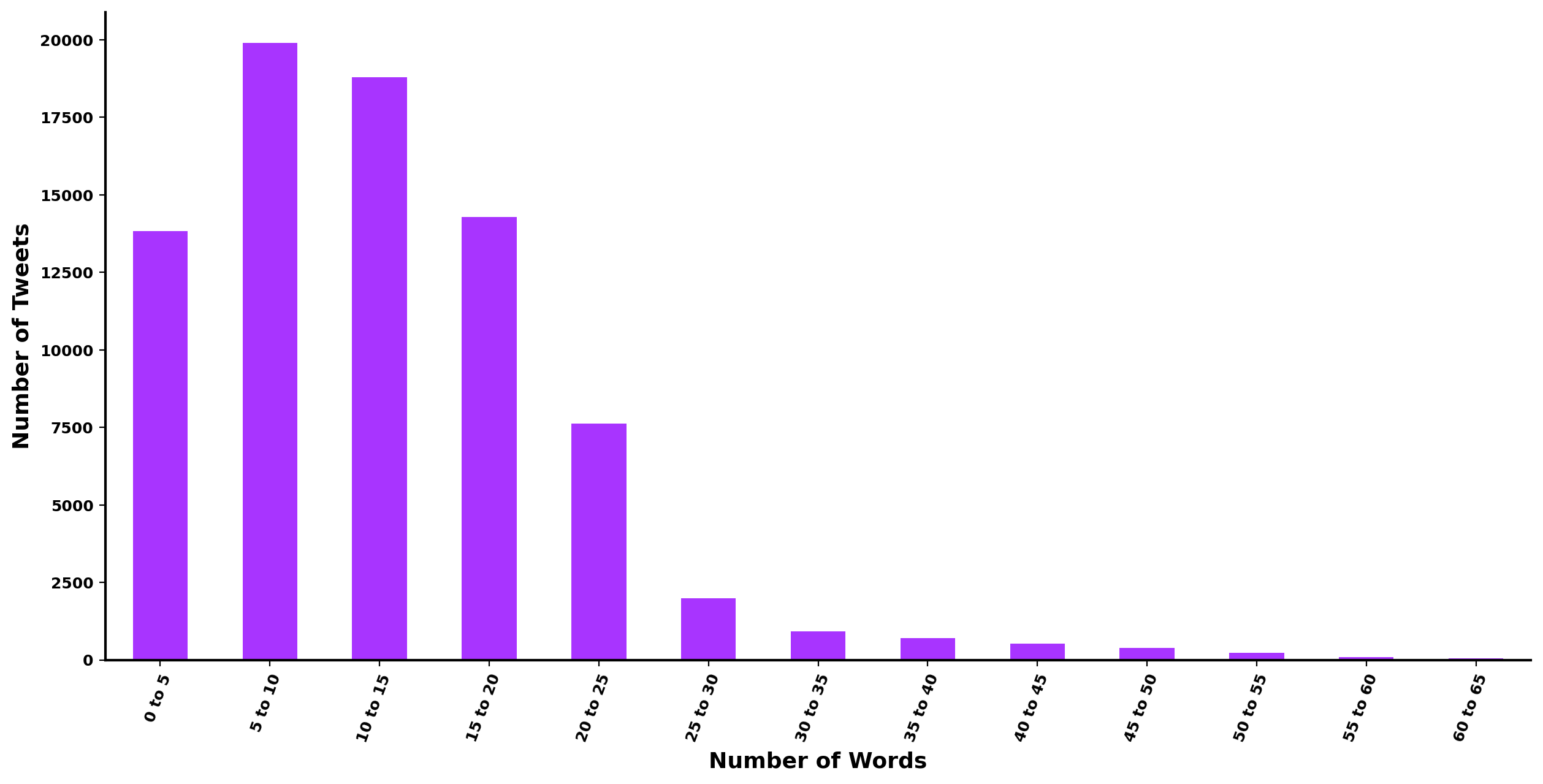}
    \end{center}
    \caption{Distribution of Pre-mortem Tweets by Word Count}
    \label{chap5:pre_mortem_word_count}  
\end{figure}

\begin{figure}
    \begin{center}
        \includegraphics[width=18cm,height=10cm]{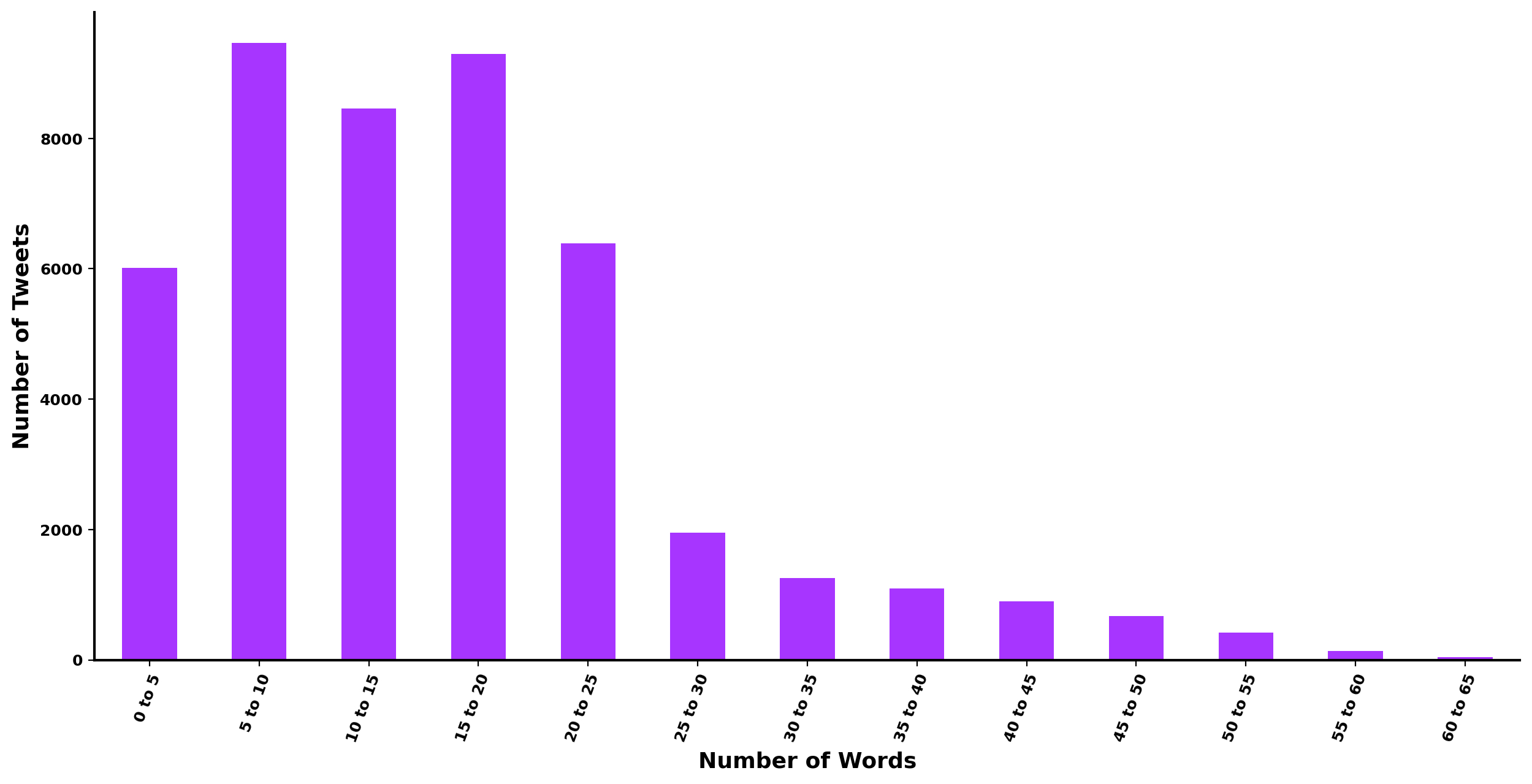}
    \end{center}
    \caption{Distribution of Post-mortem Tweets by Word Count}
    \label{chap5:post_mortem_word_count}  
\end{figure}

\begin{figure}
    \begin{center}
        \includegraphics[width=18cm,height=10cm]{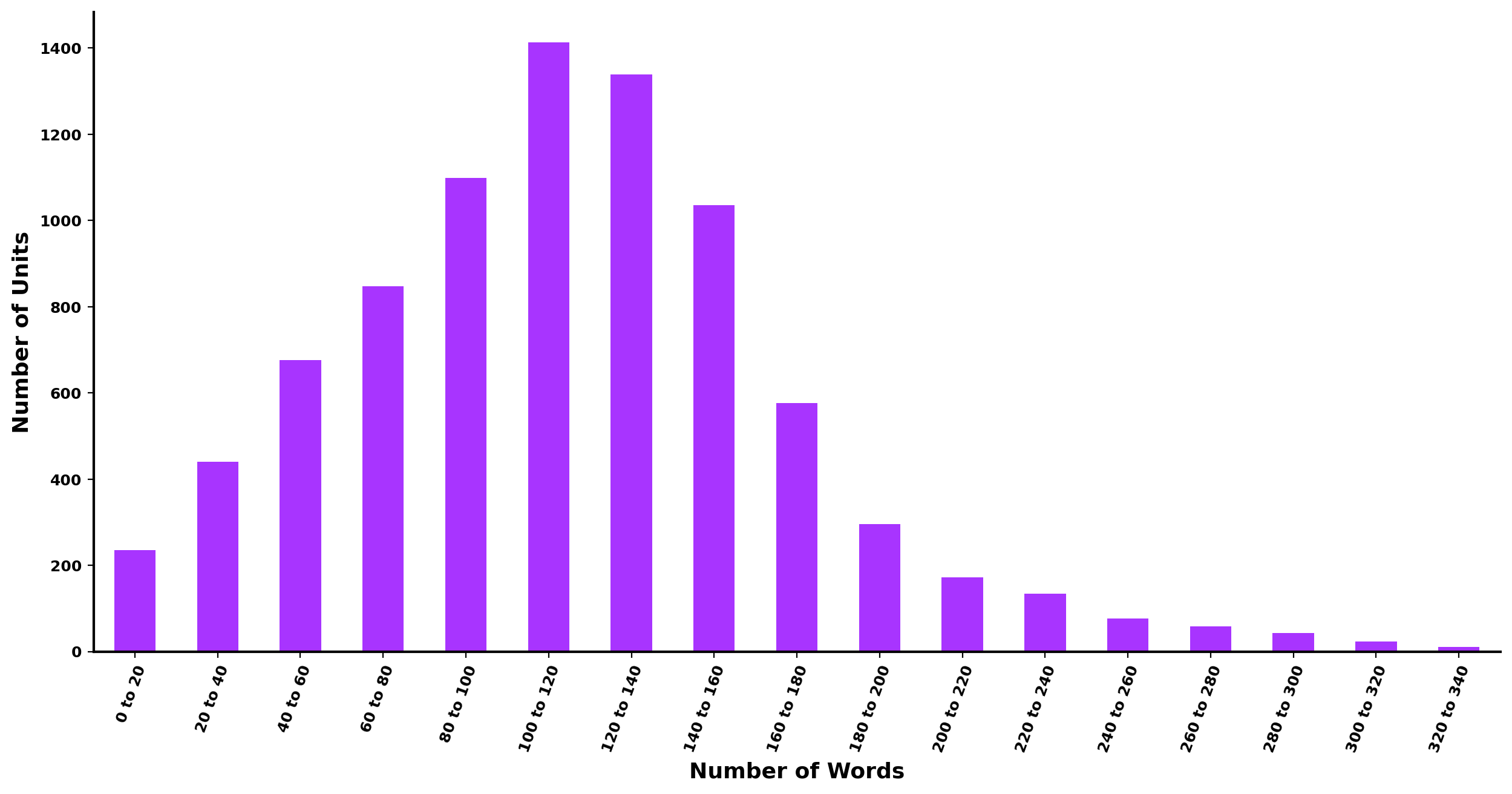}
    \end{center}
    \caption{Distribution of Pre-mortem Units of Analysis by Word Count}
    \label{chap5:dist_pre_mortem_unit_of_analysis}  
\end{figure}

\begin{figure}
    \begin{center}
        \includegraphics[width=18cm,height=10cm]{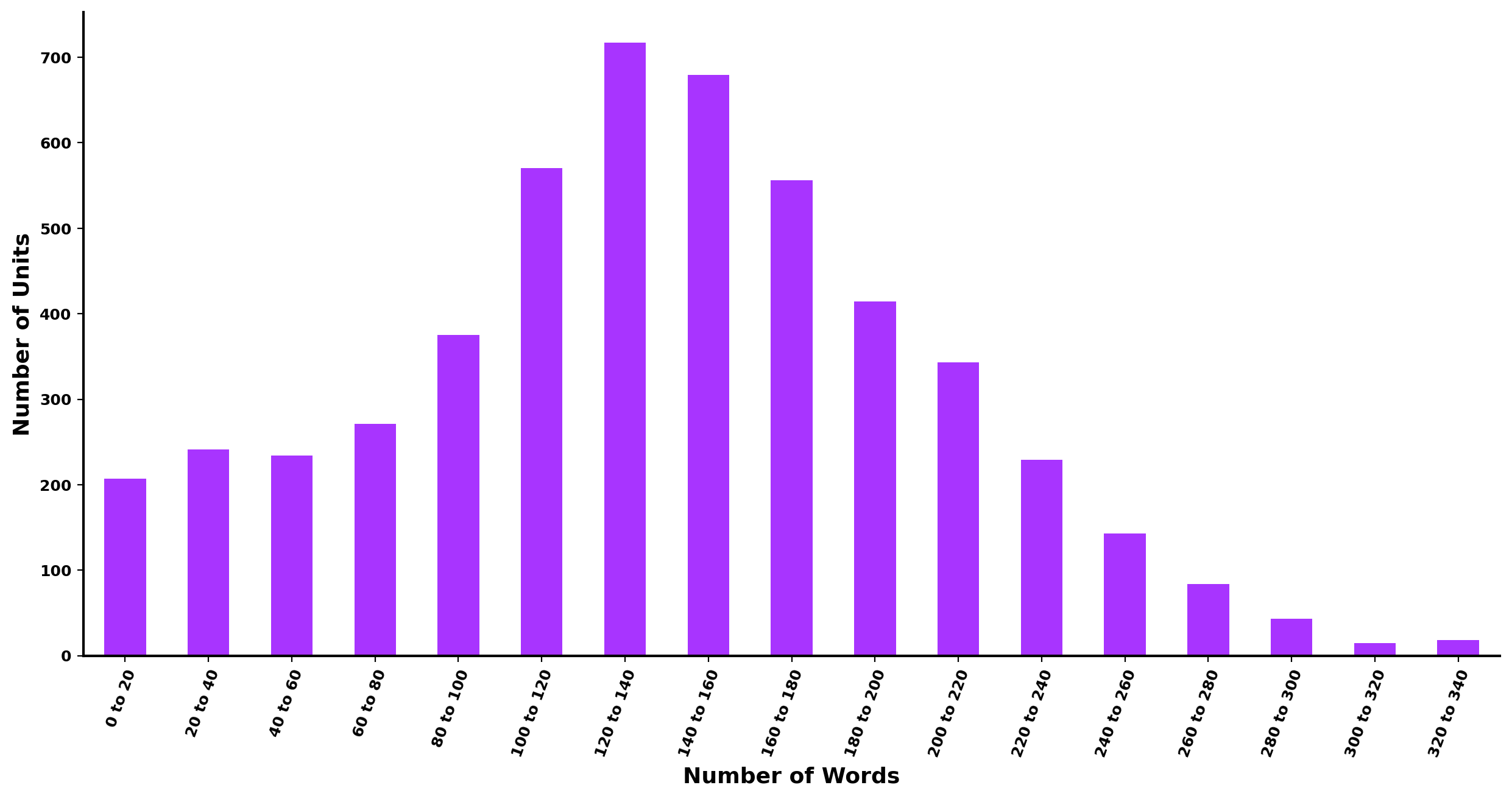}
    \end{center}
    \caption{Distribution of Post-mortem Units of Analysis by Word Count}
    \label{chap5:dist_post_mortem_unit_of_analysis}  
\end{figure}

It was also necessary to explore the characteristics of the collected tweets and silver standard based on their category, i.e., pre-mortem or post-mortem, to establish an understanding that may be useful when deciding some parameters for the text classification, sentiment analysis or emotion analysis pipelines, e.g., when determining the maximum token length in CNN or BERT. Figures \ref{chap5:pre_mortem_lenght}, \ref{chap5:post_mortem_lenght}, \ref{chap5:pre_mortem_word_count}, \ref{chap5:post_mortem_word_count}, \ref{chap5:dist_pre_mortem_unit_of_analysis}, and \ref{chap5:dist_post_mortem_unit_of_analysis} shows the distribution of pre-mortem tweets by text length, distribution of post-mortem tweets by length, distribution of pre-mortem tweets by word count, distribution of post-mortem tweets by word count, distribution of pre-mortem units of analysis by word count and distribution of post-mortem units of analysis by word count, respectively. Even though the distributions between the pre-mortem and their post-mortem counterparts in the previously mentioned figures are not identical, it would be observed that there are no significant differences between the number of words or length of text used in pre-mortem conversation and in grieving or memorialising the dead, i.e., post-mortem conversations.

\begin{figure}
    \begin{center}
        \includegraphics[width=18cm,height=10cm]{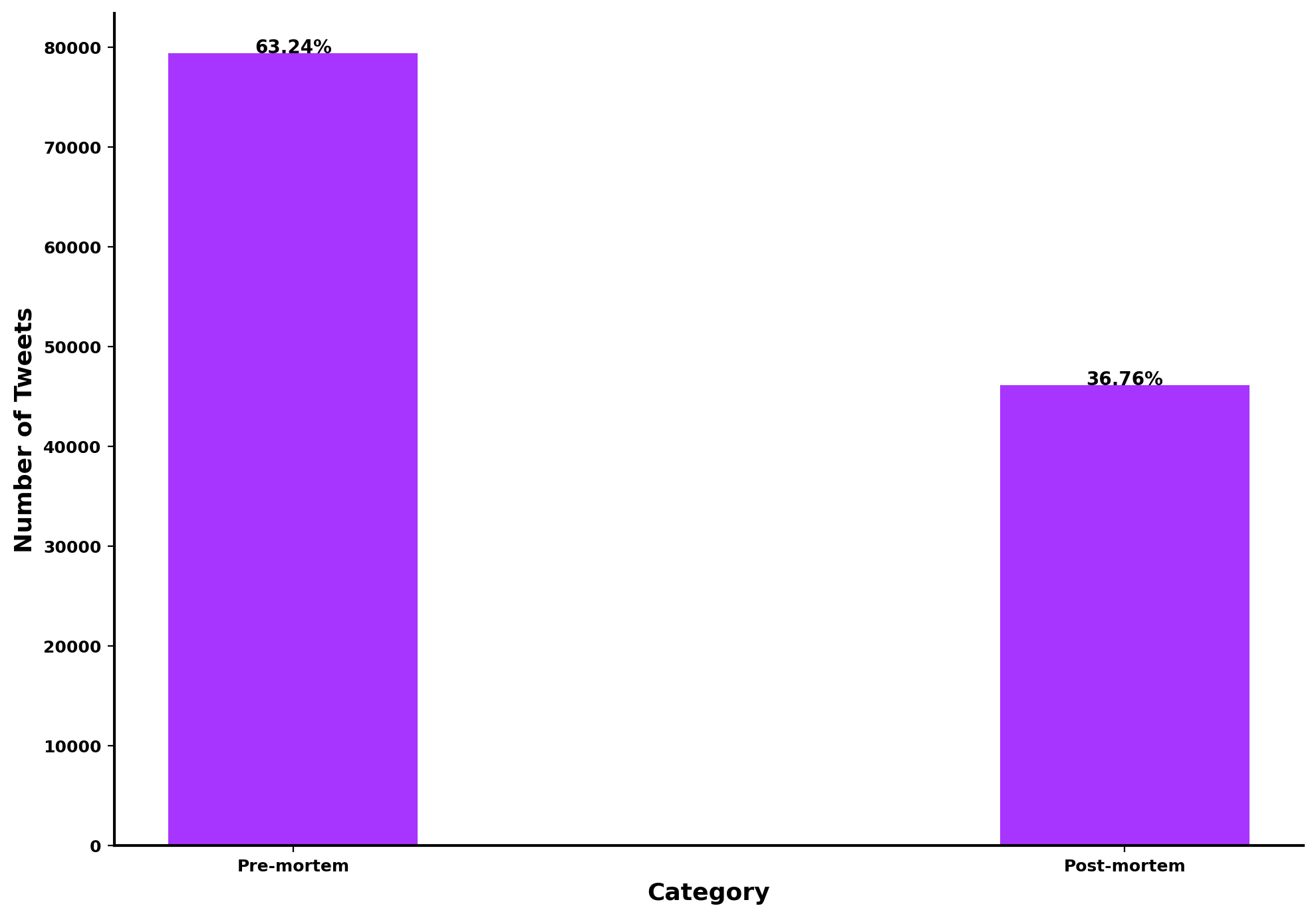}
    \end{center}
    \caption{Distribution of Collected Tweets by Category}
    \label{chap5:tweets_by_category}  
\end{figure}

\begin{figure}
    \begin{center}
        \includegraphics[width=18cm,height=10cm]{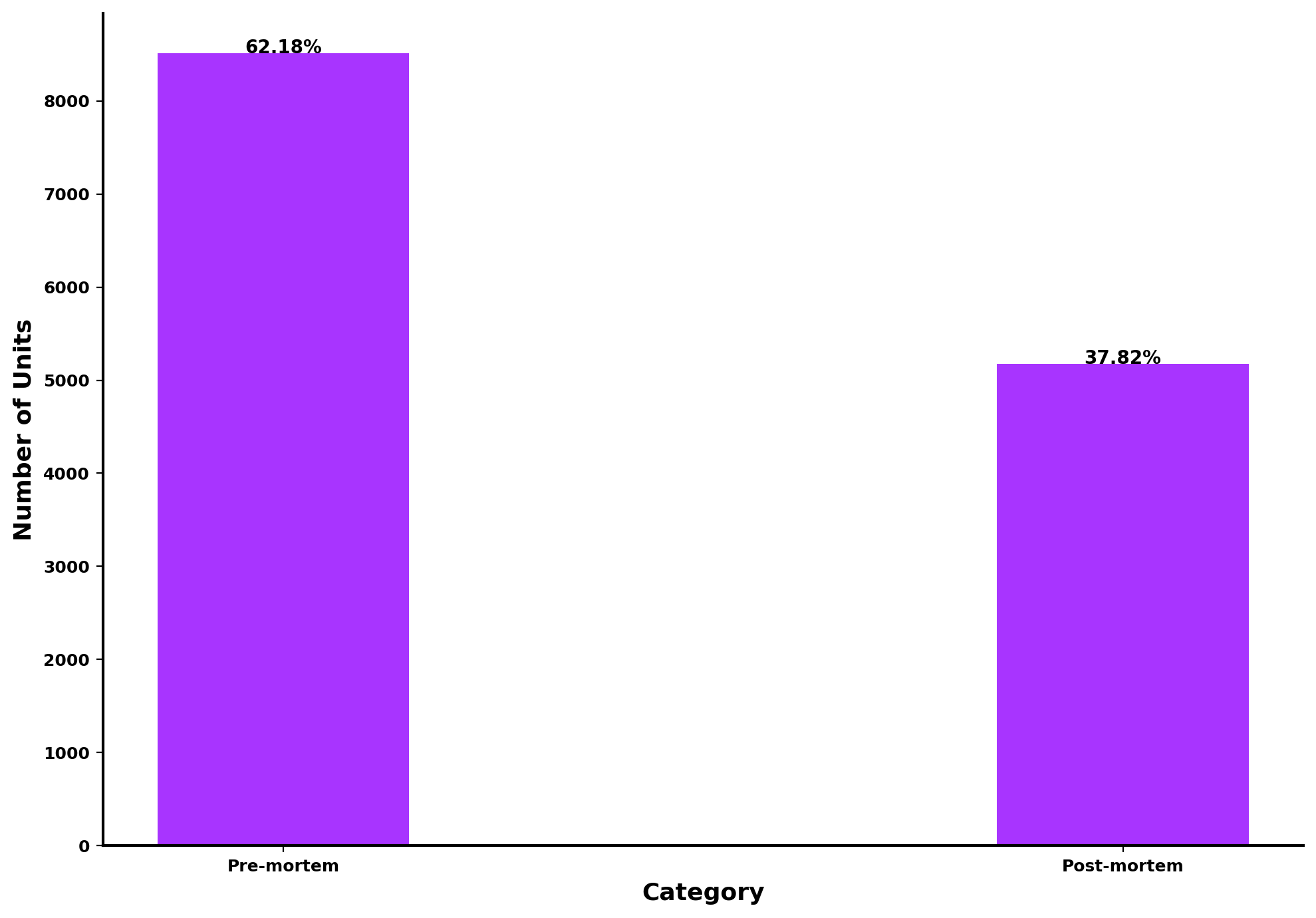}
    \end{center}
    \caption{Distribution of Units of Analysis by Category (Before Balancing)}
    \label{chap5:units_by_category_bb}  
\end{figure}

\begin{figure}
    \begin{center}
        \includegraphics[width=18cm,height=10cm]{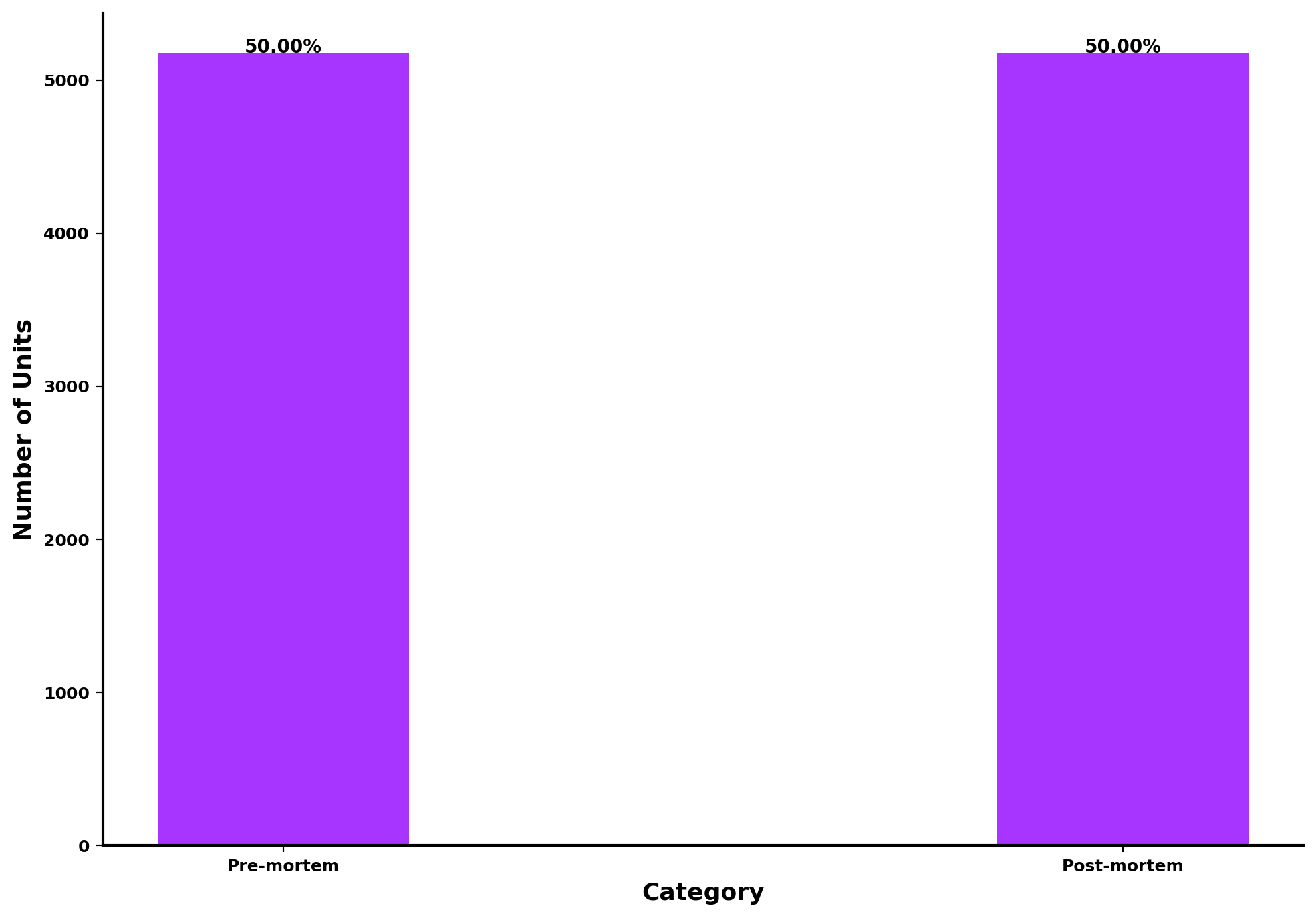}
    \end{center}
    \caption{Distribution of Units of Analysis by Category (After Balancing)}
    \label{chap5:units_by_category_ab}  
\end{figure}

In Figures \ref{chap5:tweets_by_category}, \ref{chap5:units_by_category_bb}, and \ref{chap5:units_by_category_ab}, the sizes of the collected tweets and silver standard dataset (both before and after balancing) were visualised based on the proportions of different underlying categories, i.e., pre-mortem or post-mortem. It can be observed from Figures \ref{chap5:tweets_by_category} and \ref{chap5:units_by_category_bb} that the sizes of the pre-mortem and post-mortem categories in collected tweets and the silver standard dataset are imbalanced. The imbalance informed the investigation of the impact, need and techniques that should be adopted for overcoming the imbalanced dataset problem and finally, the application of the random under-sampling approach that reduced the size of the dataset as shown in Figure \ref{chap5:units_by_category_ab} (see Section \ref{cha4:handling_imbalance_dataset} of Chapter \ref{chapter4} for more details). 

\begin{figure}
    \begin{center}
        \includegraphics[width=18cm,height=10cm]{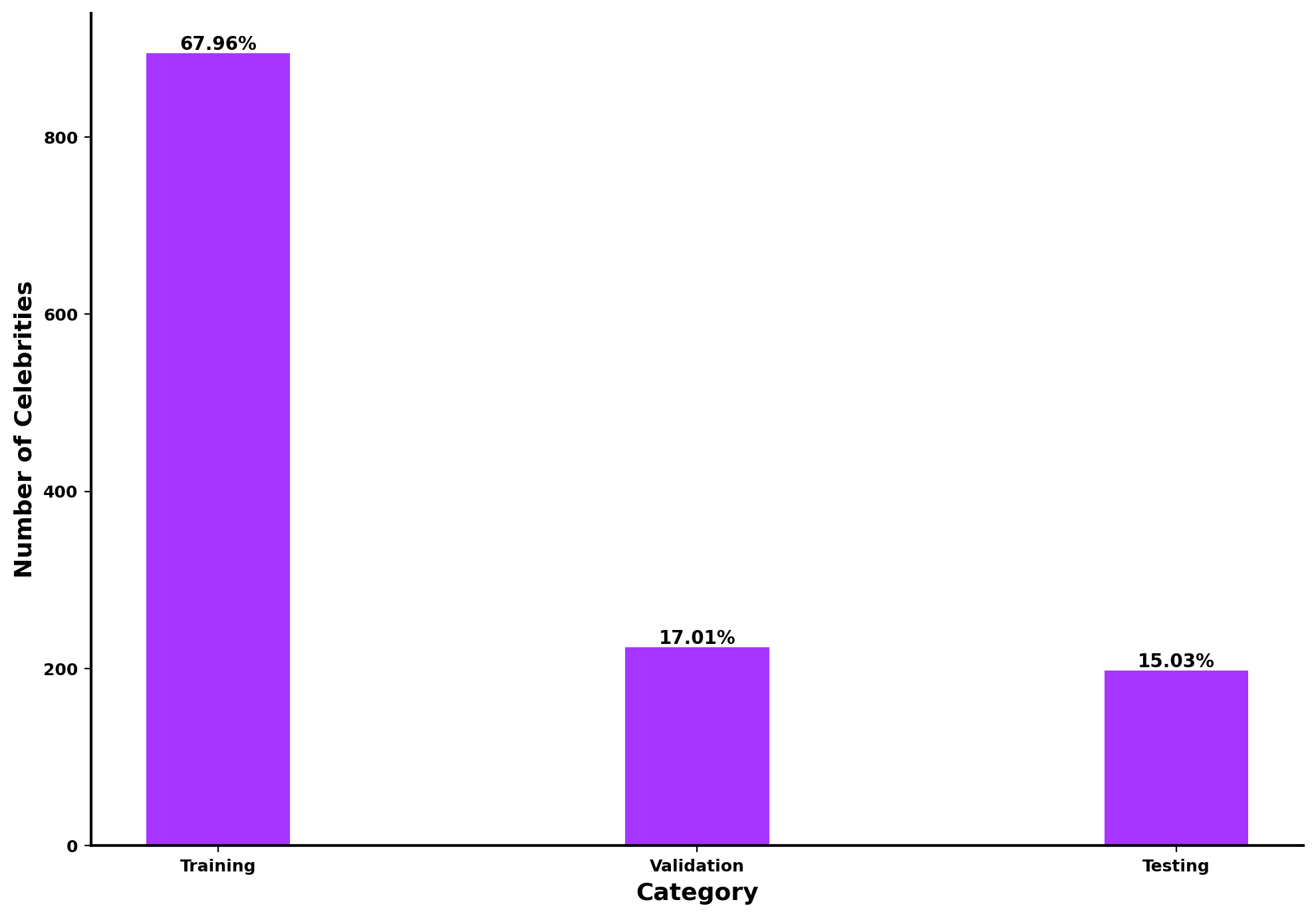}
    \end{center}
    \caption{Distribution of Celebrities for Training, Validation and Testing }
    \label{chap5:dist_celebrities_split}  
\end{figure}

\begin{figure}[!ht]
    \begin{center}
        \includegraphics[width=18cm,height=10cm]{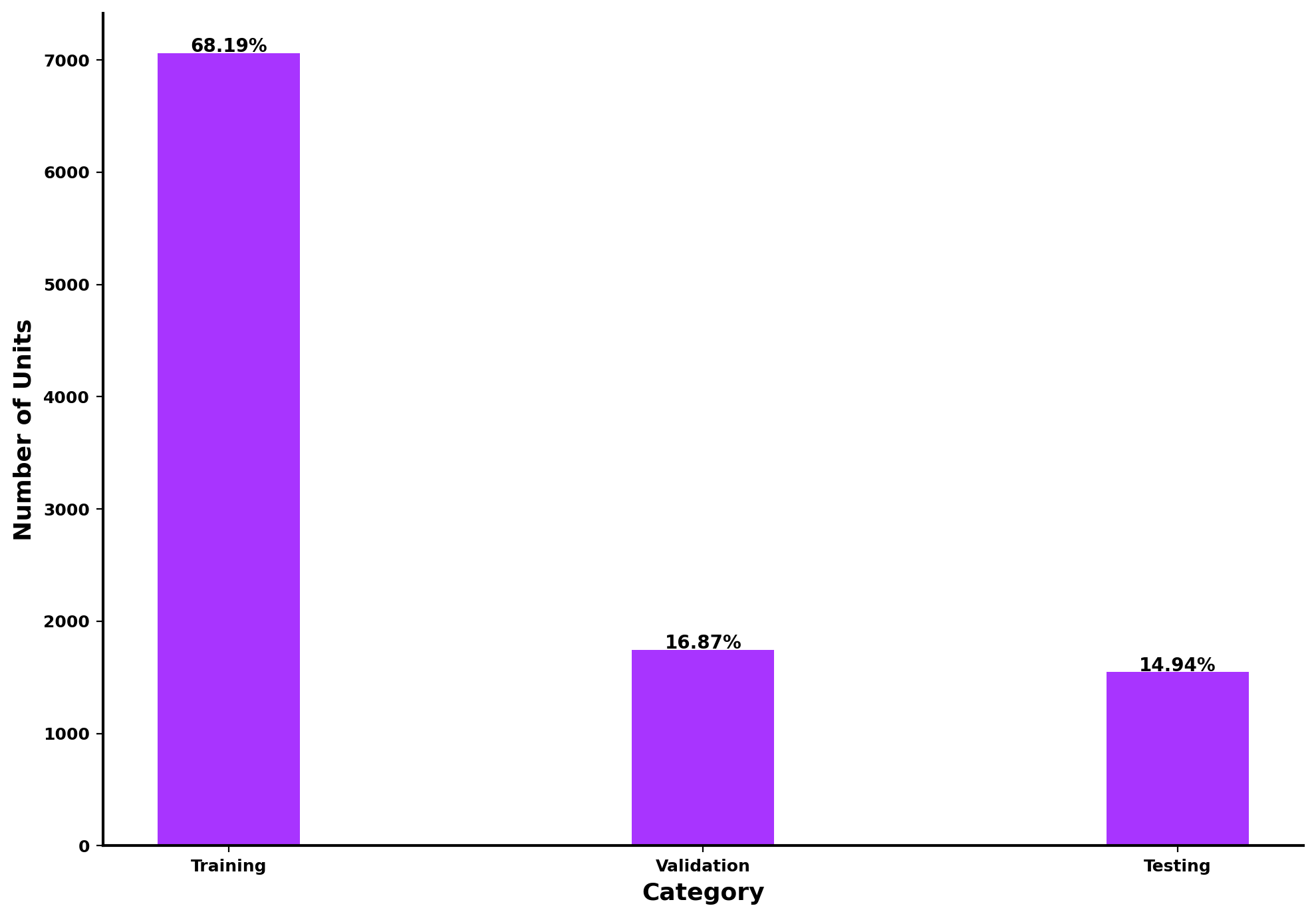}
    \end{center}
    \caption{Distribution of Dataset for Training, Validation and Testing }
    \label{chap5:dist_units_split}  
\end{figure}

The approach for splitting the dataset into training, validation and testing was discussed in Section \ref{cha4:splitting_the_dataset} of Chapter \ref{chapter4}, where it was explained that splitting was applied based on the criteria of the celebrities the units are about and not based on the criteria of the units themselves. This is to ensure that if a particular celebrity is selected to be in the testing subset, all their units, both from post- and pre-mortem categories must be used only in testing to avoid a leak of information to the machine learning algorithms during training, validation or testing stages. Figure \ref{chap5:dist_celebrities_split} shows the distribution of celebrities for testing, validation and testing and Figure \ref{chap5:dist_units_split} shows the follow-up distribution of their respective pre-mortem and post-mortem units for training, validation and testing. 

\section{Inter-annotator Agreement}
\label{cha5:iaa}
The motivation for conducting a manual annotation of a randomly selected sample of the silver standard dataset was discussed in Section \ref{cha4:iaa} of Chapter \ref{chapter4}. Thus, about 10\%, i.e., 1,200 units from the silver standard spread equally from the pre-mortem and post-mortem categories, were randomly selected and manually annotated. Following the annotation of the selected sample by two human annotators, Cohen's Kappa \cite{cohen1960coefficient} was then used to estimate the agreement between these two annotators and between the two annotators and the silver standard dataset. Table \ref{table:kappa_scores} shows the kappa scores and agreements for each comparison combination, i.e., between Annotator 1 and Annotator 2, Annotator 1 and silver standard, and Annotator 2 and silver standard. In all of the comparisons conducted, kappa scores (k) between 0.61 and 0.80 that interprets to ``substantial agreement" were obtained (see Section \ref{cha2:kappa} in Chapter \ref{chapter2} for more details on kappa scores and their respective interpretations). These scores altogether indicate strong agreement between the human annotators and between the human annotators and the automatic annotation, and demonstrate the accuracy and instill confidence in the automatic annotation procedure. 

\begin{table}[!ht]
    \begin{tabular}{|l|c|l|}
        \hline
        \multicolumn{1}{|c|}{\textbf{Annotators}} & \textbf{Kappa score (k)} & \multicolumn{1}{c|}{\textbf{Agreement}} \\ \hline
        Annotator 1 \& Annotator 2         & 0.68 & Substantial Agreement \\ \hline
        Annotator 1 \& Automatic Annotation & 0.65 & Substantial Agreement \\ \hline
        Annotator 2 \& Automatic Annotation & 0.63 & Substantial Agreement \\ \hline
    \end{tabular}
    \protect\caption{\label{table:kappa_scores}Agreements between two human annotators and automatic annotation}
\end{table}

\section{Experimental Results}
\label{cha5:experimental_results}
This section evaluates and discusses the results obtained from the different text analysis pipelines applied on the dataset: text classification, sentiment analysis and emotion analysis. Further investigations were conducted to: check the specific features that influenced the outcomes of the predictions by the machine learning models, check whether the performances of the machine learning models can be improved by performing additional tests and ablation studies, and conduct error analysis to understand the misclassified examples in the testing dataset.

\subsection{Text Classification}
As earlier discussed in Section \ref{cha4:evaluation_metrics} of Chapter \ref{chapter4}, numerous available evaluation metrics: accuracy, precision, recall, F1-Score, ROC and AUC were used to evaluate how well the machine learning models can optimally predict outputs of totally unknown input samples, i.e., the testing subset. These results will help to conclude on the model and the feature engineering technique that would be the most appropriate for automatically detecting deaths from social networking sites.

A DummyClassifier that makes predictions using simple rules (the most frequent selection strategy in this experiment) was first trained and evaluated on the testing subset to be used as a simple baseline. Afterwards, four commonly used traditional machine learning classifiers: RF, KNN, LR and SVM, two deep learning classifiers: BiLSTM and CNN, and the state-of-the-art BERT were then implemented and evaluated on the testing subset to compare with the baseline. The traditional machine learning classifiers were trained on features extracted using TF-IDF and pre-trained embeddings, i.e., Glove, Word2Vec and Fasttext and the deep learning machine learning classifiers were trained on features extracted using pre-trained embeddings, i.e., Glove, Word2Vec and Fasttext. For the state-of-the-art BERT, the bert-base-uncased that was trained on English text was fine-tuned on the dataset. 

\subsubsection{Findings}
\label{cha5:classification_results}
Table \ref{table:classification_scores} shows the performances of each machine learning model with the different feature engineering techniques. It was observed that the baseline classifier that applied the most frequent selection strategy did not perform well; it achieved precision, recall, accuracy, F1-score and AUC of 0.24952, 0.50000, 0.49903, 0.33290 and 0.50, respectively. It was also interesting that all the trained models in this experiment significantly outperformed the baseline model. The best model, i.e., BERT, achieved an outstanding accuracy of 0.91597, recall of 0.91593, F1-score of 0.91593 and AUC score of 0.92 when its predictions of previously unseen data samples were evaluated. The results of the experiment show that for all the models trained using the traditional machine learning algorithms, i.e., RF, KNN, LR and SVM, RF has the best recall, accuracy, F1-score and AUC, i.e., 0.91203, 0.91209, 0.91201, and 0.91, respectively and SVM has the best precision, i.e., 0.91808; also, all traditional machine learning models trained on features extracted using TF-IDF consistently outperformed others trained on features extracted using Glove, Word2Vec and Fastest. Also, for all the models trained using the deep learning machine learning algorithms, i.e., BiLSTM and CNN, BiLSTM has the best precision, recall, accuracy, F1-score and AUC, i.e., 0.91087, 0.91078, 0.91080, 0.91079, and 0.91, respectively; also, all deep learning models trained on features extracted using Word2Vec consistently outperformed others trained on features extracted using Glove and Fasttext. Generally, it was observed that the deep learning algorithms: BiLSTM and CNN outperformed the traditional machine learning algorithms: RF, KNN, LR and SVM when they are all applied on pre-trained embeddings: Glove, Word2Vec and Fasttext. Although the BERT model that is a deep learning concept outperformed all other models trained in this experiment, the precisions, recalls and F1-scores obtained in LR and SVM trained using TF-IDF outperformed BiLSTM and CNN trained on pre-trained word embeddings: Glove, Word2Vec and Fastext; though it was only with metrics less than 0.005 in most cases when compared to the performance obtained by the BiLSTM model trained on Word2Vec features.

\begin{table}[!ht]
\begin{tabular}{|c|c|l|l|l|l|l|l|}
\hline
\textbf{Concept} &
  \textbf{Classifier} &
  \multicolumn{1}{c|}{\textbf{Feature}} &
  \multicolumn{1}{c|}{\textbf{Precision}} &
  \multicolumn{1}{c|}{\textbf{Recall}} &
  \textbf{Accuracy} &
  \multicolumn{1}{c|}{\textbf{F1}} &
  \multicolumn{1}{c|}{\textbf{AUC}} \\ \hline
\multirow{17}{*}{Traditional}  & \multicolumn{2}{c|}{Baseline}      & 0.24952          & 0.50000          & 0.49903          & 0.33290          & 0.50          \\ \cline{2-8} 
                               & \multirow{4}{*}{RF}     & TF-IDF   & 0.91334          & 0.91203          & 0.91209          & 0.91201          & 0.91          \\ \cline{3-8} 
                               &                         & Glove    & 0.82109          & 0.81893          & 0.81900          & 0.81868          & 0.82          \\ \cline{3-8} 
                               &                         & Word2Vec & 0.86221          & 0.85703          & 0.85714          & 0.85661          & 0.86          \\ \cline{3-8} 
                               &                         & Fasttext & 0.84482          & 0.84219          & 0.84228          & 0.84196          & 0.84          \\ \cline{2-8} 
                               & \multirow{4}{*}{KNN}    & TF-IDF   & 0.74548          & 0.73517          & 0.73497          & 0.73221          & 0.74          \\ \cline{3-8} 
                               &                         & Glove    & 0.76801          & 0.74816          & 0.74790          & 0.74321          & 0.75          \\ \cline{3-8} 
                               &                         & Word2Vec & 0.78136          & 0.75659          & 0.75630          & 0.75089          & 0.76          \\ \cline{3-8} 
                               &                         & Fasttext & 0.75765          & 0.73397          & 0.73368          & 0.72750          & 0.73          \\ \cline{2-8} 
                               & \multirow{4}{*}{LR}     & TF-IDF   & 0.91758          & 0.91067          & 0.91080          & 0.91041          & 0.91          \\ \cline{3-8} 
                               &                         & Glove    & 0.83584          & 0.83582          & 0.83581          & 0.83581          & 0.84          \\ \cline{3-8} 
                               &                         & Word2Vec & 0.88643          & 0.88422          & 0.88429          & 0.88412          & 0.88          \\ \cline{3-8} 
                               &                         & Fasttext & 0.86735          & 0.86614          & 0.86619          & 0.86607          & 0.87          \\ \cline{2-8} 
                               & \multirow{4}{*}{SVM}    & TF-IDF   & \textbf{0.91808} & 0.91197          & 0.91209          & 0.91176          & 0.91          \\ \cline{3-8} 
                               &                         & Glove    & 0.84498          & 0.84484          & 0.84486          & 0.84484          & 0.84          \\ \cline{3-8} 
                               &                         & Word2Vec & 0.90201          & 0.89777          & 0.89787          & 0.89759          & 0.90          \\ \cline{3-8} 
                               &                         & Fasttext & 0.88551          & 0.88424          & 0.88419          & 0.88419          & 0.88          \\ \hline
\multirow{7}{*}{Deep Learning} & \multirow{3}{*}{BiLSTM} & Glove    & 0.90334          & 0.90234          & 0.90239          & 0.90233          & 0.90          \\ \cline{3-8} 
                               &                         & Word2Vec & 0.91087          & 0.91078          & 0.91080          & 0.91079          & 0.91          \\ \cline{3-8} 
                               &                         & Fasttext & 0.90262          & 0.90242          & 0.90239          & 0.90238          & 0.90          \\ \cline{2-8} 
                               & \multirow{3}{*}{CNN}    & Glove    & 0.87190          & 0.87133          & 0.87136          & 0.87131          & 0.87          \\ \cline{3-8} 
                               &                         & Word2Vec & 0.90285          & 0.90169          & 0.90167          & 0.90175          & 0.90          \\ \cline{3-8} 
                               &                         & Fasttext & 0.89011          & 0.88876          & 0.88871          & 0.88882          & 0.89          \\ \cline{2-8} 
                               & BERT                    & bert-base-uncased & 0.91662          & \textbf{0.91593} & \textbf{0.91597} & \textbf{0.91593} & \textbf{0.92} \\ \hline
\end{tabular}
\protect\caption[Performance metrics of the trained machine learning models.]{\label{table:classification_scores}Performance metrics of the trained machine learning models. Model with the best in every metric is shown in \textbf{bold}.}
\end{table}

\subsubsection{Feature Importance and Model Explainability}
\label{cha5:feature_importance}
\begin{figure}[!ht]
    \begin{center}
        \includegraphics[width=18cm,height=10cm]{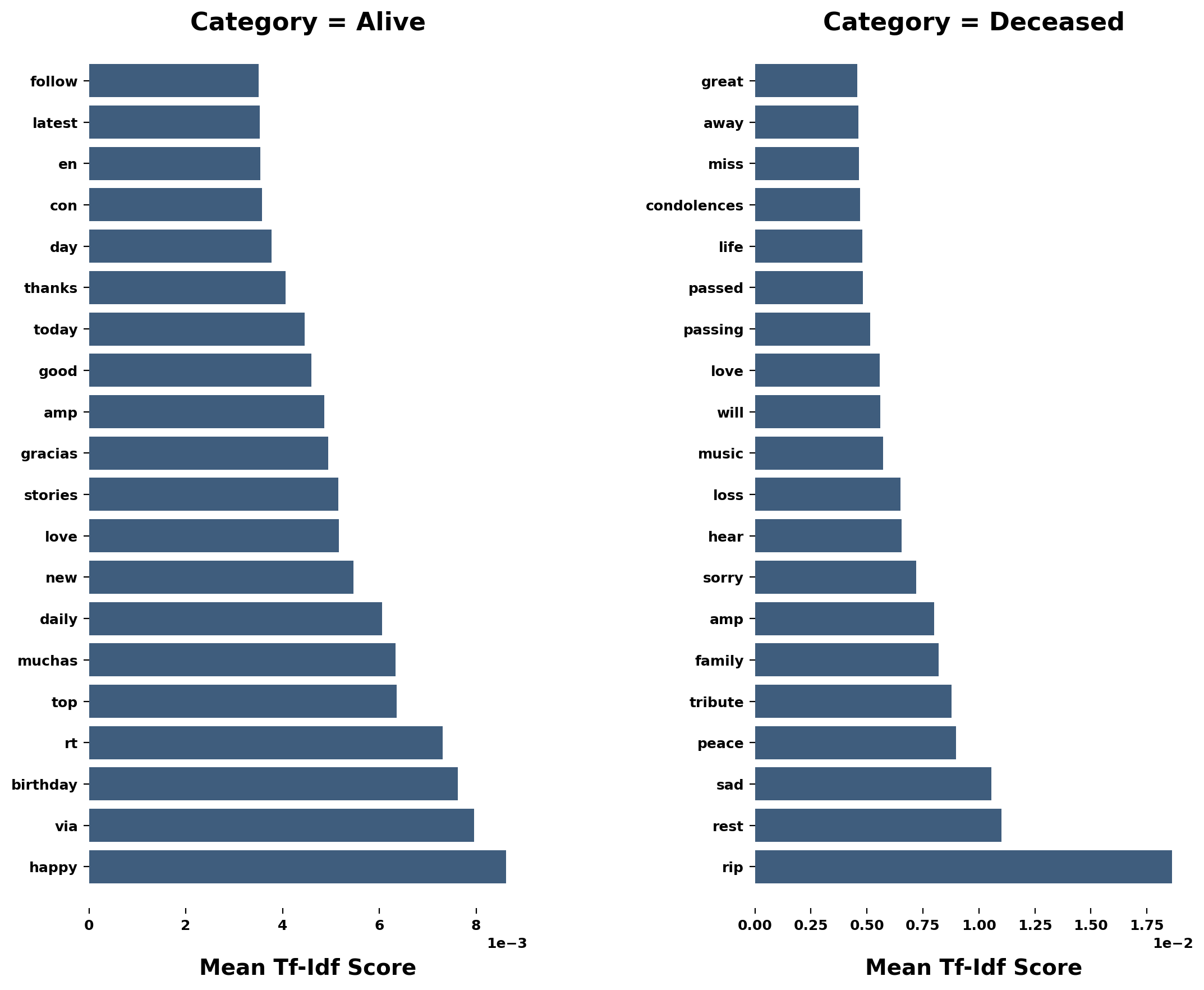}
    \end{center}
    \caption{Top 20 features from each of the pre-mortem and post-mortem categories in the training subset.}
    \label{chap5:tf_idf}  
\end{figure}

To understand and investigate whether there is a pattern in the TF-IDF matrix generated from the training dataset, the training subset was divided based on categories (i.e., pre-mortem and post-mortem) and the top 20 features in the TF-IDF matrix of each category were selected and visualised. Figure \ref{chap5:tf_idf} shows the top 20 features in each of the pre-mortem and post-mortem categories. As may be observed from the top 20 features that are visualised, words like ``rip", ``rest", ``sad", ``tribute", etc., are associated with a higher likelihood of being post-mortem and words like ``happy", ``birthday", ``love", ``thanks", etc., are associated with higher likelihood of being pre-mortem.

\begin{figure}[ht]
     \centering
     \begin{subfigure}[b]{0.42\textwidth}
        \centering
        \includegraphics[width=\textwidth, height=4cm]{{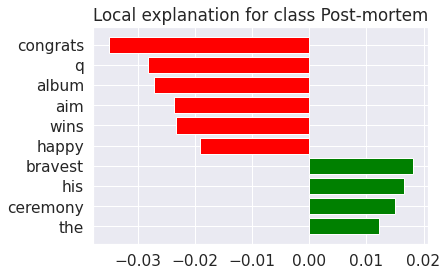}}
        \caption[position=bottom]{}
        \label{chap5:lime_feature_pre1}  
     \end{subfigure}
     \hfill
     \begin{subfigure}[b]{0.42\textwidth}
        \centering
        \includegraphics[width=\textwidth, height=4cm]{{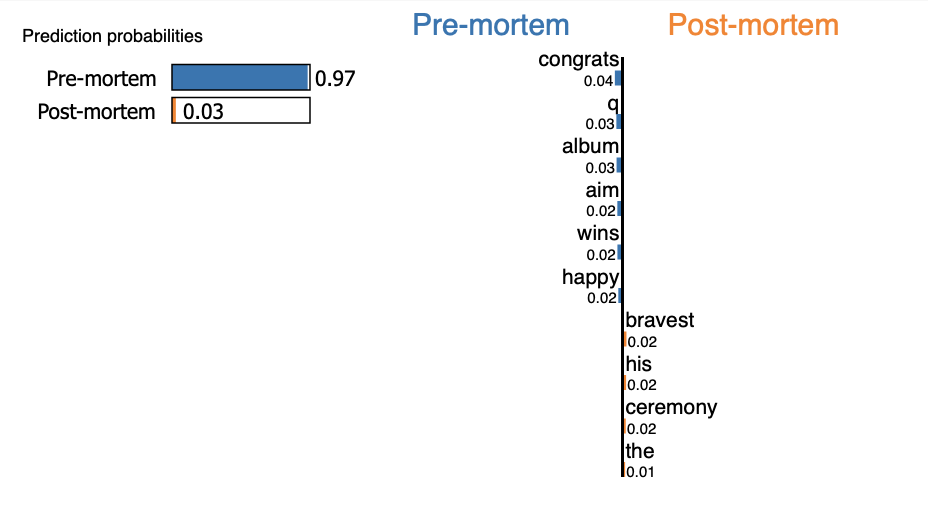}}
        \caption[position=bottom]{}
        \label{chap5:lime_feature_pre2} 
     \end{subfigure}
     \caption{Sample insights generated by LIME when a pre-mortem testing sample was evaluated on a SVM Model trained on Word2Vec Features.}
\end{figure}

\begin{figure}[ht]
     \centering
     \begin{subfigure}[b]{0.42\textwidth}
        \centering
        \includegraphics[width=\textwidth, height=4cm]{{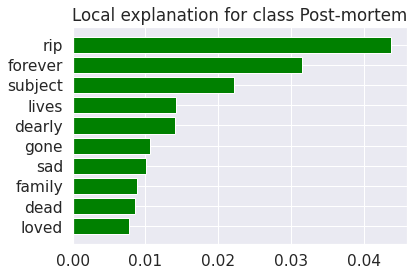}}
        \caption[position=bottom]{}
        \label{chap5:lime_feature_post1}  
     \end{subfigure}
     \hfill
     \begin{subfigure}[b]{0.42\textwidth}
        \centering
        \includegraphics[width=\textwidth, height=4cm]{{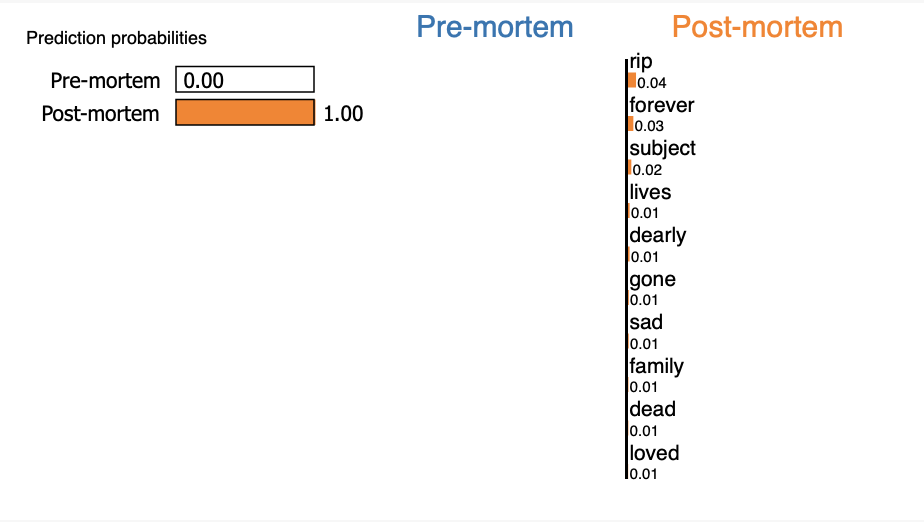}}
        \caption[position=bottom]{}
        \label{chap5:lime_feature_post2} 
     \end{subfigure}
     \caption{Sample insights generated by LIME when a post-mortem testing sample was evaluated on a SVM Model trained on Word2Vec Features.}
\end{figure}

Further, more steps were taken to understand how the value of a specific feature influenced the outcome of an individual prediction. These steps were taken to investigate whether the linearly-nature traditional machine learning algorithms and the black-box-nature deep learning algorithms trained on different feature engineering techniques have assigned similar weights to particular features for scenarios where they reached a similar outcome for an individual prediction. Local Interpretable Model-agnostic Explanations (LIME) \cite{zhang2019should} by Zhang et al., a model explanation algorithm that provides insights into how much each feature contributed to a machine learning model's outcome for an individual prediction was used. It was observed that even when different machine learning algorithms from either of the deep learning- or traditional- machine learning algorithms were trained using different feature engineering techniques, they assigned similar weights to the same features in particular testing samples when they make identical predictions. Figures \ref{chap5:lime_feature_pre1} and \ref{chap5:lime_feature_pre2} show examples of the explanation insights generated by LIME when a post-mortem test sample was assessed on a SVM model trained using features extracted from Word2Vec pre-trained embedding and Figures \ref{chap5:lime_feature_post1} and \ref{chap5:lime_feature_post2} show example explanations generated when a pre-mortem sample was evaluated with the same model. In Figures \ref{chap5:lime_feature_pre1} and \ref{chap5:lime_feature_post1}, the bars in red and green show the extent to which tokens suggest that a text is pre-mortem or post-mortem, respectively. Also, when LIME assigns negative values to a token, it means the token has a higher likelihood of being pre-mortem, whereas positive values indicate the token has a higher likelihood of being post-mortem.

\subsubsection{Additional Tests and Ablation Study}
\label{cha5:additional_tests}
Further experiments that involved altering some of the parameters and settings of the machine learning models and removing some ``feature(s)" of the deep learning algorithms were performed and the following were observed: 

\begin{enumerate}
    \item The BERT and CNN models achieved highest performances when the model's sequence length is obtained from averaging the number of tokens in all the training samples rather than using the maximum sequence length or randomly selecting sequence lengths.
    
    \item Although the \emph{``bert-large-uncased"} is trained on higher parameters, i.e., ``layer", ``hidden", ``heads", ``parameters", and it occupies more than two times memory size compared to the \emph{``bert-base-uncased"}, the \emph{``bert-base-uncased"} attained higher performances than the  \emph{``bert-large-uncased"} when fine-tuned on the dataset.
    
    \item Two \emph{1D} convolutional layers give the CNN model a better chance of learning the underlying representations in the features generated from the dataset; hence helping the trained models to reach their highest performances. However, increasing the layers to three does not increase the performance of the model; rather, it dropped slightly.
    
    \item The number of epochs used in training does not matter a lot as early stopping and a checkpoint were implemented to ensure that training stops when five consecutive epochs do not improve the model's performance and that the model with the best validation loss during all the epochs is saved and returned to be used to evaluate the training data subset.
\end{enumerate}

\subsubsection{Error Analysis}
\label{cha5:error_analysis}
Error analysis was conducted to investigate the misclassified data observations in the testing subset. Because many different models were obtained, only the best models obtained for every machine learning algorithm were closely investigated. The misclassified test samples for each model were obtained and closely observed; the observations are described below. Importantly, it was ensured that only tweets or a collection of tweets that cannot be easily traced back through web or Twitter search to the Twitter usernames they originated from are presented. Also, the names, usernames or hashtags in the presented tweets have been omitted according to ethical requirements discussed in Section \ref{cha3:ethical_considerations} of Chapter \ref{chapter2}.

Firstly, the test samples that the state-of-the-art BERT model misclassified were first analysed because it achieved the best performance amidst all the models that were trained in the experiment. It was observed that most of the pre-mortem samples that the BERT model misclassified as post-mortem resemble post-mortem messages because they use words that indicate death, and most of the samples that the BERT model misclassified as pre-mortem are either very short or use words that do not clearly show that the subject is being grieved or memorialised. For instance, this test sample \emph{``\#SUBJECT realizing that you're a hater is the worst \_*\_ i love you mum. i will spend the rest of my life trying to be as amazing as u. RIP. \_*\_ Happy Birthday booboo I love you, I hope it's one for the books. Hopefully I'll see you soon \_*\_ this will make your whoooole daaaay. BRO. \_*\_ \#SUBJECT i was in the shower, imma call after i get dressed :)"} was misclassified as post-mortem. The reason for the misclassification may be because it contain the word ``RIP" that strongly suggest that the subject the texts were written about is deceased. However, this is particularly a scenario in which the author of this tweet is the subject celebrity and they are memorialising their mum. Additionally, the sample \emph{``\#SUBJECT you always made sure I was a good person \_*\_ Just left the loves \#SUBJECT and many more. I had a good time! \_*\_ I am glad I was able to experience you before you went back home \#SUBJECT"} was misclassified as pre-mortem, just like in most other misclassified post mortem samples, it does not contain words that suggest that the subject they were written about is deceased; thus, even a human annotator may be unsure.

Furthermore, the misclassified examples in other high performing models, i.e., Random Forest trained on TF-IDF features and BiLSTM trained on Word2Vec features, were also investigated. It was observed that out of the around 140 testing samples misclassified by each of these models (state-of-the-art BERT, Random Forest trained on TF-IDF features and BiLSTM trained on Word2Vec features), slightly over 60\% were similarly misclassified by all the models. This indicates that although these models were trained using different algorithms and feature engineering techniques, they misclassified similar testing samples. This shows that most of the misclassifications are not a result of poor algorithm architecture, hyperparameter tuning or feature engineering technique but likely from the data. It may be recalled from Section \ref{cha4:silver_standard} of Chapter \ref{chapter4} that the dataset used in this experiment was automatically labelled; thus, this may be some errors resulting from mislabelling. Although inter-annotator agreements obtained in Section \ref{cha5:iaa} show strong agreements between the human annotators and the automatic annotation, an 100\% agreement was not obtained, and so, achieving a 100\% performance evaluation by the machine learning algorithms trained with the dataset would be unlikely.

\subsection{Analysis of Linguistic Characteristics and Practices}
The collected post-mortem and pre-mortem tweets were analysed separately using various techniques and tools to detect and discuss the differences between the embedded linguistic characteristics and practices. The results obtained are discussed below.

\subsubsection{Sentiment Analysis}
\label{cha5:sentiment_analysis}
As earlier discussed in Section \ref{cha4:sentiment_analysis} of Chapter \ref{chapter4}, VADER was used to separately estimate the sentiments expressed in the collected 79,431 pre-mortem and 46,180 post-mortem tweets to enable a comprehensive comparison between the similarities and differences in the linguistic practices in both. 

\begin{figure}[!ht]
     \centering
     \begin{subfigure}[b]{0.42\textwidth}
        \centering
        \includegraphics[width=\textwidth, height=4cm]{{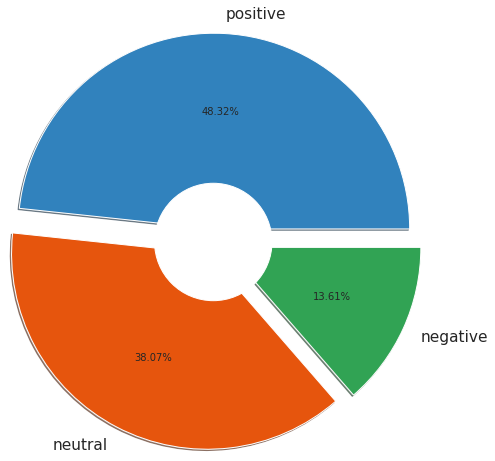}}
        \caption[position=bottom]{}
        \label{chap5:sentiment_analysis_pre}  
     \end{subfigure}
     \hfill
     \begin{subfigure}[b]{0.42\textwidth}
        \centering
        \includegraphics[width=\textwidth, height=4cm]{{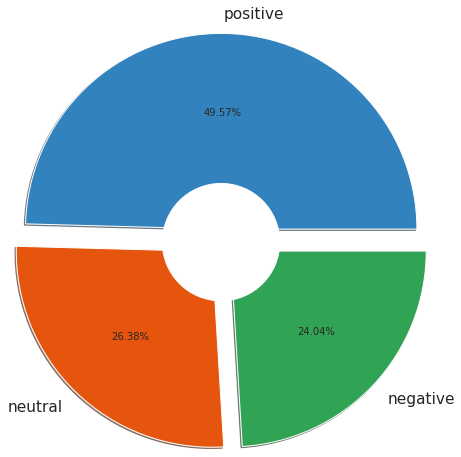}}
        \caption[position=bottom]{}
        \label{chap5:sentiment_analysis_post} 
     \end{subfigure}
     \caption{(a) Proportion of sentiments expressed in pre-mortem tweets. (b) Proportion of sentiments expressed in post-mortem tweets.}
\end{figure}

\paragraph{Findings} \mbox{}\\
The sentiments expressed in the extracted 79,431 pre-mortem and 46,180 post-mortem tweets were estimated separately, and the extent of positive, negative and neutral sentiments was visualised using a pie chart and can be seen in Figures \ref{chap5:sentiment_analysis_pre} and \ref{chap5:sentiment_analysis_post}. It was found that the post-mortem tweets have a significantly higher proportion of negative sentiments, i.e., 24.04\%, than the the pre-mortem counterparts, i.e., 13.61\%. Additionally, although the difference between the proportion of positive sentiments expressed in the pre-mortem tweets and that expressed in the post-mortem tweet is just slightly over 1\%, the difference between the proportion of neutral sentiments expressed in both categories is quite significant, with the pre-mortem and post-mortem tweets expressing neutral sentiments of 38.07\% and 26.36\%, respectively. This shows that although pre-mortem and post-mortem tweets express almost the same extent of positive sentiments, post-mortem tweets express higher negative sentiments while the pre-mortem tweets express higher neutral sentiments.

\subsubsection{Emotion Analysis}
\label{cha5:emotion_analysis}
As earlier discussed in Section \ref{cha4:emotion_analysis} of Chapter \ref{chapter4}, beyond estimating the sentiments, i.e., positive, negative or neutral in texts, Text2emotion was used to separately extract five different feelings, happy, sad, angry, surprise and fear, from the 79,431 pre-mortem and 46,180 post-mortem tweets. This is to enable a comprehensive comparison between the linguistic practices used in pre-mortem and post-mortem tweets.

\begin{figure}[!ht]
     \centering
     \begin{subfigure}[b]{0.42\textwidth}
        \centering
        \includegraphics[width=\textwidth, height=4cm]{{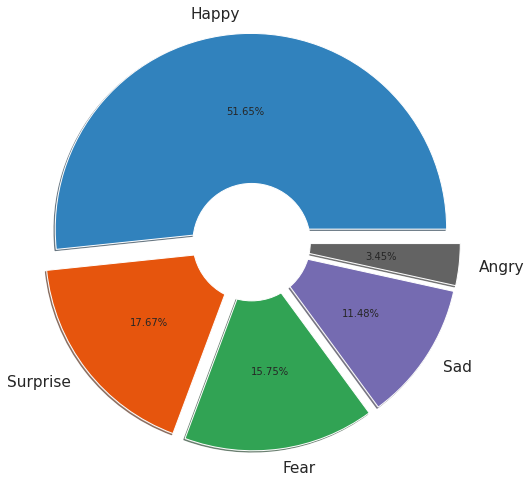}}
        \caption[position=bottom]{}
        \label{chap5:emotion_analysis_pre}  
     \end{subfigure}
     \hfill
     \begin{subfigure}[b]{0.42\textwidth}
        \centering
        \includegraphics[width=\textwidth, height=4cm]{{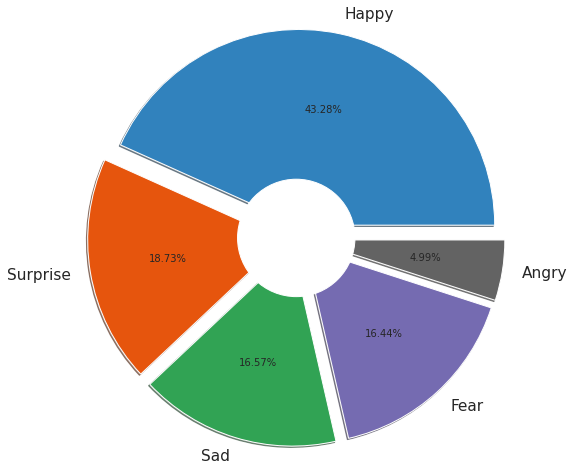}}
        \caption[position=bottom]{}
        \label{chap5:emotion_analysis_post} 
     \end{subfigure}
     \caption{(a) Proportion of the five feelings, happy, sad, angry, surprise and fear, expressed in pre-mortem tweets. (b) Proportion of the five feelings, happy, sad, angry, surprise and fear, expressed in  post-mortem tweets.}
\end{figure}

\paragraph{Findings} \mbox{}\\
The feelings expressed in the extracted 79,431 pre-mortem and 46,180 post-mortem tweets were estimated separately, and the extent of happy, sad, angry, surprise and fear feelings were visualised using a pie chart and can be seen in Figures \ref{chap5:emotion_analysis_pre} and \ref{chap5:emotion_analysis_post}. It was found that the proportion of sad, angry, surprise and fear feelings expressed in the post-mortem tweets, i.e., 16.57\%, 4.99\%, 18.73\%, and 16.44\%, respectively, are higher than those expressed in the pre-mortem tweets, i.e., 11.48\%, 3.45\%, 17.67\%, and 15.75\%, respectively. As expected, among all the five feelings that were estimated, only the proportion of happy feelings expressed in the pre-mortem tweets, i.e., 51.65\%, is higher than that expressed in the post-mortem tweets, i.e., 43.28\%. Thus, this shows that feelings that suggest negativity, i.e., sad, angry, surprise, and fear, are more predominant in post-mortem tweets than in pre-mortem tweets and the feeling that suggests positivity, i.e., happy, is more predominant in pre-mortem tweets than in post-mortem tweets. 

\subsubsection{Language Inquiry and Word Count}
\label{cha5:liwc}
As earlier discussed in Section \ref{cha4:liwc} of Chapter \ref{chapter4}, LIWC was used to separately analyse the 79,431 pre-mortem and 46,180 post-mortem tweets. This is to enable a comprehensive comparison between the linguistic practices used in pre-mortem and post-mortem tweets.

\paragraph{Findings} \mbox{}\\
The mean of every LIWC metric for all the tweets in every category (pre-mortem and post-mortem) was estimated, and only the metrics that are related to the conducted experiment in this research and shows a significant difference between linguistic characteristics in pre-mortem and post-mortem tweets were selected. Table \ref{table:liwc} juxtaposes the results obtained for both pre-mortem and post-mortem tweets categories in the selected LIWC metrics. It was found that the number of words, personal pronouns, verbs, family words, religious words, death words, and swear words in post-mortem tweets are higher than those in pre-mortem tweets; whereas, the number of impersonal pronouns and informal words in pre-mortem tweets are higher than those in post-mortem tweets. Additionally, analytical thinking is more expressed in post-mortem conversations than in pre-mortem conversations.

\begin{table}[!ht]
\begin{tabular}{|l|c|c|}
\hline
\multicolumn{1}{|c|}{\textbf{LIWC Metric}} & \textbf{\begin{tabular}[c]{@{}c@{}}Pre-mortem Tweets\\ (Mean)\end{tabular}} & \textbf{\begin{tabular}[c]{@{}c@{}}Post-mortem Tweets\\ (Mean)\end{tabular}} \\ \hline
Word count                                 & 13.22                                                                       & \textbf{16.41}                                                               \\ \hline
Analytical Thinking                        & 55.27                                                                       & \textbf{58.12}                                                               \\ \hline
Personal Pronouns                          & 5.65                                                                        & \textbf{7.07}                                                                \\ \hline
Impersonal Pronouns                        & \textbf{15.54}                                                              & 12.19                                                                        \\ \hline
Verb                                       & 9.98                                                                        & \textbf{11.32}                                                               \\ \hline
Family words                               & 0.32                                                                        & \textbf{0.55}                                                                \\ \hline
Friend words                               & 0.39                                                                        & \textbf{0.59}                                                                \\ \hline
Religious words                            & 0.478                                                                       & \textbf{0.65}                                                                \\ \hline
Death Words                                & 0.19                                                                        & \textbf{1.01}                                                                \\ \hline
Informal                                   & \textbf{2.99}                                                               & 2.05                                                                         \\ \hline
Swear words                                & 0.27                                                                        & \textbf{0.31}                                                                \\ \hline
Exclamations                               & \textbf{5.57}                                                               & 2.5                                                                          \\ \hline
\end{tabular}
\protect\caption[LIWC analysis of post-mortem and pre-mortem tweets.]{\label{table:liwc}LIWC analysis of post-mortem and pre-mortem tweets. Categories with higher LIWC metric is shown in \textbf{bold}.}
\end{table}

\section{Discussion}
\label{cha5:discussion}
This section aims to use the results obtained in the conducted experiments to propose answers to the research questions that were introduced in Section \ref{cha1:research_questions} of Chapter \ref{cha:intro}.

\paragraph{Are there differences between the linguistic characteristics and practices in pre- and post-mortem posts, and to what extent can humans and text classification models detect the post-mortem posts from their pre-mortem counterparts?}

The experimental results obtained in Sections \ref{cha5:sentiment_analysis}, \ref{cha5:emotion_analysis}, \ref{cha5:liwc} show that, similar to the results obtained by Getty et al. \cite{getty2011said} and Jiang \& Brubaker \cite{jiang2018tending, brubaker2018describing}, post-mortem tweets express higher negative sentiments while the pre-mortem tweets express higher neutral sentiments. Similarly, feelings that suggest negativity, i.e., sad, angry, surprise, and fear, are more dominant in post-mortem tweets than in pre-mortem tweets and the feeling that suggests positivity, i.e., happy, is more dominant in pre-mortem tweets than in post-mortem tweets. Additionally, the number of words, personal pronouns, verbs, family words, religious words, death words and swear words in post-mortem tweets are higher than those in pre-mortem tweets; whereas, the number of impersonal pronouns and informal words in pre-mortem tweets are higher than those in post-mortem tweets. Also, analytical thinking is more expressed in post-mortem conversations than in pre-mortem conversations. Thus, there are differences in the linguistic characteristics and practices of pre-mortem and post-mortem posts as they contain different extents of feelings, sentiments, word counts, etc.

In Section \ref{cha4:iaa} of Chapter \ref{chapter4}, the motivation for conducting a manual annotation of a randomly selected sample of the silver standard dataset was discussed, and in Section \ref{cha5:iaa}, the results obtained from estimating inter-annotator agreement between the two manual annotators and between the two manual annotators and the automatic annotation were presented. The Kappa interpretation of the agreement obtained by the manual annotators on the sample of the annotated dataset is \emph{``strong agreement"}, and the accuracy obtained is slightly over 88\%, i.e., they assigned similar labels to slightly over 88\% of the annotated samples. Thus, this shows that humans can detect online post-mortem contents from their pre-mortem counterparts to a vast extent.

In this experiment, various machine learning algorithms, both traditional: RF, KNN, LR and SVM, and deep learning: BiLSTM, CNN, and the state-of-the-art BERT, that have been reported to perform well on text classification problems were trained on features extracted using varieties of techniques: TF-IDF and famous pre-trained embeddings (Glove, Word2Vec, and Fasttext) and evaluated using various metrics: accuracy, precision, recall, F1-Score, ROC and AUC to classify post-mortem units from their pre-mortem counterparts. The results obtained from this activity is presented in Section \ref{cha5:classification_results}. The BERT model has the best performance with an outstanding accuracy of 0.91597, recall of 0.91593, F1-score of 0.91593 and AUC score of 0.92 when its predictions of previously unseen data samples were evaluated. Besides, many other trained models also achieved outstanding results. This also shows that to a reasonably acceptable extent, text classification models successfully detected online post-mortem contents from their pre-mortem counterparts.

\paragraph{Between text classification models in either the traditional- or deep learning- based machine learning algorithms group and between famous pre-trained word embeddings, which ones will perform better in classifying the post-mortem posts from their pre-mortem counterparts?}

From the results obtained in Section \ref{cha5:classification_results}, it was found that all the traditional machine learning models trained on features extracted using TF-IDF consistently outperformed those trained on features extracted using pre-trained word-embeddings: Glove, Word2Vec and Fastest. Also, it was observed that the deep learning algorithms: BiLSTM and CNN outperformed the traditional machine learning algorithms: RF, KNN, LR and SVM when applied on pre-trained embeddings: Glove, Word2Vec and Fasttext. However, the precisions, recalls and F1-scores obtained in LR and SVM trained using TF-IDF outperformed BiLSTM and CNN trained on pre-trained word embeddings: Glove, Word2Vec and Fastext; though it was only with metrics less than 0.005 in most cases when compared to the performance obtained by the BiLSTM trained on Word2Vec features. TF-IDF was a state-of-the-art feature extraction technique for many years before the discovery of the pre-trained word embeddings \cite{dessi2021tf}; the results obtained from comparing TF-IDF and pre-trained word embeddings: Glove, Word2Vec and Fastext in this experiment show that TF-IDF is still highly relevant for consideration in text classification problems. Finally, the BERT model, which is considered to be deep-learning-based outperformed all other models trained in this experiment, although it is not with significant metrics scores in all cases compared to the results obtained by other high performing models like LR and SVM trained on TF-IDF and BiLSTM trained on Word2Vec.

\paragraph{Will the famous state-of-art Bidirectional Encoder Representations from Transformers (BERT) perform better than the other text classification models (traditional- and deep learning machine learning-based) in classifying post-mortem posts from their pre-mortem counterparts?}

Since the discovery of the BERT model, it has obtained state-of-the-art results on various ``language processing tasks, including pushing the GLUE score to 80.5\% (7.7\% point absolute improvement), MultiNLI accuracy to 86.7\% (4.6\% absolute improvement), SQuAD v1.1 question answering Test F1 to 93.2 (1.5 point absolute improvement) and SQuAD v2.0 Test F1 to 83.1 (5.1 point absolute improvement)" \cite{devlin2018bert}. Similar to these achievements, the ``bert-base-uncased" variant of the BERT model that was pre-trained on English language using a masked language modelling objective achieved the highest recall, accuracy, F1-score and AUC after it was fine-tuned on the dataset developed in this experiment. The BERT model consistently pushed the best recall, accuracy, F1-score, and AUC from 0.91203 to 0.91593, 0.91209 to 0.91597, and 0.91 to 0.92, respectfully. Thus, the BERT model trained in this experiment outperforms all other text classification models, both traditional- and deep learning- machine learning-based in classifying post-mortem posts from their pre-mortem counterparts.

\chapter{Conclusion \& Future Work}
\section{Conclusion}
This dissertation analysed and discussed the differences between the linguistic characteristics and practices in pre-mortem social media contents and their post-mortem counterparts and reported machine learning classifiers that achieved high performance in automatically detecting deaths of users of social networking sites from the posts associated with their profiles. 

To avoid similar limitations as in Ma et al. \cite{ma2017write} and Jian \& Brubakar \cite{jiang2018tending, jiang2018describing}, where the datasets used for training classifiers are not as diverse as in practical social networking sites, are drawn from so many years ago and may not capture the recent linguistic practices used on social networking sites or do not include pre-mortem contents, this dissertation developed a new dataset from open-source platforms by using SPARQL over Wikidata and Twitter academic research API over Twitter. The newly developed dataset was then evaluated through numerous exploratory data analysis steps to ensure that it covers a wide range of people from different backgrounds and age groups and captures linguistic practices from at least ten years ago until recently; this is to create confidence regarding whether the trained classifiers will be able to generalise and remain accurate in different contexts. Because of the excessive labour, cost and time required to develop gold standard datasets through manual annotation, automatic annotation approaches were adopted to develop a silver standard dataset. Two Human annotators were then recruited to annotate a sample of this developed silver standard dataset to establish confidence in the automatic annotation procedure; the agreements between the human annotators and between the human annotators and the automatic annotation were estimated using Cohen's Kappa \cite{cohen1960coefficient} and a result that suggests a strong agreement between all parties was obtained. 

Machine learning algorithms, both traditional: RF, KNN, LR and SVM, and deep learning: BiLSTM, CNN, and the state-of-the-art BERT, that have been reported to perform well on text classification problems were trained on features extracted using varieties of techniques: TF-IDF and famous pre-trained embeddings (Glove, Word2Vec, and Fasttext) and evaluated using various metrics: accuracy, precision, recall, F1-Score, ROC and AUC to classify post-mortem units from their pre-mortem counterparts. The best model, i.e., BERT achieved an outstanding accuracy of 0.91597, recall of 0.91593, F1-score of 0.91593 and AUC score of 0.92 when its predictions of previously unseen data samples were evaluated. It was found that for all the models trained using the traditional machine learning algorithms, i.e., RF, KNN, LR and SVM, RF has the best recall, accuracy, F1-score and AUC, i.e., 0.91203, 0.91209, 0.91201, and 0.91, respectively and SVM has the best precision, i.e., 0.91808; also, all traditional machine learning models trained on features extracted using TF-IDF consistently outperformed those that were extracted using Glove, Word2Vec and Fastest. Also, for all the models trained using the deep learning machine learning algorithms, i.e., BiLSTM and CNN, BiLSTM has the best precision, recall, accuracy, F1-score and AUC, i.e., 0.91087, 0.91078, 0.91080, 0.91079, and 0.91, respectively; also, all deep learning models trained on features extracted using Word2Vec consistently outperformed others trained on features extracted using Glove and Fasttext. Generally, it was observed that the deep learning algorithms: BiLSTM and CNN outperformed the traditional machine learning algorithms: RF, KNN, LR and SVM when they are all applied on pre-trained embeddings: Glove, Word2Vec and Fasttext. Although the BERT model that is a deep learning concept outperformed all other models trained in this experiment, the precisions, recalls and F1-scores obtained in LR and SVM trained using TF-IDF outperformed BiLSTM and CNN trained on pre-trained word embeddings: Glove, Word2Vec and Fastext; though it was only with metrics less than 0.005 in most cases when compared to the performance obtained by the BiLSTM trained on Word2Vec features.

The extent of sentiments (i.e., positive, negative or neutral), feelings (i.e., happy, sad, angry, surprise or fear), and other linguistic characteristics expressed in the pre-mortem and post-mortem tweets that were collected were separately estimated using a sentiment analysis tool, VADER, emotion analysis tool, Text2Emotion, and text analysis tool, LIWC, respectively; this is to enable a comprehensive comparison between the similarities and differences of the linguistic practices in post-mortem tweets and their pre-mortem counterparts. It was found that although pre-mortem and post-mortem tweets express almost the same extent of positive sentiments, post-mortem tweets express higher negative sentiments while the pre-mortem tweets express higher neutral sentiments. Also, feelings that suggest negativity, i.e., sad, angry, surprise, and fear, are more predominant in post-mortem tweets than in pre-mortem tweets and the feeling that suggests positivity, i.e., happy, is more dominant in pre-mortem tweets than in post-mortem tweets. It was also found that the number of words, personal pronouns, verbs, family words, religious words, death words, and swear words in post-mortem tweets are higher than those in pre-mortem tweets; whereas, the number of impersonal pronouns and informal words in pre-mortem tweets are higher than those in post-mortem tweets. Additionally, analytical thinking is more expressed in post-mortem conversations than in pre-mortem conversations.

This experiment's significant contribution is the successful development of an incredibly high performing technique for automatically detecting deaths of users of social networking sites from the posts associated with their profiles. This technique is a potential solution that can be used as part of the tools required for the creation and adoption of an international standard for transferring digital estates to the next-of-kin of Internet users who die a sudden death; hence, reducing the risks of subscribers dying and leaving their digital estates that are supposedly important to their relatives or friends in the coffers of online-service providers.

\section{Future Work}
Although this experiment succeeded in training high performing machine learning models to classify post-mortem contents from their pre-mortem counterparts, and analysed and presented comprehensive differences between pre-mortem and post-mortem linguistic characteristics, there are still existing limitations in the methods adopted in the experiments. These limitations are used to propose future works that can be applied to improve the results obtained in this experiment.

\begin{enumerate}
    \item \textbf{Train classifiers on Tweets in all languages.} This experiment only collected tweets written in English and the text classifiers trained in this experiment can only classify post-mortem contents from their pre-mortem counterparts if they are written in English. This is why for now, this solution only serves as a step in the right direction, but there is still a need to collect post-mortem and pre-mortem text contents written using many more languages to be used to train machine learning classifiers that would be able to generalise well on texts that span many different languages.
    
    \item \textbf{Train machine learning classifiers to detect post-mortem language contextually.} There is need to train the machine learning classifiers to detect post-mortem language contextually, so that death is not reported only because of the presence of post-mortem words but because of how those post mortem words have been used. For example, the best model trained in this experiment would classify the text sample ``[subject]'s performance in his most recent series `RIP to him' is phenomenal. I love how he acted at the funeral of his child, and I will miss the series so much now that it has come to an end" as post-mortem because of the presence of the post-mortem words ``RIP", ``funeral", and ``miss"; however, a human would be able to detect that the text is only describing a scenario in which the ``[subject]" acted a series and not expressing the death of the ``[subject]".
    
    \item \textbf{Test developed classifiers on test sets collected from other social networking sites.} Although the text classification models trained in the experiments conducted in this research achieved outstanding performance in many different competitive metrics on text data collected from Twitter, there is a need to evaluate their performance on text data collected from other social networking platforms like Facebook, Instagram, etc.; this is to assess whether the trained machine learning classifiers would generalise well and remain accurate when evaluated on testing subset collected from a different platform from where the training and validation subsets were collected.
\end{enumerate}

\bibliography{others/refs}    % this causes the references to be listed

\bibliographystyle{IEEEtran}
%% the bibliography style determines the format  in which both citations and references are printed,
%% other possible values are plain and abbrv
%%
%% If you want more control of the format of your citations you might want to take a look at
%% natbib.sty, which should be part of any standard LaTeX installation
%%
%% University regulations simply require that your citation style be consistent, so see what style
%% your supervisor recommends.

% Appendices start here

\appendix
\chapter{Annotation Guideline}
\label{appendix:annotation guide}

\section{Introduction}
Social Networking Sites (SNSs), e.g. Twitter, Facebook, Instagram e.t.c. have continued to gain significant popularity from the late 20th through to the 21st century \cite{hillis2018digitalizing}. SNSs are used to grieve or disclose death because; they help to avoid the discomfort of having to individually announce the loss to every single person or tell the news over and over again \cite{gathman2014everybody}, allow others to partake in grief which helps in consoling those in grief \cite{dickinson2011shared}, and help the bereaved to feel that they are not alone \cite{katims2010grieving}.

This project aims to train machine learning models and use the appropriate Natural Language Processing pipelines to automatically detect the deaths of social media users from the posts associated with their profiles.

\section{Dataset Description}
The unit of analysis in this project will be a collection or aggregation of tweets that are related to a particular subject. The aggregation will be based on tweets that contain the Twitter handle of a particular subject between a given period. Tweets usually contain hashtags, username mentions, urls, emojis, e.t.c., that torments text-processing tools \cite{sproat2001normalization}; so, a cascade of text pre-processing steps have been performed to bring the texts into a form that is predictable and analyzable. Some of these steps include:

\begin{enumerate}
    \item removing all hashtags, urls, and XML/HTML tags;
    \item replacing emojis with their grammatical equivalents;
    \item replacing all subjects' Twitter usernames in a unit (collection of tweets) with a token \#SUBJECT and all other usernames with a token \#OTHERS; and
    \item using the special character ``\_*\_" to signify the end of a particular tweet within the unit.
\end{enumerate}

\begin{figure}[!ht]
    \includegraphics[width=16cm, height=12cm]{{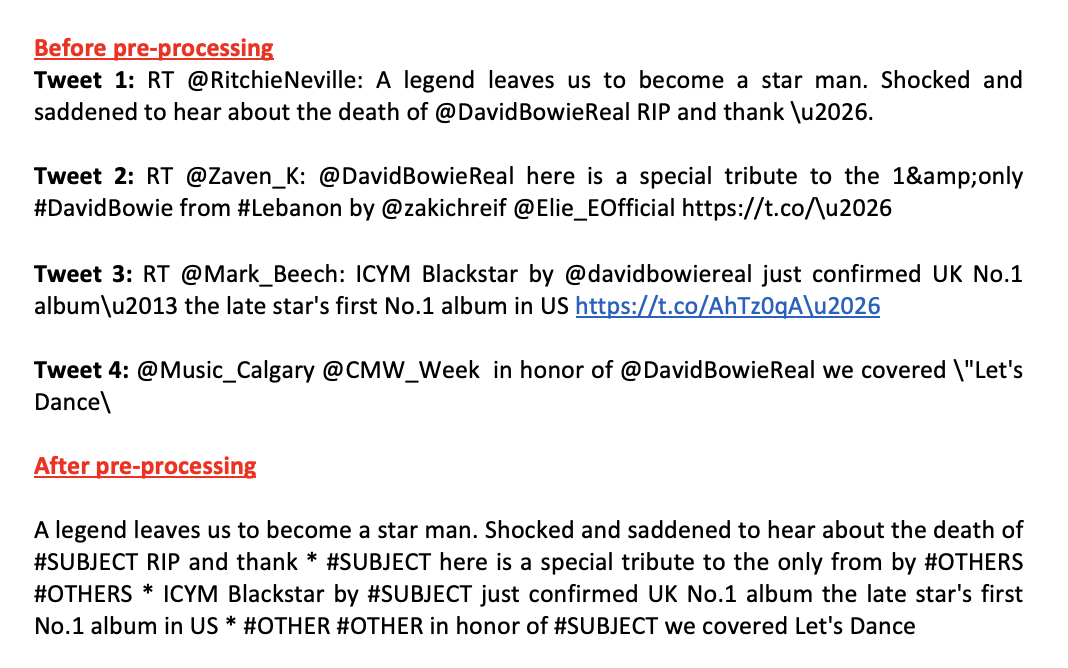}}
    \caption[Sample of a unit before and after preprocessing.]{Sample of a unit before and after preprocessing.}
    \label{fig:unit_sample}
\end{figure}

In addition, every unit will begin with a unique code which should not be considered during annotation as they are only added to ease their identification with subject owners after annotation. This is because all the tweets have been pre-processed and Twitter usernames have been masked. Figure \ref{fig:unit_sample} shows an example of a unit related to a particular subject before and after the text pre-processing steps are applied.

\section{Annotation Procedure}
\begin{figure}[!ht]
    \includegraphics[width=16cm, height=12cm]{{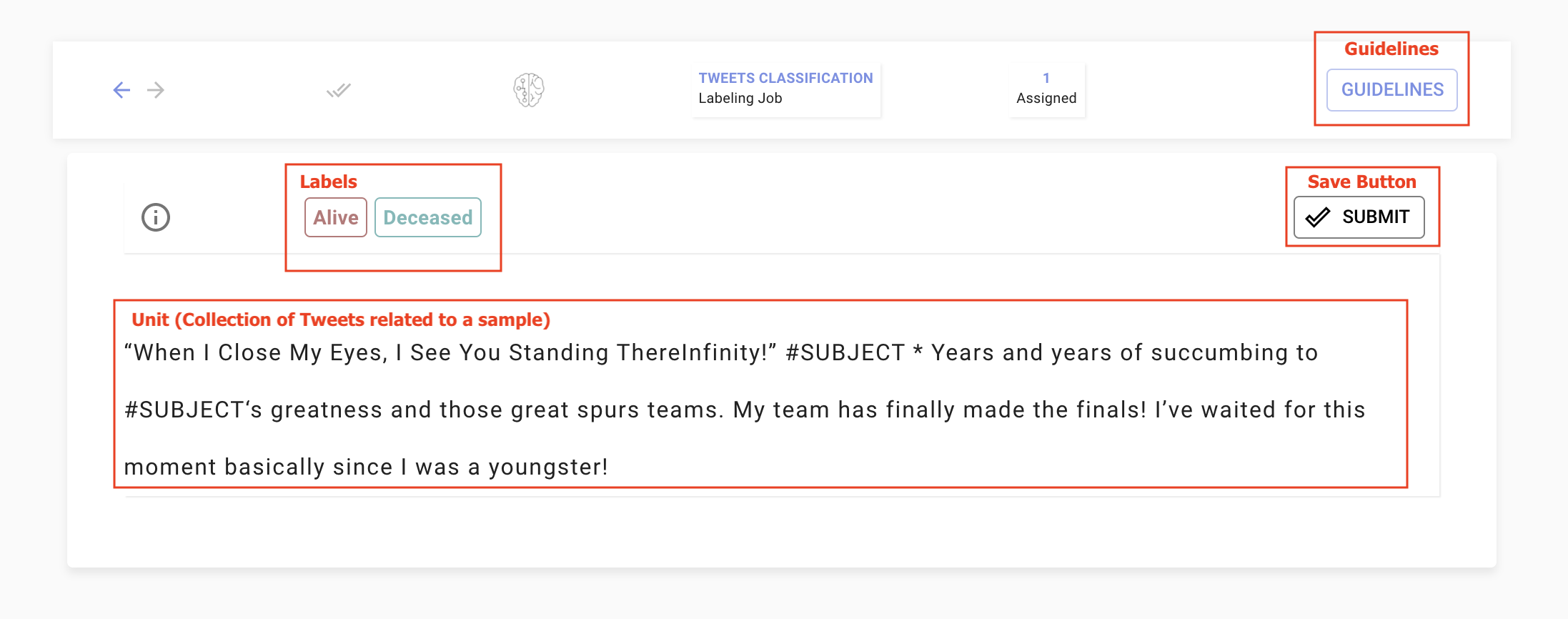}}
    \caption[Labelled user interface of the annotation tool, LightTag]{Labelled user interface of the annotation tool, LightTag }
    \label{fig:ligthtagsample}
\end{figure}

There are two labels, ``Alive" and ``Deceased":
\begin{enumerate}
    \item \textbf{Alive: }to be chosen when the annotator is convinced that the \#SUBJECT token as used in the text to represent the subject's Twitter username does not suggest that the subject is deceased.
    \item \textbf{Deceased: }to be chosen when the annotator is convinced that the \#SUBJECT token as used in the text to represent the subject's Twitter username obviously suggests that the subject is deceased.
\end{enumerate}
For each collection of tweets about a subject, annotators should first read the complete text in order to get an understanding and then choose either of the two labels, ``Alive" or ``Deceased" that best applies, from above the text, see figure \ref{fig:ligthtagsample}.\\\\
\textbf{Note: }In scenarios of doubt, i.e., when some tweets in a unit suggest that the subject is deceased and others suggest otherwise, annotators should always prioritise the ``Deceased" label. This means that the ``Alive" label should only be used when none of the tweets in the collection suggests that the subject is deceased.

\section{Recruitment Requirements}
Annotators need to only have a good understanding of the English Language. Knowledge of NLP is not required.
\end{document}